\documentclass[12pt]{report}
\usepackage{epsf}
\def\ltap{\raisebox{-.4ex}{\rlap{$\sim$}} \raisebox{.4ex}{$<$}}
\def\gtap{\raisebox{-.4ex}{\rlap{$\sim$}} \raisebox{.4ex}{$>$}}

\def\bea{\begin{eqnarray}}
\def\eea{\end{eqnarray}}
\def\beq{\begin{equation}}
\def\eeq{\end{equation}}

\def\[{\left[}
\def\]{\right]}
\def\({\left(}
\def\){\right)}

\makeatletter
 
\if@twoside
\else
\fi
 
\leftmargin \leftmargini

\def\singlespace{
  \let\oldstretch=\baselinestretch
  \def\baselinestretch{1}
  \@normalsize
}
\def\doublespace{
  \let\oldstretch=\baselinestretch
  \def\baselinestretch{2.0}
  \@normalsize
}
\def\oldspacing{
  \let\baselinestretch=\oldstretch
  \@normalsize
}

\if@twoside
  \def\scchapter#1{\markboth{\sc \@chapapp\ \thechapter. \ #1}{}}
  \def\scsection#1{\markright{\sc \thesection. \ #1}}
  \def\rightmarginpagenumber{\rlap{\kern 30pt \rm\thepage}}
  \def\leftmarginpagenumber{\llap{\rm\thepage \kern 30pt}}
  \def\numbertop{
    \gdef\ps@plain{
      \def\@oddhead{\hfil\rightmarginpagenumber}
      \def\@oddfoot{}
      \def\@evenhead{\leftmarginpagenumber\hfil}
      \def\@evenfoot{}
    }
    \gdef\ps@headings{
      \let\chaptermark=\scchapter
      \let\sectionmark=\scsection
      \def\@oddhead{\hfil\rightmark\hfil\rightmarginpagenumber}
      \def\@oddfoot{}
      \def\@evenhead{\leftmarginpagenumber\hfil\leftmark\hfil}
      \def\@evenfoot{}
    }
  }
  \def\numberbottom{
    \gdef\ps@plain{
      \def\@oddhead{}
      \def\@oddfoot{\hfil\rm\thepage\hfil}
      \let\@evenhead=\@oddhead
      \let\@evenfoot=\@oddfoot
    }
    \gdef\ps@headings{
      \let\chaptermark=\scchapter
      \let\sectionmark=\scsection
      \def\@oddfoot{\hfil\rm\thepage\hfil}
      \let\@evenfoot=\@oddfoot
    }
  }
\else
  \def\smallcapsheading#1{\markboth{\sc \@chapapp\ \thechapter. \ #1}{}}
  \def\marginpagenumber{\rlap{\kern 30pt \rm\thepage}}
  \def\numbertop{
    \gdef\ps@plain{
      \def\@oddhead{\hfil\marginpagenumber}
      \def\@oddfoot{}
    }
    \gdef\ps@headings{
      \let\chaptermark=\smallcapsheading
      \def\@oddhead{\hfil\leftmark\hfil\marginpagenumber}
      \def\@oddfoot{}
    }
  }
  \def\numberbottom{
    \gdef\ps@plain{
      \def\@oddhead{}
      \def\@oddfoot{\hfil\rm\thepage\hfil}
    }
    \gdef\ps@headings{
      \let\chaptermark=\smallcapsheading
      \def\@oddfoot{\hfil\rm\thepage\hfil}
    }
  }
\fi
 
\def\@makechapterhead#1{             
  {
    \centering 
    \Large\bf\@chapapp{} \thechapter 
    \par 
    \vskip 20pt                      
    \Large\bf#1\par                  
    \nobreak                         
    \vskip 40pt                      
  }
}
\def\@makeschapterhead#1{            
  {
    \centering
    \Large\bf#1\par                  
    \nobreak                         
    \vskip 40pt                      
    \markboth{\sc #1}{\sc #1}
  }
}
\def\section{
  \@startsection
    {section}{1}{\z@}{-3.5ex plus-1ex minus-.2ex}{2.3ex plus.2ex}{\large\bf}
}
 
\def\tableofcontents{
  \@restonecolfalse
  \if@twocolumn \@restonecoltrue \onecolumn \fi
  \chapter*{Contents}
  \@starttoc{toc}
  \if@restonecol \twocolumn \fi
}
 
\def\@begintheorem#1#2{
  \list{}{\leftmargin 0pt\labelwidth 0pt}\item[]{\bf #1\ #2. }%
}
\let\@endtheorem=\endlist
 
\def\l@figure{\@dottedtocline{1}{0pt}{2.3em}}
\let\l@table=\l@figure
 
\def\@ident#1{#1\endgroup}
\def\ident{\begingroup \tt \catcode`_=11 \catcode`^=11 \@ident}
 
\def\l@subsection#1#2{}
 
\def\@afterheading{}
 
\makeatother

\def\preliminaries{
  \clearpage
  \pagenumbering{roman}
  \numberbottom
}
\def\endpreliminaries{
  \clearpage
  \pagenumbering{arabic}
 \numberbottom
  \pagestyle{plain}
}
 
\doublespace
\input epsf

\preliminaries

\begin{document}

\begin{titlepage}
\begin{center}
\hfill    LBNL-41874 \\
\hfill    UCB-PTH-98/29 

\vskip 0.03in

{\large \bf 
Naturalness and Supersymmetry}\footnote{This work was supported in part
by the Director, Office of Energy
Research, Office of High Energy and Nuclear Physics, Division of High
Energy Physics of the U.S. Department of Energy under Contract
DE-AC03-76SF00098 and in part by the National Science Foundation
under grant PHY-90-21139, and also by the Berkeley Graduate Fellowship.}
\vskip .1in
(Ph.D. Dissertation, University of California, Berkeley, May 1998)  
\vskip .15in
Kaustubh Agashe \footnote{current email: agashe@oregon.uoregon.edu}\\
\vskip .1in
{\em
    Theoretical Physics Group\\
    Lawrence Berkeley National Laboratory\\
    University of California,
    Berkeley, California 94720 \\}
and \\
{\em
    Department of Physics \\
    University of California,
    Berkeley, California 94720}

\end{center}

\vspace{-2in}

\begin{abstract}
In the Standard Model of elementary particle physics,
electroweak symmetry breaking 
is achieved by a Higgs scalar doublet
with a negative (mass)$^2$.
The Standard Model
has the well known gauge hierarchy problem: quadratically divergent
quantum corrections drive the Higgs mass and thus the weak scale
to the scale of 
new physics. Thus, if the scale of new physics is say the Planck scale,
then correct electroweak 
symmetry breaking requires a fine tuning between the 
bare Higgs mass and the quantum corrections.

Supersymmetry, a symmetry between
fermions and bosons,
solves the gauge hierarchy problem of the Standard Model:
the quadratically divergent corrections to
the Higgs mass cancel between fermions and bosons.
The remaining corrections to
the Higgs mass are proportional to the supersymmetry
breaking
masses for the supersymmetric partners (the {\it s}particles) of the
Standard Model particles.
The large top quark Yukawa coupling
results in a negative Higgs $(\hbox{mass})^2$.
Thus, 
electroweak symmetry breaking 
occurs naturally
at the correct scale
if the masses of 
the sparticles
are close to the weak scale.
  
In this thesis, we argue that the supersymmetric Standard Model, 
while avoiding the fine tuning in electroweak
symmetry breaking,
requires unnaturalness/fine
tuning in some (other) sector of the theory.
For example, Baryon and Lepton number
violating operators are allowed which lead to proton
decay and flavor changing neutral currents. We study
some of the constraints from the latter in this thesis.
We have to impose an $R$-parity for the theory to be both natural
and viable. 

In the absence of flavor symmetries, the 
supersymmetry breaking masses for the squarks and sleptons
lead to too large flavor changing neutral currents.
We show that two of the solutions to
this problem, gauge mediation of supersymmetry breaking
and making the scalars of the first two generations
heavier than a few TeV,
reintroduce fine tuning in electroweak symmetry breaking.
We also construct a
model of low energy gauge mediation
with a non-minimal messenger sector
which improves the
fine tuning and 
also generates required
Higgs mass terms.
We show that this model can be derived from
a Grand Unified Theory despite the non-minimal spectrum.

\end{abstract}
\end{titlepage}

\newpage



\tableofcontents

\listoffigures

\listoftables

\newpage
\section*{Acknowledgements}
I would like to express my sincere gratitude to my advisors,
Professor Mahiko Suzuki and Dr. Ian Hinchliffe, for their guidance,
help and support. I appreciate the freedom they gave me in choosing
research topics. I thank my fellow student, Michael Graesser,
for many wonderful collaborations and discussions.

I would also like to thank other members of the Theoretical
Physics group, especially Professor Hitoshi Murayama, Nima Arkani-Hamed,
Chris Carone, Takeo Moroi, John Terning, Csaba Cs\'aki and Jonathan Feng
for useful
discussions and the whole group for a pleasant
experience of
being at LBNL. I am grateful to Anne Takizawa and Donna Sakima of the
Physics department and Luanne Neumann, Barbara Gordon and Mary Kihanya
at LBNL for help with administrative work.

I am indebted to my roommates and other friends for making my stay
at Berkeley thoroughly enjoyable. I thank my parents and my brother
for their support and encouragement. 

\endpreliminaries

\chapter{Introduction}
A Standard Model (SM) \cite{smref,GIM} of elementary 
particle physics 
has developed over the last 
twenty five years
or so.
It describes the
interactions of the elementary particles using gauge theories.
The elementary particles are the matter fermions (spin half particles) 
called the
quarks and the leptons,
and the gauge bosons (spin one particles) which are the carriers
of the interactions.
There are three generations, with 
identical quantum numbers, of quarks and leptons:
up ($u$) and  down ($d$) quarks, electron ($e$) 
and it's neutrino ($\nu$) (the leptons)
in the first generation, charm ($c$) and  strange ($s$) quarks, muon ($\mu$)
and it's neutrino in the second,
and top ($t$) and  bottom 
($b$) quarks, tau ($\tau$) lepton and it's neutrino in the third.
The $W$, $B$ (the hypercharge gauge boson)
 and the gluon ($g$) are the gauge bosons. 
There is also one Higgs scalar.
The particle content of the SM is summarized in Table \ref{partcont}.

\renewcommand{\arraystretch}{0.3}
\begin{table}
\begin{center}
\begin{tabular}{||c|c||c|c|c||}\hline 
 & & & &  \\
particle & sparticle & $SU(3)_c$ & $SU(2)_w$ & $U(1)_Y$ \\ 
 & & & &  \\
\hline
 & & & &  \\
$\left(
\begin{array}{c}
u \\
d \\
\end{array} \right) _i$
&
$\left(
\begin{array}{c}
\tilde{u} \\
\tilde{d} \\
\end{array} \right) _i$
& ${\bf 3}$ & ${\bf 2}$ & $\frac{1}{6}$  \\ 
 & & & & \\  
\hline 
 & & & & \\
$u^c_i$ & $\tilde{u}^c_i$ & ${\bf \bar{3}}$  
&  $ {\bf 1}$  & $ - \frac{2}{3}$   \\ 
 & & & & \\  \hline 
 & & & & \\
$d^c_i$ & $\tilde{d}^c_i$ & 
${\bf \bar{3}}$ &  ${\bf 1}$ &  $\frac{1}{3}$  \\ 
 & & & & \\  \hline 
 & & & & \\
$\left(
\begin{array}{c}
\nu \\
e \\
\end{array} \right) _i$
&
$\left(
\begin{array}{c}
\tilde{\nu} \\
\tilde{e} \\
\end{array} \right) _i$
& ${\bf 1}$ & ${\bf 2}$ & $-\frac{1}{2}$  \\ 
 & & & & \\  \hline 
 & & & & \\
$e^c_i$ & $\tilde{e}^c_i$ & ${\bf 1}$ &  ${\bf 1}$ &  $1$  \\ 
 & & & & \\  \hline 
 & & & & \\
$W$ & $\tilde{W}$ & ${\bf 1}$ & ${\bf 3}$ & $0$ \\
 & & & & \\  \hline
 & & & & \\
$g$ & $\tilde{g}$ & ${\bf 8}$ & ${\bf 1}$ & $0$ \\
 & & & & \\  \hline
 & & & & \\
$B$ & $\tilde{B}$ & ${\bf 1}$ & ${\bf 1}$ & $0$ \\
 & & & & \\  \hline
 & & & & \\
$\left(
\begin{array}{c}
H_u^+\\
H_u^0\\
\end{array} \right) $
&
$\left(
\begin{array}{c}
\tilde{H}_u^+ \\
\tilde{H_u^0} \\
\end{array} \right) $
& ${\bf 1}$ & ${\bf 2}$ & $\frac{1}{2}$  \\  
 & & & & \\  \hline
 & & & & \\
$\left(
\begin{array}{c}
H_d^0\\
H_d^-\\
\end{array} \right) $
&
$\left(
\begin{array}{c}
\tilde{H}_d^0 \\
\tilde{H_d^-} \\
\end{array} \right) $
& ${\bf 1}$ & ${\bf 2}$ & $ - \frac{1}{2}$  \\
 & & & & \\ \hline 
\end{tabular}
\end{center}
\caption{The particle content of the SM (left column)
and it's supersymmetric extension (the sparticles). 
The fermions are left-handed
Weyl spinors. So, $e^c$ stands for the left-handed positron which is
the antiparticle of the right-handed electron. $i=1,2,3$ denotes the 
generation, for example, $u_3$ is the top $(t)$
quark and $e^c_2$ is the anti-muon ($\bar{\mu}$). 
The electric charge is given by $Q= T_3 + Y$, where
$T_3$ is the third component of the $SU(2)_w$ isospin and $Y$ is 
the hypercharge.}
\label{partcont} 
\end{table}

The gauge theory of the interactions of the quarks,
Quantum Chromodynamics (QCD) \cite{QCD}, is based on the gauge group
$SU(3)_c$ where the ``c'' stands for ``color'' which is the charge
under QCD in analogy to 
electric charge. The interaction
is mediated by eight massless gauge bosons called gluons.
This theory is asymptotically free, {\it i.e.}, it
has the property that it's gauge coupling becomes
weak at high energies (much larger than $\sim 1$ GeV)
and becomes strong at energies below $\sim 1$ GeV. Thus,
at low energies the theory confines, {\it i.e.}, the strong interactions
bind the quarks into color singlet
states called hadrons, for example the proton
and the pion. 
So, we observe only these bound states of quarks and not
the elementary quarks.
However, when the proton is probed at 
high energies (large momentum transfers)
or when the quarks are produced in high energy collisions, 
the quarks should behave as if they do not feel the strong
interactions. 
This is indeed confirmed in a large number of
experiments 
at high energies (see, for example, review of QCD in \cite{pdg}).

The weak and electromagnetic interactions of quarks and leptons are
unified into the electroweak theory based on the gauge group
$SU(2)_w \times U(1)_Y$ \cite{smref}. This theory has four gauge bosons.
This electroweak symmetry is broken to the $U(1)$
of electromagnetism (Quantum Electrodynamics, QED). 
Three of the gauge bosons 
(called the $W$ and $Z$ gauge bosons)
get a mass in this process whereas the photon (the carrier
of electromagnetism) is massless. 
The theory predicts the relations between the $W$ and $Z$
masses and couplings of
quarks and leptons to these gauge bosons.\footnote{
We assume that the mechanism for the symmetry
breaking has a custodial
$SU(2)$ symmetry.} The stringent tests of these
predictions
at the electron-positron collider at CERN (LEP)
and 
at the proton-antiproton collider at Fermilab 
(up to energies of a few $100$ GeV)
have been highly successful.

One of the central issues of particle physics today is the
mechanism of Electroweak Symmetry Breaking (EWSB), {\it i.e.},
how is $SU(2)_w \times U(1)_Y$ broken to $U(1)_{em}$?
In the SM, this is achieved by the Higgs scalar, $H$,  which is
a doublet of $SU(2)_w$. The Higgs scalar has the following
potential:
\begin{equation}
V_{Higgs} = m^2 \left|H\right|^2 + \lambda \left|H\right|^4.
\label{vhiggs}
\end{equation} 
If $m^2 <0$, then at the minimum of the potential,
the Higgs scalar acquires a vacuum expectation value (vev):
\begin{equation}
\langle H \rangle = \left( \begin{array}{c}
v \\
0 
\end{array} \right),
\end{equation}
where $v = \sqrt{-m^2/(2 \lambda)}$.
Thus two of the generators of the $SU(2)_w$ gauge group
and also one combination of the third $SU(2)_w$ generator
and $U(1)_Y$ are broken. The corresponding gauge bosons acquire
masses given by $\sim g_2 \; v
$ and $\sim \sqrt{g_2^2 + g_Y^2} \; v$
respectively and are the $W$ and the $Z$.
The Higgs vev and thus, if $\lambda \sim O(1)$, the 
mass parameter $m^2$ has to be of the order of ($100$
GeV)$^2$ to give the experimentally measured $W$ and $Z$ gauge boson
masses.
The other combination of the third $SU(2)_w$ generator
and $U(1)_Y$ is still a good symmetry and the corresponding
gauge boson is massless and is the photon ($\gamma$).
There is also a physical electrically neutral Higgs scalar left
after EWSB. This is the only particle of the SM which has not
been discovered.

To generate masses for the quarks and leptons, we add the following
Yukawa couplings (the quark and lepton $SU(2)_w$ doublets
are denoted by $q$ and $l$ and $i,j$ are generation indices):
\begin{equation}
{\cal L}_{Yukawa} = \lambda^u_{ij} H q_i u^c_j +  \lambda ^d _{ij} 
H^{\dagger} q_i d^c_j
+ \lambda ^l _{ij} H^{\dagger} l_i e^c_j,
\label{yukawa}
\end{equation}
where repeated indices are summed over.
These couplings 
become mass terms for the fermions when the Higgs develops a vev.
There are 13 physical parameters in the above Lagrangian:
6 masses for the quarks, 3 masses for the leptons and 3 mixing angles
and a phase in the quark sector. The 3 mixing angles and the phase
appear at the $W$ vertex involving the quarks and constitute
the $3 \times 3$ matrix called the Cabibbo-Kobayashi-Maskawa 
(CKM) matrix \cite{kobayashi}. These 13 parameters can be, a priori, arbitrary
and are fixed only by measurements of the quark and lepton masses
and the mixings (the latter using decays of quarks through
a virtual $W$). 
In the SM, processes involving 
conversion of one flavor of 
quark into another flavor with the same electric
charge,
for example, conversion of a strange quark into a down quark
resulting in mixing between the $K$-meson and it's antiparticle,
do not occur at tree level, but
occur at one loop due to the mixings. The experimental observations of these
flavor changing neutral currents (FCNC's) are consistent with
the mixing angles (as measured using decays of quarks).
Since there is no right-handed neutrino in the SM, we cannot
write a Dirac mass term for the neutrino and at the renormalizable
level, we cannot write a Majorana mass term since we do not have a 
$SU(2)_w$ triplet Higgs. So,
neutrinos are massless in the SM.
\footnote{There is some evidence for non-zero neutrino masses,
but it is not conclusive.}
This results in conservation laws for the individual
lepton numbers, {\it i.e.,} electron, muon and tau numbers.
Thus, the FCNC decay, $\mu \rightarrow e\; \gamma$ is forbidden in the
SM and the experimental limits on such processes are indeed
extremely small \cite{pdg}. 

The SM, thus, seems to describe the observed
properties of the elementary particles
remarkably well, up to energies $\sim$ few $100$ GeV.
Of course, the Higgs scalar remains to be found.
But, the SM has some aesthetically unappealing features which
we now discuss.

The SM particle content and gauge group naturally raise
the questions: Why are there three gauge groups (with  
different strengths for the couplings) and three generations
of quarks and leptons 
with the particular quantum numbers? Attempts have been made to simplify
this structure by building Grand Unified Theories (GUT's).
The gauge coupling strengths depend on the energy/momentum
scale at which they are probed (this was already mentioned for QCD 
above). In the GUT's it is postulated that these three couplings
are equal at some very high energy scale called the GUT scale
so that at that energy scale the three gauge groups can be
embedded into one gauge group with one coupling constant. The
GUT gauge group gets broken at that scale to the SM gauge groups
resulting in different evolutions for the three gauge couplings
below the GUT scale.
Also, in the GUT's, the quarks and leptons can be unified into
the same representation of the gauge group. In the simplest
GUT, based on the $SU(5)$ gauge group \cite{splitting}, the
$d^c$ and the lepton doublet ($l$) form an
anti-fundamental $({\bf \bar{5}})$ under the gauge group.
The Higgs doublets are in a 
${\bf 5}$ representation of $SU(5)$ and so have $SU(3)_c$
triplet partners
which are required to be heavy since they mediate proton decay 
\cite{splitting}.
When the three coupling constants were measured in the late 1970's,
and evolved with the SM particle content to high energies,
they appeared to
meet at an energy scale of
$\sim 10^{14}$ GeV \cite{quinn}. 
But, the more accurate measurements in the 1990's
show that this convergence is not perfect \cite{langacker}.

The 13 parameters of the Yukawa Lagrangian of Eqn.(\ref{yukawa})
exhibit hierarchies or patterns, for example the ratio of the mass
of the heaviest (top) quark and the lightest lepton (electron) is 
about $10^{-6}$. 
One would like to have a more fundamental theory of these
Yukawa couplings which can explain these hierarchies
in terms of fewer parameters. A GUT can make some progress in this direction
by relating the quark masses to the lepton masses since they are in 
the same representation of the GUT group \cite{splitting}. For example 
in many GUT's we get the relation
$m_b = m_{\tau}$.

Perhaps the most severe ``problem'' of the SM is the gauge hierarchy
problem \cite{susskind} which we now explain.
It concerns the Higgs mass parameter,
$m^2$, of Eqn.(\ref{vhiggs}). There are two issues here.
The first issue is the origin of this mass parameter.
As mentioned above, $m^2 \sim (100$ GeV)$^2$.
We would like to have one ``fundamental'' mass scale in our theory
and ``derive'' all other mass scales from this scale.
Particle physicists like to think that this scale should
be the Planck scale, $M_{Pl} \sim 10^{18}$ GeV, which is the scale at which
the gravitational interactions have to be quantized.
There is one other scale in the SM besides
the Higgs mass parameter.
It is the strong interaction scale of QCD denoted by $\Lambda _{QCD}$.
Naively, this is the scale at which the $SU(3)_c$ coupling constant
becomes strong binding quarks into hadrons. Thus, this scale
can be related to the Planck scale and the $SU(3)_c$ coupling constant
at the Planck scale by the logarithmic Renormalization Group (RG)
evolution of the gauge coupling as follows:
\begin{equation}
\Lambda _{QCD} \sim  M_{Pl} \exp \left(-\frac{8 \pi ^2}{g^2(M_{Pl})} \right).
\end{equation}
This relation is valid, strictly speaking, at one loop.
Thus, if
$g(M_{Pl}) \ltap 1$, there is a natural explanation for the hierarchy
$\Lambda _{QCD}/M_{Pl}$. We would like to have a similar
explanation for the hierarchy $m/M_{Pl}$.

\begin{figure}
\centerline{\epsfxsize=.8\textwidth \epsfbox{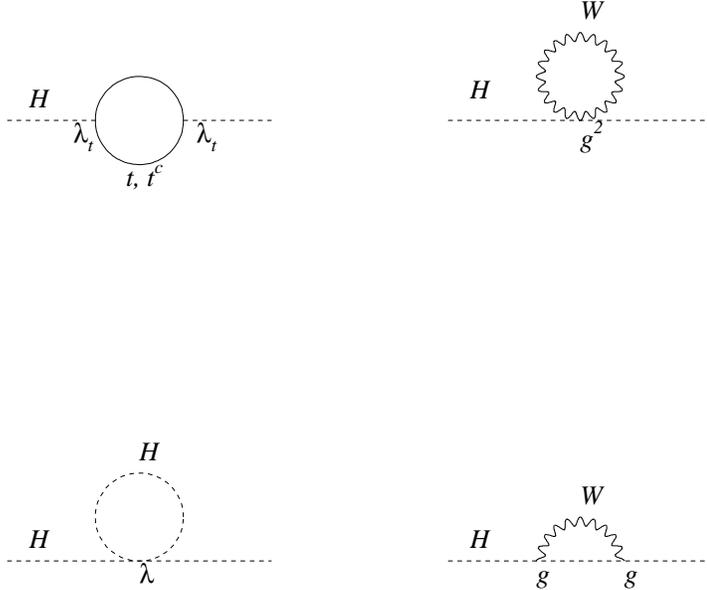}}
\caption{The Feynman diagrams which give quadratically
divergent contributions to the Higgs mass in the SM.}
\vspace{0.5in}
\protect\label{sm}
\end{figure}

The second issue is whether the mass scale $m$ is stable to quantum
corrections. In the SM, the Feynman diagrams in Fig.\ref{sm}
give 
quadratically divergent contributions to $m^2$, since
the corresponding integrals over the loop momentum $k$
are $\sim \int d^4k / (k^2 - m^2)$.
The corrections due to the top quark in the loop
are important due to the large Yukawa coupling of the
top quark.
Thus, the renormalized Higgs mass parameter is given by:
\begin{equation}
m^2_{ren.} \sim m^2_{bare} + \frac{1}{16 \pi^2} \Lambda ^2,
\label{qdigt}
\end{equation}
for all dimensionless couplings of order one.
$\Lambda$ is the cut-off for the quadratically divergent integral.
We know that the SM cannot describe quantum gravity. Thus, we certainly
expect some new physics (string theory?) at $M_{Pl}$. There could, of course,
be some new physics at lower energy scales as well, for example the GUT scale.
In some such extension to the SM, it turns out that the
scale $\Lambda$ is the scale of new physics. 
Thus, in the SM, the Higgs mass gets driven due to quantum
corrections all the way to some high energy scale
of new physics (see Eqn.(\ref{qdigt})).
We need $m^2_{ren.} \sim (100$ GeV$)^2$ so that EWSB occurs
correctly. 
We can achieve this by a cancellation between
$m^2_{bare}$ and the quantum corrections, 
which is of the order of one part in
$\Lambda ^2 / (100$ GeV$)^2$. For $\Lambda = M_{Pl}$, this is 
enormous. 
Thus, in the SM, the bare Higgs mass parameter has to 
be fine tuned to give the correct $W$ and $Z$ masses.
Such a problem does not occur for dimensionless couplings,
since the quantum corrections are proportional to the logarithm of the
cut-off or for fermion masses which are protected by
chiral symmetries. 

Supersymmetry (SUSY) \cite{nilles} provides a solution to the gauge hierarchy
problem of the SM. SUSY is a symmetry between fermions and bosons,
{\it i.e.}, a Lagrangian is supersymmetric if it is invariant under
a (specific) transformation between fermions and bosons.
In particular, the fermion and the boson in a
representation of the SUSY algebra have the same 
interactions.
So, to make the SM supersymmetric, we add to the SM particle content
fermionic (spin half) partners of the gauge bosons called 
``gauginos'' (for example
the partner of the gluon is the gluino) and scalar (spin zero) partners of 
the quarks and leptons called ``squarks'' and ``sleptons'',
respectively (for
example selectron is the partner of the electron). Similarly,
the fermionic partners of the Higgs scalars are called Higgsinos.
We have to add another 
Higgsino doublet (and a Higgs scalar doublet) 
to cancel
the $SU(2)_w^2 \times U(1)_Y$ anomaly.
We denote the superpartners by a tilde over the corresponding
SM particle. 
The supersymmetric SM (SSM) has the particle content shown in
Table \ref{partcont}.
The irreducible representation
of the SUSY algebra containing a matter fermion and it's scalar partner
is called a chiral superfield. 
We denote
the components of a chiral superfield
by lower case letters and the superfields
by upper case letters except for the Higgs (and in some cases for 
other fields
which acquire a vev)
for which both 
the superfield and components are denoted by upper case letters.\footnote{The
chiral superfields appear in the ``K\"ahler'' potential and the 
``superpotential'' (to be defined later) and the component fields
appear in the Lagrangian.}
The Yukawa couplings for the fermions
can be written in a supersymmetric way in terms of a ``superpotential'':
\begin{equation}
W = \lambda ^u _{ij} H_u Q_i U^c_j + \lambda ^d _{ij} H_d Q_i D^c_j
+ \lambda ^l _{ij} H_d L_i E^c_j + \mu H_u H_d,
\label{WMSSM}
\end{equation}
where $Q$ and $L$ are the quark and lepton $SU(2)_w$ doublets.
The $\mu$ term is a mass term for the Higgs doublets.
The superpotential gives the following terms in the Lagrangian:
\begin{eqnarray}
{\cal L} & = &\sum_i \left |
\frac{\partial W}{\partial \Phi_i}\right |^2_{\Phi = \phi}
+ \left. \sum_{i,j}
\psi_i \psi_j \frac{\partial W}{\partial \Phi_i} 
\frac{\partial W}{\partial \Phi_j}
\right| _{\Phi = \phi} + \; \hbox{h.c.} \nonumber \\ 
 & & \hbox{(where} \; \phi 
\; \hbox{and} \; \psi \; \hbox{are scalar and fermionic components of}
\; \Phi).
\label{LfromW}
\end{eqnarray}
Thus, to get a term in the Lagrangian with fermions
from a term of the superpotential,
we pick fermions from two of the chiral superfields and scalars from
the rest (if any).
This gives the Yukawa couplings of Eqn.(\ref{yukawa}).
We get the following terms 
in the scalar
potential from the first term of Eqn.(\ref{LfromW}):
\begin{equation}
V  =  \mu^2 (\left| H_u \right|^2 + \left| H_d \right|^2) + 
\lambda _t^2 \left|
H_u \right| ^2(\left|
\tilde{q}_t \right|^2 + 
\left| \tilde{t}_c \right|^2) + ....
\end{equation}

SUSY requires that the hermitian conjugates of the chiral superfields
(anti-chiral superfields) cannot appear in the superpotential.
Thus, we cannot use $H_u^{\dagger}$ in Eqn.(\ref{WMSSM}) 
to give mass to the down quarks.
This is another reason for adding the second Higgs doublet.
For the same reason, the $\mu$ term 
is the only gauge invariant mass term for the Higgs chiral superfields
and we cannot write down a term in the superpotential which will
give a quartic Higgs scalar term in the Lagrangian.

In addition to the terms from the above superpotential, there are
kinetic terms for gauge fields (which can also be derived from a
superpotential) and 
kinetic terms for
matter fields which can be written in a supersymmetric 
and gauge invariant way in
terms
of a ``K\"ahler'' potential: $\sum_{\Phi} \Phi^{\dagger} e^V \Phi$
(where $V$ is the gauge multiplet). 
The K\"ahler potential and the gauge superpotential
generate two kinds of terms in the
SSM (in the supersymmetric limit) which are relevant for us. 
The first one is a coupling between a matter fermion, a gaugino and the
scalar partner of the fermion, for example, a quark-squark-gluino
coupling, {\it i.e.,} $q \tilde{q}^{\dagger} \tilde{g}$. 
The second term is called the $D$-term which gives a quartic 
coupling between the scalars proportional to the gauge coupling squared:
$\sum_a g^2/2 \left( \sum_{\phi} \phi^{\dagger} T^a \phi \right) ^2$
for each gauge group, where $T^a$ is a generator of the gauge group.
This gives, in particular, a quartic term for the Higgs scalars.

\begin{figure}
\vspace{-.6in}
\centerline{\epsfxsize=.75\textwidth \epsfbox{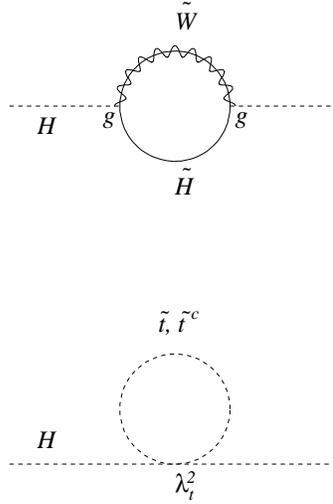}}
\vspace{-1.1in}
\caption{The Feynman diagrams in SSM
which cancel the quadratic
divergences of the SM contributions to the Higgs mass.}
\vspace{0.4in} 
\protect\label{ssm}
\end{figure}

We now discuss how SUSY solves the gauge hierarchy problem.
In a supersymmetric theory, there is a cancellation between
fermions and bosons in the quantum corrections since a Feynman diagram
with an internal fermion has an opposite sign relative to the one with an 
internal boson. Thus, there is a non-renormalization theorem
in a supersymmetric theory which says that the superpotential terms
are not renormalized \cite{nonrenorm}.
This means that the mass term for the Higgs, the $\mu$ term, does not
receive any corrections in the supersymmetric limit. In other words,
due to supersymmetry, the chiral symmetry protecting the Higgsino
mass also protects the Higgs scalar mass.
The quantum corrections due to the Feynman diagrams of Fig.\ref{sm}
are exactly cancelled by their supersymmetric analogs, Fig.\ref{ssm}.
It is crucial for this cancellation that the quartic
interaction of the Higgs scalars is given by the gauge coupling since
it is due to the $D$-terms mentioned above, {\it i.e.,} the quartic
coupling $\lambda \sim g^2$ in the SSM.

We know that SUSY cannot be an exact
symmetry of nature since we have not
observed a selectron degenerate with the electron. So, we add SUSY breaking
terms to the Lagrangian which give a large mass to the unobserved
superpartners (gauginos, sleptons and squarks) and the Higgs scalars:
\begin{equation}
{\cal L}_{SUSY breaking} = \sum_{i,j} 
m_{ij}^2 \phi^{\dagger}_i \phi_j + 
B \mu H_u H_d + 
\sum_A M_A \lambda^A_a \lambda^A_a
\end{equation}
where $\phi _i$ denotes a scalar and $\lambda^A_a$ a gaugino
of the gauge group $A$
and $B\mu$ is a SUSY breaking mass term for the Higgs scalars.\footnote{These
terms, along with the trilinear scalar terms, $A \phi_i\phi_j\phi_k$, break
SUSY softly, {\it i.e.,} do not reintroduce quadratic divergences.}
Since SUSY is broken, {\it i.e.,}
fermions and their partner bosons no longer have the same mass,
the cancellation between fermions and bosons
in the quantum corrections to the Higgs masses is no longer exact.
The quadratically divergent corrections to the Higgs masses still cancel
(between the diagrams of Fig.\ref{sm} and Fig.\ref{ssm}),
but the logarithmically divergent corrections do not and are
proportional to the SUSY breaking masses.
This gives:
\begin{equation}
m^2_{H_u,ren.} \sim m^2_{H_u,bare} + \left(
-\frac{\lambda _t^2}{16 \pi^2} 
m_{\tilde{t}}^2 + \frac{g^2}{16 \pi^2} M^2 \right) \log \Lambda, 
\end{equation}
where the first one loop correction on the right is due
top squarks and the second is due to gauginos.
Here, $\Lambda$ is the scale at which the SUSY breaking
masses are generated. Even if it is the Planck scale,
the logarithm is $O(10)$.
It turns out that for a large part
of the parameter space, the Higgs (mass)$^2$ renormalized at the weak scale
is negative due to the stop contribution
($\lambda _t$ is larger than
$g$)
and is of the order
of the stop (mass)$^2$ \cite{ross,wise}. The down type Higgs (mass)$^2$ is also 
negative if the bottom Yukawa coupling is large.
The Higgs scalar potential is:
\begin{eqnarray}
V_{Higgs} & = & (m_{H_u}^2 + \mu ^2 ) \left|H_u\right|^2 + 
(m_{H_d}^2 + \mu^2 )  \left|H_d\right|^2
\nonumber \\
 & & - B \mu H_u H_d + 
\frac{g^2_Z}{8} \left( \left|H_u\right|^2 - 
\left| H_d\right|^2 \right)^2 \; \;(D-\hbox{terms}),
\end{eqnarray}  
where $g_Z^2 = g_2^2 + g_Y^2$.
Using this potential, we can show that 
the negative Higgs (mass)$^2$
results, for a large part of the parameter space,
in a vev for both the Higgs doublets, breaking electroweak
symmetry. 
In particular, the $Z$ mass is:
\begin{equation}
\frac{1}{2} m_Z^2 = - \mu^2 + \frac{m_{H_u}^2 \tan ^2 \beta - m_{H_d}^2}
{1 - \tan ^2 \beta},
\end{equation}
where $\tan \beta = v_u/v_d$ is the ratio of vevs for the two Higgs scalars.
Thus, to get the correct $Z$ mass, we need the $\mu$ term and the
renormalized Higgs masses (and in turn the stop mass) to be of the order
of the weak scale. If the stop mass is larger, say greater than
$\sim 1$ TeV, it drives the Higgs (mass)$^2$ to too large (negative) values, 
$\sim (500)$ GeV$^2$. We can still get the correct $Z$ mass ($\sim
100$ GeV) by choosing the $\mu$ term to cancel the negative Higgs (mass)$^2$.
But, this requires a fine tuning, naively of 1 part in 
$(500 \;\hbox{GeV})^2/
(100 \; \hbox{GeV})^2 \sim 25$ (for large $\tan \beta$).
Thus, EWSB is natural in the SSM due to the large top
quark Yukawa coupling provided the stop masses are less than 
about 1 TeV \cite{barbieri1,anderson}. 
This solves the second part of the gauge hierarchy problem:
in the SSM, the weak scale is naturally stabilized at the scale
of the superpartner masses. 
Thus in the SSM, the first part of the gauge hierarchy problem, {\it i.e.,} 
what is the origin of the weak scale, can be rephrased as:
what is the origin of
these soft mass terms, {\it i.e.}, how is
SUSY broken? As mentioned before, we do not want to put
in the soft masses by hand, but rather derive them from a more
fundamental scale, for example the Planck scale.
If SUSY is broken spontaneously
in the SSM with no extra gauge group and no higher dimensional
terms in the K\"ahler potential, then, at tree level,
there is a
colored scalar
lighter than the up or down quarks \cite{georgi}.
So, the superpartners have to
acquire mass through
radiative corrections or non-renormalizable terms in the
K\"ahler potential.
For these effects to dominate over the tree level renormalizable
effects,
a ``modular'' structure is necessary, {\it i.e.,} we need a ``new''
sector where
SUSY is broken spontaneously and then
communicated to the SSM by some
``messenger'' interactions. 

There are two problems here: how
is SUSY broken in the 
new
sector at the right scale  and what are the messengers?
There are models in which a
dynamical superpotential
is generated by non-perturbative effects
which breaks SUSY \cite{affleck}. The SUSY
breaking scale is related to the Planck scale by
dimensional transmutation and thus can be naturally smaller
than the Planck scale (as in QCD).
Two possibilities have been
discussed in the literature for
the messengers. One is gravity
which couples to both the sectors \cite{lykken}.
In a supergravity (SUGRA)
theory, there are non-renormalizable
couplings between the two sectors which generate
soft SUSY breaking operators in the
SSM
once SUSY is broken in the ``hidden'' sector.
The other messengers
are the SM gauge interactions \cite{gm}.
Thus, dynamical SUSY breaking with superpartners at 
$\sim 100$ GeV$-1$ TeV
can explain the gauge hierarchy: SUSY stabilzes the weak scale
at the scale of the superpartner masses which in turn can be derived
from the more ``fundamental'' Planck scale.
Also, with the superpartners
at the weak scale, the gauge coupling unification works well
in a supersymmetric GUT \cite{langacker}.

If SUSY solves the fine tuning problem of the Higgs 
mass, {\it i.e.}, EWSB is natural in the SSM, 
does it introduce any other fine tuning or 
unnaturalness? This is the central issue
of this thesis. We show that 
consistency of the SSM with phenomenology (experimental observations)
requires that,
unless we impose additional symmetries,
we have to introduce some degree of fine tuning or unnaturalness
in some sector of the theory (in some cases, reintroduce
fine tuning in EWSB). The phenomenological constraints
on the SSM that we study all result (in one way or another) from
requiring consistency with FCNC's.

We begin with a ``problem'' one faces right away when one 
supersymmetrizes the
SM and adds all renormalizable
terms consistent with SUSY
and gauge invariance. Requiring the Lagrangian to
be gauge invariant does not uniquely determine the form of the
superpotential. In addition to
Eqn.(\ref{WMSSM}) the following renormalizable
 terms
\begin{equation}
\lambda_{ijk}L_iL_jE_k^c + \bar{\lambda}_{ijk}L_iQ_jD^c_k +
\lambda^{\prime\prime}_{ijk}U^c_iD^c_jD^c_k
\label{rpw}
\end{equation}
are allowed.\footnote{A term $\mu_iL_iH_u$ is also allowed.
This may be rotated away through a redefinition of the $L$ and
$H_d$
fields
\cite{suzuki}.} Unlike the interactions of Eqn.(\ref{WMSSM}), these
terms violate lepton number ($L$) and baryon number ($B$).
Thus, a priori, SSM has $L$ and $B$ violation at the renormalizable level
unlike the SM where no $B$ or $L$ violating terms can be written
at the renormalizable level.
These terms  
are usually
forbidden by imposing a discrete symmetry, $R$-parity,
which is ${(-1)}^{3B+L+2S}$ on a component field with baryon
number $B$, lepton number $L$ and spin $S$.
If we do not impose $R$-parity, what are the constraints
on these $R$-parity violating couplings?
If both lepton and baryon number violating
interactions
are present, then limits on the proton lifetime place
stringent constraints on the products of most of
these couplings (the limits are $\sim 10^{-24}$).
So,
it is usually
assumed that if $R$-parity is violated, then either
lepton or baryon number violating interactions, but
not both, are
present. If either
$L_iQ_jD_k^c$ or $U^c_iD^c_jD^c_k$ terms are present, flavor
changing neutral current (FCNC) processes are induced. It has been
assumed that if only one $R$-parity violating ($\not \!\! R_p\,$)
coupling with a particular
flavor structure is non-zero, then these flavor changing processes
are avoided. In this \it single coupling scheme \rm \cite{hall}
then,
efforts at constraining $R$-parity violation have concentrated on
flavor conserving processes
\cite{barger,godbole,dawson,choudhury,ellis,mohapatra}.

In chapter 2, we demonstrate that the \it single coupling scheme
\rm cannot be realized in the quark mass basis.
Despite
the general values the couplings may have in the weak basis,
after electroweak symmetry
breaking there is at least one large $\not \!\! R_p\,$
coupling and many other $\not \!\! R_p\,$
couplings with different flavor
structure. Therefore, in the mass basis the $R$-parity
breaking couplings \it cannot \rm be diagonal in generation space.
Thus, flavor changing neutral current processes are always
present
in either the charge $2/3$ or the charge $-1/3$ quark sectors.
We use these processes to place constraints on $R$-parity breaking.
We find
constraints on the first and the second generation couplings that are
much stronger than existing limits.
Thus, we show that $R$-parity violation always leads to FCNC's, 
even with the assumption
that there is (a priori) a ``single''
$R$-parity violating coupling (either $L$ or $B$ violating),
unless this ``single'' coupling is small.
Thus, either we impose $R$-parity (or $L$ {\it and} $B$ conservation)
or introduce some degree of unnaturalness in the form of small couplings
in order not to be ruled out by phenomenology.
If we introduce flavor symmetries to explain
the hierarchies in the Yukawa couplings, it is possible that 
the same symmetries can also explain why the $R$-parity violating couplings
are so small.
However, it turns out that, in general, the suppression is not sufficient to
evade the proton decay limits.
The SSM with the particle content of Table \ref{partcont}
and with $R$-parity is called the minimal supersymmetric
Standard Model (MSSM).

\begin{figure}
\centerline{\epsfxsize=1.0\textwidth \epsfbox{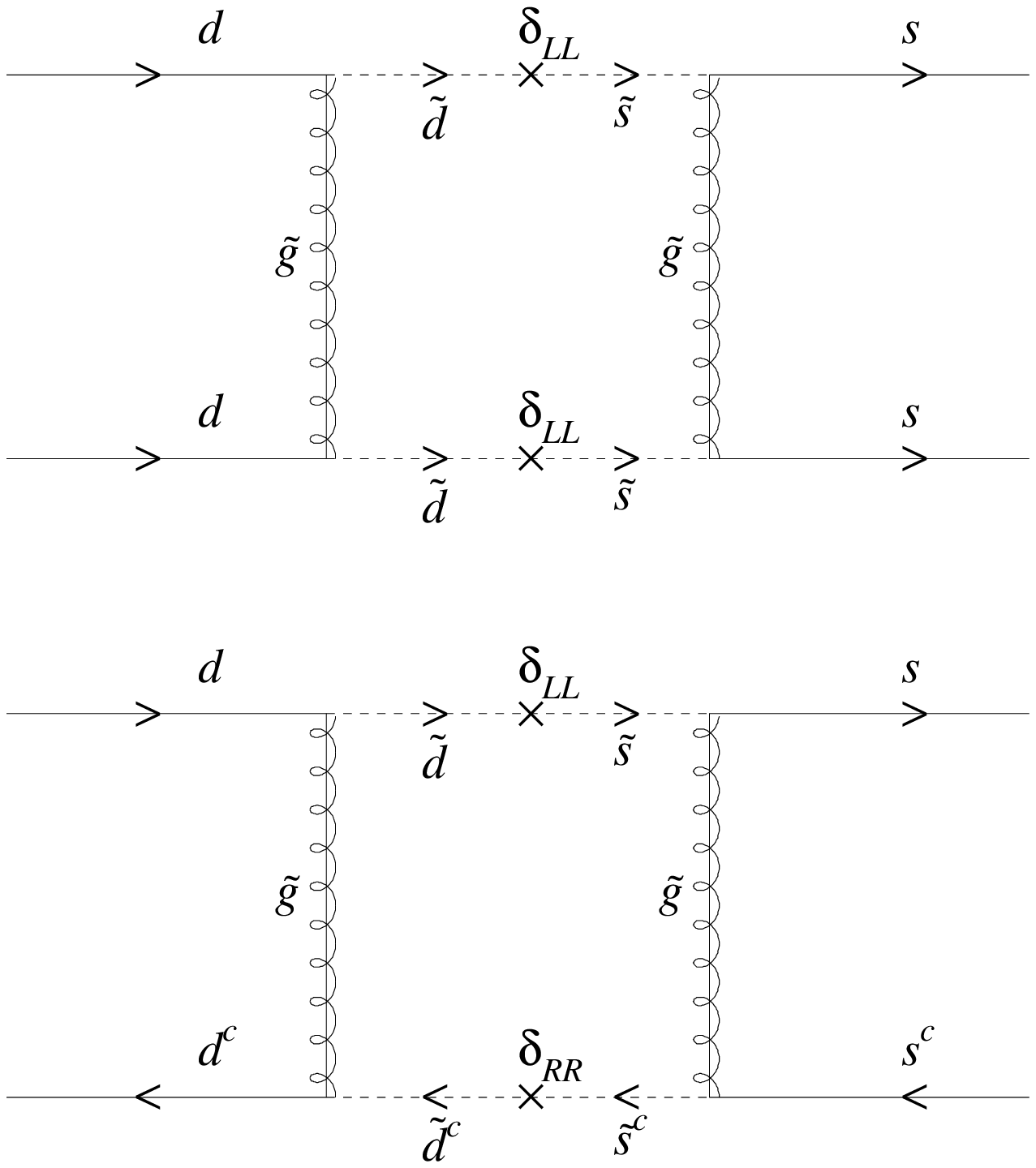}}
\caption{Some of the SUSY contributions to the $\Delta S= 2$
four fermion operator. $\tilde{d}(\tilde{s})$ is the scalar partner of
left-handed down (strange) quark and $\tilde{d}^c (\tilde{s}^c)$
is the scalar partner of the antiparticle of the right-handed down (strange)
quark.}
\protect\label{kk}
\end{figure}

The second problem we discuss is the SUSY flavor problem \cite{georgi}.
As mentioned before, we have to add soft SUSY breaking masses for all 
squarks and sleptons. If these mass matrices are generic in flavor
space, {\it i.e.,} they are not at all correlated with the fermion 
Yukawa couplings, we get large SUSY contributions to the
FCNC's. 
To give a quantitative discussion, we need to define a basis for the
squark and slepton mass matrices. We first rotate the quarks/leptons
to their
mass basis by a unitary transformation, $U$. We do the
same transformation on the squarks/sleptons (thus, it is a superfield
unitary transformation). In this basis for the quarks and squarks, the 
neutral gaugino vertices are flavor diagonal. The squark/slepton
mass matrix in this basis
can be arbitrary since, a priori, there is no relation between the 
squark/slepton and quark/lepton mass matrices so that they need
not be diagonalized
by the same $U$. Thus, there are off-diagonal (in flavor space)
terms in the squark mass matrix in this basis and we get flavor violation.
For concreteness, we discuss the $K\!-\!\bar{K}$ mixing (see Fig.\ref{kk}).
For simplicity, consider the $2 \times 2$ 
mass matrices for the ``left'' and ``right'' 
down and strange squarks (which are the partners
of the left and right handed quarks)
and neglect left-right mixing
(which is likely to be suppressed by the small
Yukawa couplings). We denote the diagonal
elements of the mass matrix by $M_S^2$ and the off-diagonal element,
which converts a down squark to a strange squark, by 
$\Delta$ and define $\delta \sim  \Delta / M_S^2$. 
A posteriori, we know that $\delta$ has to be small
and so we work to first order in $\delta$.
We have to diagonalize this squark mass matrix to get
the mixing angles and the mass eigenvalues. Then, 
$\delta$ is also roughly the product of the 
squark mixing angle and the degeneracy (ratio of 
the difference in the mass eigenvalues to the
average mass eigenvalue).
We then get 
contributions to $K\!-\!\bar{K}$ mixing shown in Fig.\ref{kk}.

In the first diagram the flavor violation comes from using (twice)
the off diagonal element of the left squark mass matix, {\it i.e.,}
$\delta_{LL}$ (there is a similar diagram with insertion of 
two $\delta_{RR}$'s) 
and in the second diagram both
$\delta _{RR}$ and $\delta _{LL}$ are used.
We can estimate the coefficient of the four fermion $\Delta S=2$ operator
to be:
\begin{equation}
\frac{g_3^4}{16 \pi^2} \frac{1}{M_S^2} f({M_{\tilde{g}},M_S^2}) \delta^2,
\label{susykk}
\end{equation}
where the function $f$ comes from the loop integral.
Recall that the SM contribution to this operator
(which already gives a contribution 
to $K\!-\!\bar{K}$ mass difference ($\Delta m_K$) close to the experimental 
value)
is
\begin{equation}
\frac{g_2^4}{16\pi^2} \frac{1}{M_W^2} \frac{m_c^2}{M_W^2}
\end{equation} 
due to the Glashow-Iliopoulos-Maiani (GIM) suppression \cite{GIM}.
Thus with weak scale values of $M_S$ and $M_{\tilde{g}}$ and $\delta \sim
O(1)$,
the SUSY contribution is huge.
Similarly, there are contribution to other FCNC's, for example
$\mu \rightarrow \hbox{e}
\gamma$. Recall that in the SM, there is no contribution
to this process.
So, in order not to be ruled out by FCNC's, 
the $\delta$'s have to be very small if the scalar masses
are $\sim 100$ GeV-
$1$ TeV, {\it i.e.,} the squarks and
sleptons of the first two generations have to be degenerate
to within $\sim$ few GeV \cite{gabbiani} if the mixing angles
are $\sim O(1)$.

The SUSY contribution to FCNC's thus depends on how the soft masses
are generated. In SUGRA, 
unless one makes assumptions about the K\"ahler potential
terms, the squark masses are arbitrary resulting in $\delta \sim O(1)$.
Thus with weak scale values of the superpartner masses,
we either fine tune the $\delta$'s to be small
or 
introduce
approximate
non-abelian or abelian flavor symmetries \cite{flavor} to
restrict the form of the scalar mass matrices so that
the $\delta$'s are small. 
These flavor symmetries can also simultaneously explain
the Yukawa couplings.
A related idea is
squark-quark
mass matrix alignment \cite{seiberg} in which
the quark and squark mass matices are aligned so that the same unitary matrix
diagonalises both of them, resulting in $\delta \sim 0$.

In the other mechanism for communicating SUSY breaking 
mentioned above, {\it i.e.,} gauge mediated SUSY breaking (GMSB),
the scalars of the first two generations
are naturally degenerate since they
have the same gauge quantum numbers, thus giving 
$\delta \sim 0$.
This is
an attractive feature of these models,
since
the FCNC
constraints are naturally avoided and no
fine tuning between the masses of the first two
generation scalars is required. Since this lack
of fine tuning is a compelling argument in favor
of these models, it
is important to
investigate
whether other sectors of these models are fine tuned.
We will argue, in chapter 3, (and this is also discussed in
\cite{nirshirman,arkani,strumia}) that the
minimal model (to be defined in chapter 3) of  
gauge mediated SUSY breaking with a low messenger scale
requires 
fine tuning to generate a correct vacuum ($Z$ mass).
Further, if a gauge-singlet and extra vector-like quintets
are introduced to generate
the ``$\mu$'' and ``$B\mu$'' terms, 
the fine tuning required to 
correctly break the electroweak symmetry is more severe.
These fine tunings
make it difficult to understand, within the
context of these models,
how SUSY can provide
some understanding of the origin of electroweak symmetry
breaking and the scale of the $Z$ and $W$ gauge boson
masses. 
It turns out that in models of
gauge mediation with a high messenger scale
the fine tuning is not much better than in the case of
low messenger scale \cite{alex}.

Typically, the models of gauge mediation have vector-like fields
with SM quantum numbers and with a non-supersymmetric spectrum.
These fields communicate SUSY breaking to the SSM fields and are
therefore called ``messengers''. 
In the minimal model of gauge mediation, 
the messengers form 
complete $SU(5)$ representations in order to preserve
the gauge
coupling unification.
In chapter 3, 
we construct a model of low energy gauge mediation
with 
split $({\bf 5} + {\bf \bar{5}})$ 
messenger fields
that improves the fine tuning.
This model has additional color triplets in the low
energy theory (necessary to maintain gauge coupling
unification) which get a mass of $O(500)$ GeV
from a coupling to a gauge-singlet.
The same model with the singlet coupled to the
Higgs doublets
generates the $\mu$ term.
The improvement in fine tuning is quantified
in these models and the 
phenomenology 
is discussed in detail. 
We show how to derive these split messenger $({\bf{5+\bar{5}}})$'s
from a GUT using a known doublet-triplet splitting mechanism.
A complete
model, including the 
doublet-triplet splitting of the usual Higgs multiplets, is presented and some
phenomenological constraints are discussed.

An obvious solution to the SUSY flavor
problem, from Eqn.(\ref{susykk}), is 
raising the soft
masses of the first two generation scalars
to the tens of TeV range 
so that even if $\delta \sim O(1)$, the SUSY contribution
to FCNC's is small
\cite{dine,pomoral,dvali,nelson2,nelson3,nelson,riotto,dimopoulos}.
Thus, the fine tuning of $\delta$'s is avoided.
The phenomenological viability and naturalness
of this
scenario is the subject of 
chapter 4. We assume that there
is some natural model to make these scalars heavy. We want to investigate
if this leads to unnaturalness in some {\em other} sector. 
To suppress flavour changing processes, the heavy scalars must
have masses between a 
few TeV and a hundred TeV.
The actual value depends on the degree of mass degeneracy and mixing
between the 
first two generation scalars.\footnote{Once the amount 
of fine tuning ({\it i.e.,}
how small $\delta$) we are willing to tolerate 
is given, we 
can estimate the
$M_S$ required from Eqn.(\ref{susykk}).}
As we discussed before, only the stop masses have to be smaller
than about 1 TeV to get natural EWSB.
However, as discussed in reference \cite{nima}, the masses of
the heavy scalars cannot be made arbitrarily large without
breaking colour and charge.
This is because
the heavy scalar masses contribute to the two loop
Renormalization
Group Equation (RGE) for the soft masses of the light scalars, such
that
the stop soft (mass)$^2$ become
more negative in RG scaling
to smaller energy scales. This negative contribution is large if
the scale at which supersymmetry breaking is communicated
to the visible sector is close to the GUT scale \cite{nima}.
With the first two generation soft scalar masses
$\approx$ 10 TeV, the initial value of the
soft masses for the light stops must be $\approx
\hbox{few} $ TeV
to cancel this negative contribution \cite{nima} to
obtain the correct vaccum. This requires, however, an unnatural amount
of fine tuning to correctly break the electroweak
symmetry \cite{barbieri1,anderson}.

In chapter 4, we analyze these issues and include two new items:
the effect of the large top quark Yukawa coupling, $\lambda_t$,
in the RG evolution, that drives the stop soft (mass)$^2$ more negative,
and QCD
radiative corrections in the $\Delta m_K$
constraint \cite{bagger}.
This modifies the bound on the heavy scalar masses which is
consistent with the measured value of $\Delta m_K$. This, in
turn, affects the minimum value of the initial scalar masses that is
required to keep the scalar (mass)$^2$ positive at the weak scale.

We note that the severe constraint obtained for the initial
stop masses
assumes that supersymmetry breaking occurs at a high scale.
This
leaves open the possibility that requiring positivity
of the scalar (mass)$^2$ is not a strong constraint if the
scale of supersymmetry breaking is not much larger than the
mass scale of the heavy scalars. In chapter 4
we investigate this possibility
by computing the finite parts of the same two loop diagrams
responsible for the negative contribution to the light scalar
RG equation, and use these results as an {\em estimate}
of the two loop contribution in an actual model of low energy
supersymmetry breaking.
We find that in certain
classes of models of this kind, requiring positivity of the soft (mass)$^2$
may place strong necessary conditions that such models must
satisfy in order to be phenomenologically viable.


\chapter{
$R$-parity Violation in Flavor Changing Neutral Current
Processes}
In a supersymmetric extension of the SM without $R$-parity,
we show 
that even
with a ``single'' coupling scheme, {\it i.e.,} with only ``one''
$R$-parity violating coupling (either $L$ or $B$ violating) with
a particular flavor structure being non-zero,
the flavor changing neutral current processes can be avoided only 
in
either the charge $+2/3$ or the charge $-1/3$ quark sector,
but not both.
We use the processes $K-\bar{K}$ mixing,
$B-\bar{B}$ mixing and $K^+ \rightarrow \pi^+ 
\nu \bar{\nu}$ (in the down sector) and  $D-\bar{D}$ mixing (in the up sector)
to place constraints on $\not \!\! R_p\,$
couplings.
The constraints on the first and the second generation couplings are better
than
those existing in the literature.    

Flavor changing 
neutral current processes are more clearly seen by 
examining the structure of the
interactions in the quark mass basis. In this basis, the 
$\bar{\lambda}_{ijk}$ interactions of Eqn.\ref{rpw} are 
\begin{equation}
\lambda^{\prime}_{ijk}(N^m_i(V_{KM})_{jl}D^m_l 
- E^m_iU^m_j)D^{cm}_k,
\label{physbasis}
\end{equation}
where
\begin{equation}
\lambda^{\prime}_{ijk}=
\bar{\lambda}_{imn}U_{Lmj}D_{Rnk}^{\ast},
\label{up}
\end{equation}
and $N$ is the neutrino chiral superfield.
The superfields in 
Eqn.(\ref{physbasis}) have their fermionic 
components in the mass basis so that the 
Cabibbo-Kobayashi-Maskawa (CKM) matrix \cite{kobayashi}
$V_{KM}$ 
appears explicitly. The rotation matrices $U_L$ and $D_R$ 
appearing in the previous equation are defined by
\begin{eqnarray}
u_{Li}=U_{Lij}u^m_{Lj}, \\
d_{Ri}=D_{Rij}d^m_{Rj},
\end{eqnarray}
where $q_i\;(q_i^m)$ are quark fields in the weak (mass) basis. 
Henceforth, all the fields will be in the mass basis and we drop
the superscript $m$.

Unitarity of the rotation matrices implies that the couplings 
$\lambda_{ijk}^{\prime}$ and $\bar{\lambda}_{ijk}$ satisfy
\begin{equation}
\sum_{jk} {\left|\lambda^{\prime}_{ijk}\right|}^2 =  \sum_{mn} 
{\left|\bar{\lambda}_{imn}\right|}^2.
\end{equation}
So any constraint on the $\not \!\! R_p\,$
 couplings in the quark mass basis also 
places a bound on the $\not\!R_p$
couplings
in the weak basis.

In terms of component fields, the interactions, in Dirac
notation, are
\begin{equation}
\lambda^{\prime}_{ijk}
[(V_{KM})_{jl}({\tilde{\nu}}^i_L{\bar{d}}^k_Rd^l_L 
+ {\tilde{d}}^l_L{\bar{d}}^k_R {\nu}^i_L +
(\tilde{d} ^k_R
)^* \overline{(\nu ^i_L)^c}d_L^l) 
- {\tilde{e}}^i_L{\bar{d}}^k_Ru^j_L 
- {\tilde{u}}^j_L{\bar{d}}^k_Re^i_L - (\tilde{d} ^k_R
)^* \overline{(e^i_L)^c}u_L^j],
\label{eq:RparityLag}
\end{equation}
where $e$ denotes 
the electron and ${\tilde{e}}$ it's scalar partner 
and similarly for the other particles.

The contributions of the 
$R$-parity violating interactions to low 
energy processes involving no sparticles
in the final state arise
from using the $\not \!\! R_p\,$
 interactions an even number of times. If two 
$\lambda^{\prime}$ 's or $\lambda^{\prime\prime}$ 's with
 different
flavor structure are 
non-zero, flavor changing low energy processes 
can occur. These processes are considered in references
\cite{suzuki} and \cite{barbieri}, respectively.
Therefore, it is 
usually assumed that either only one $\lambda^{\prime}$ 
with a
 particular
flavor structure
is non-zero, or that the 
$R$-parity breaking couplings are diagonal in 
generation space. However, 
Eqn.(\ref{eq:RparityLag})
indicates that this does not 
imply that there is only one set of 
interactions with a
particular flavor 
structure, or even that they are diagonal in flavor 
space. In fact, in this case of one 
$\lambda^{\prime}_{ijk}\neq 0$, 
the CKM matrix generates couplings 
involving each of the three down-type quarks. 
Thus, flavor violation occurs in the down quark sector, though 
suppressed by the small values of the off-diagonal CKM elements.
Below, we use these 
processes to obtain constraints on $R$-parity 
breaking, assuming only one
$\lambda^{\prime}_{ijk}\neq 0$.


It would seem that the flavor changing neutral current 
processes may be 
``rotated'' away by making a different physical assumption 
concerning
which $\not\!\!R_p$ coupling is
non-zero. For example, while leaving the quark fields 
in the mass 
basis, Eqn.(\ref{physbasis}) gives
\begin{eqnarray}
W_{\not\!R_p}&=&\lambda^{\prime}_{ijk}(N_i(V_{KM})
_{jl}D_l - E_iU_j)D^c_k \\
&=&(\lambda^{\prime}_{ijk}V_{KMjl})
(N_iD_l - E_i(V_{KMlp}^{-1})U_p)D^c_k \\
\label{twiddle}
&=&\tilde{\lambda}_{ijk}(N_iD_j - E_i(V_{KMjp}^{-1})U_p)D^c_k,
\end{eqnarray}
where
\begin{eqnarray}
\tilde{\lambda}_{ijk} & \equiv & \lambda^{\prime}_{imk}(V_{KM})_{mj} 
\nonumber \\
 & = & \bar{\lambda}_{imn}D_{Lmj}D_{Rnk}^{\ast}.
\label{down}
\end{eqnarray}
With the
assumption that the $\lambda^{\prime}_{ijk}$ coefficients have 
values such 
that only one $\tilde{\lambda}_{ijk}$
is non-zero, 
there is only one interaction of the form $N_LD_LD^c$.
There is then no longer any flavor violation
in the 
down quark sector. In particular, there are no $\not \!\! R_p\,$
 contributions to 
the processes discussed below.
But now there 
are couplings involving each of the three up type quarks. So 
these interactions contribute to
FCNC in the up sector; 
for example, $D^0\!\!-\!\!\bar{D}^0$ mixing. We use 
$D^0\!\!-\!\!\bar{D}^0$ mixing to place 
constraints on $R$-parity violation assuming only one 
$\tilde{\lambda}_{ijk} \neq 0$.
Thus, there is no basis in
which FCNC can be avoided in both sectors. 

It might be more natural to assume that there is only one large 
$\not \!\! R_p\,$
coupling in the
\it weak \rm  basis, {\it i.e.}, only one
$\bar{\lambda}_{ijk}\neq0$.
In general, there will be a rotation in both the up
and the down quark sectors to go to the mass basis, {\it i.e.,}
$U_L$, $D_L$ and $D_R$ are not equal to the identity matrix.
Then, from Eqns.(\ref{up}) and (\ref{down}), we see that
there are many $\lambda^{\prime}$'s and $\tilde{\lambda}$'s
even if one $\bar{\lambda}$ is non-zero leading to FCNC's
in both the sectors. It is possible that $D_R$ and either $U_L$ {\it or}
$D_L$ are identity matrices, but both $D_L$ and $U_L$
cannot be the identity matrix since their
product is $V_{KM}$. So, with one $\bar{\lambda} \neq 0$, we get
FCNC's in at least one of (and in general both) up and down quark
sectors.

The conclusion that FCNC constraints always exist in either the 
charged $-1/3$ or 
charged $2/3$ quark 
sectors follows solely from requiring consistency with electroweak
symmetry breaking, and
is not specific 
to $R-$parity violation. For example, a similar conclusion
about leptoquark interactions, which
are similar to $\not\!\!R_p$ interactions, is reached 
in reference \cite{leurer}.

\section{$K^0\!\!-\!\!\bar{K}^0$ Mixing}

\begin{figure}
\vspace{0.2in}
\setlength{\unitlength}{0.6pt}
\begin{picture}(610,100)
\put(0,0){
\begin{picture}(300,100)
\put(0,90){\vector(1,0){45}}
\put(45,95){$s_L$}
\put(45,90){\line(1,0){45}}
\put(90,0){\vector(0,1){45}}
\put(95,45){$d_{kL}^c$}
\put(90,90){\line(0,-1){45}}
\put(90,0){\vector(-1,0){45}}
\put(45,+5){$d_L$}
\put(0,0){\line(1,0){45}}
\multiput(90,90)(20,0){2}{\line(1,0){10}}
\put(140,90){\vector(-1,0){10}}
\put(130,95){$\tilde{\nu}_i$}
\multiput(150,90)(20,0){2}{\line(1,0){10}}
\put(180,90){\vector(1,0){45}}
\put(225,95){$d_L$}
\put(225,90){\line(1,0){45}}
\put(180,90){\vector(0,-1){45}}
\put(185,45){$d_{kL}^c$}
\put(180,0){\line(0,1){45}}
\put(270,0){\vector(-1,0){45}}
\put(225,+5){$s_L$}
\put(180,0){\line(1,0){45}}
\multiput(90,0)(20,0){2}{\line(1,0){10}}
\put(130,0){\vector(+1,0){10}}
\put(140,+5){$\tilde{\nu}_i$}
\multiput(150,0)(20,0){2}{\line(1,0){10}}
\put(290,45){$+$}
\end{picture}
}
  
\put(310,0){
\begin{picture}(330,100)
\put(0,90){\vector(1,0){45}}
\put(45,95){$s_L$}
\put(45,90){\line(1,0){45}}
\put(180,90){\vector(-1,0){45}}
\put(135,95){${\nu}_i$}
\put(90,90){\line(1,0){45}}
\put(90,0){\vector(-1,0){45}}
\put(45,+5){$d_L$}
\put(0,0){\line(1,0){45}}
\multiput(90,90)(0,-20){2}{\line(0,-1){10}}
\put(90,40){\vector(0,1){10}}
\put(95,50){$\tilde{d}_{kL}^c$}
\multiput(90,30)(0,-20){2}{\line(0,-1){10}}
\put(180,90){\vector(1,0){45}}
\put(225,95){$d_L$}
\put(225,90){\line(1,0){45}}
\put(90,0){\vector(1,0){45}}
\put(135,5){${\nu}_i$}
\put(180,0){\line(-1,0){45}}
\put(270,0){\vector(-1,0){45}}
\put(225,+5){$s_L$}
\put(180,0){\line(1,0){45}}
\multiput(180,90)(0,-20){2}{\line(0,-1){10}}
\put(180,50){\vector(0,-1){10}}
\put(185,40){$\tilde{d}_{kL}^c$}
\multiput(180,30)(0,-20){2}{\line(0,-1){10}}
\put(280,55){$+$ crossed}
\put(280,35){diagrams}
\end{picture}
}
\end{picture}
\caption{$\not\!\!R_p$ contributions to $K^0\!\!-\!\!\bar{K}^0$
mixing with 
one ${\lambda}^{\prime}_{ijk} \neq 0$. Arrows indicate 
flow of propagating left handed fields.}
\label{kkrp}
\vspace{0.4in}
\end{figure}
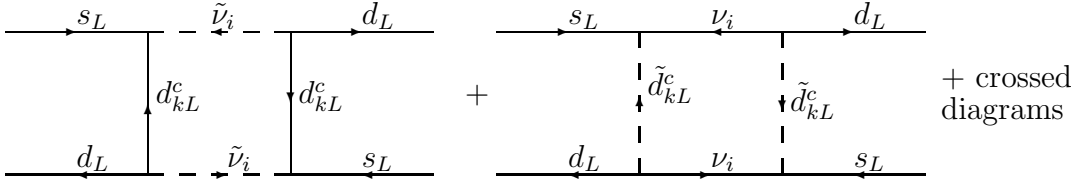

With one $\lambda_{ijk}^{\prime} \neq 0$, the interactions of 
Eqn.$(\ref{eq:RparityLag})$
involve down and strange quarks. So, there are contributions to 
$K^0\!\!-\!\!\bar{K}^0$
mixing through the box diagrams shown in Fig.\ref{kkrp}.
A constraint on the $\not \!\! R_p\,$
 couplings is obtained by constraining the
sum of the $\not \!\! R_p\,$
 and Standard Model contributions to the
$K_L-K_S$ mass difference to be less than the measured value.
 
Evaluating these diagrams at zero external
momentum and neglecting 
the down quark masses, the following effective 
Hamiltonian is generated  
\begin{equation}
{\cal H}^{\Delta
S=2}_{\not\!R_p}=
\frac{1}{128{\pi}^2}{\left|{\lambda}^{\prime}_{ijk}\right|}^4
\left(\frac{1}{m_{{\tilde{\nu}}_i}^2}+\frac{1}{m_{{\tilde{d}}
_{Rk}}^
2}\right)\big((V_{KM})_{j2} 
(V_{KM})_{j1}^*\big)
^2 (\bar{d}_L \gamma^{\mu} s_L)^2,
\end{equation}
where $m_{{\tilde{\nu}}_i}$ is the sneutrino mass and 
$m_{{\tilde{d}}_{Rk}}$ 
is the right-handed down squark mass. As this operator
is suppressed by the CKM angles, it is largest when 
$\lambda_{ijk}^{\prime}$ 
is non-zero for $j=1$ or $j=2$.  

The Standard Model effective Hamiltonian is \cite{lee}
\begin{equation}
{\cal H}^{\Delta S=2}_{SM} = 
\frac{G_{F}^2}{4{\pi}^2}{m_c}^2
\big((V_{KM})_{12}(V_{KM})_{11}^*\big)^2
  (\bar{d}_L \gamma^{\mu} s_L)^2, 
\end{equation}
where the 
CKM suppressed top quark contribution, the up quark mass, 
QCD radiative 
corrections, and long distance effects have been ignored.

The $\Delta S=2$ effective Hamiltonian is then
\begin{eqnarray}
{\cal H}^{\Delta S=2}&
=&{\cal H}^{\Delta S=2}_{SM}+{\cal H}^{\Delta
S=2}_{\not\!R_p} \\
&\equiv&G(\tilde{\lambda}_{ijk},m_{{\tilde{l}}_i},
m_{{\tilde{d}}_{Rk}},V_{KM})
(\bar{d}_L \gamma^{\mu} s_L)^2.
\end{eqnarray}

In the vacuum saturation approximation, 
this effective Hamiltonian contributes
an amount
\begin{equation}
(\Delta m)_{th}\equiv m_{K_L}-m_{K_S}=\frac{2}{3}f_K^2m_KB_K
ReG(\tilde{\lambda}_{ijk},
m_{{\tilde{l}}_i},m_{{\tilde{d}}_{Rk}})
\end{equation}
to the $K_L-K_S$ mass difference.
With $f_K=160\, \hbox{MeV}$ \cite{lattice}, 
$B_K\sim0.6$ \cite{bernard},
$m_K=497\, \hbox{MeV}$ \cite{pdg1}, and
$\left|(\Delta m)_{exp}\right|
=3.510\times10^{-12}\,\hbox{MeV}$ \cite{pdg1}, and
$m_c\geq 1.0\, \hbox{GeV}$,   
the constraint is
\begin{equation}
\left|{\lambda}_{ijk}^{\prime}\right|
\leq 0.11 {\left(\frac{1}{{z_i}^2} 
+ \frac{1}{{w_k}^2}\right)}^{-\frac{1}{4}},
\end{equation}
where $z_i=m_{{\tilde{\nu}}_i}/(100\,\hbox{GeV})$ and 
$w_k=m_{{\tilde{d}}_{Rk}}/(100\,\hbox{GeV})$.
This constraint applies for $j=1$ or $j=2$ and 
for any $i$ or $k$. The
constraint for $j=3$ is not interesting as the 
CKM angles suppress the 
$\not \!\! R_p\,$ operator relative to the 
Standard
Model operator.

\section{$B^0\!\!-\!\!\bar{B}^0$ Mixing}

The $\not \!\! R_p\,$ interactions also contribute to both 
$B^0\!\!-\!\!\bar{B}^0$ mixing 
and $B_s^0\!\!-\!\!\bar{B}_s^0$
mixing through box diagrams similar to those 
given in the previous section. As $B_s^0\!\!-\!\!\bar{B}_s^0$ 
mixing 
is expected 
to be nearly maximal, it is not possible at
present to place a constraint on any 
non-Standard Model effects that would 
\it add \rm more mixing. However, 
$B^0\!\!-\!\!\bar{B}^0$ mixing has been 
observed \cite{argus} with a moderate 
$x_d\sim 0.7$ \cite{pdg1}. 
    
The effective Hamiltonian generated by these 
$\not \!\! R_p\,$ processes is
\begin{equation}
{\cal H}_{\not\!R_p}=\frac{1}{128{\pi}^2}
{\left|{\lambda}^{\prime}_{ijk}
\right|}^4
\left(\frac{1}{m_{{\tilde{\nu}}_i}^2}
+\frac{1}{m_{{\tilde{d}}_{Rk}}^2}
\right)
\big((V_{KM})_{j3}(V_{KM})_{j1}^*\big)^2
 (\bar{d}_L \gamma^{\mu} b_L)^2.
\end{equation}
This is largest when $\lambda_{i3k}^{\prime}$ is non-zero.

The dominant contribution to $B^0\!\!-\!\!\bar{B}^0$ mixing 
in the 
Standard Model is \cite{inami}
\begin{equation}
{\cal H}
_{SM} = 
\frac{G_F^2m_t^2}{4{\pi}^2}\big((V_{KM})_{33}
(V_{KM})_{31}^*\big)^2G(x_t) 
 (\bar{d}_L \gamma^{\mu} b_L)^2, 
\end{equation}
where $x_t=m_t^2/m_W^2$, and 
\begin{equation}
G(x)=\frac{4-11x+x^2}{4(x-1)^2}-\frac{3x^2\ln{x}}{2(1-x)^3}.
\end{equation}
For a top mass of $176$ $\hbox{GeV}$, $G(x_t)=0.54$.

A constraint for $\lambda_{i3k}^{\prime}$ is obtained by 
demanding 
that the sum of
the Standard Model and $\not \! \! R_p$ contributions to the 
$B_L-B_S$ mass
difference not exceed the measured value. 
With $f_B=200\, \hbox{MeV}$ \cite{lattice}, 
$B_B\sim1.2$ \cite{martinelli},
$m_B=5279\, \hbox{MeV}$ \cite{pdg1}, $\left|(\Delta m)
_{exp}\right|=3.3\times10^{-10}\,\hbox{MeV}$ \cite{pdg1}
 and $\left|V_{KM13}\right|\geq0.004$ \cite{pdg1},
a conservative constraint is
\begin{equation}
\left|{\lambda}_{i3k}^{\prime}\right|
\leq 1.1 {\left(\frac{1}{{z_i}^2} 
+ \frac{1}{{w_k}^2}\right)}^{-\frac{1}{4}}
\end{equation}
with $z_i$ and $w_k$ as previously defined. In this case the 
$\not \! \! R_p$ couplings
are only weakly constrained.

In addition to inducing $B^0\!\!-\!\!\bar{B}^0$ mixing, these 
interactions also contribute to the $b\rightarrow s+\gamma$ 
amplitude.
However, with reasonable values for squark and sneutrino masses, 
the constraint is weak.

\section{$K^{+}\rightarrow \pi^{+} \nu \bar{\nu}$}

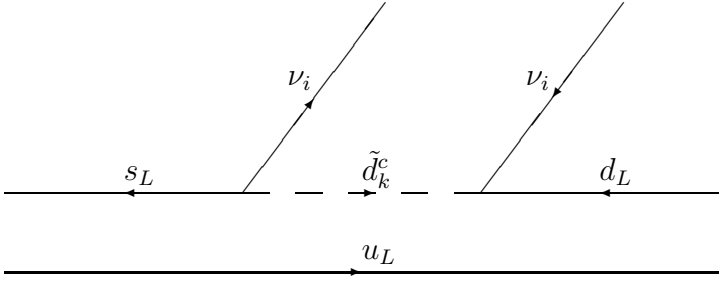
\begin{figure}
\begin{picture}(300,110)
\put(0,30){\line(1,0){45}}
\put(90,30){\vector(-1,0){45}}
\put(45,35){$s_L$}
\put(90,30){\vector(3,4){27}}
\put(107,71){${\nu}_i$}
\put(117,66){\line(3,4){27}}
\multiput(90,30)(20,0){2}{\line(1,0){10}}
\put(130,30){\vector(1,0){10}}
\put(135,35){$\tilde{d}_k^c$}
\multiput(150,30)(20,0){2}{\line(1,0){10}}
\put(180,30){\line(1,0){45}}
\put(270,30){\vector(-1,0){45}}
\put(225,35){$d_L$}
\put(234,102){\vector(-3,-4){27}}
\put(197,71){${\nu}_i$}
\put(180,30){\line(3,4){27}}
\put(0,0){\vector(1,0){135}}
\put(135,5){$u_L$}
\put(135,0){\line(1,0){135}}
\end{picture}
\caption{$\not \!\! R_p\,$ contribution to 
$K^{+}\rightarrow \pi^{+} \nu \bar{\nu}$
with one ${\lambda}^{\prime}_{ijk} \neq 0$.}
\label{kpinu}
\vspace{0.4in}
\end{figure}
 
The tree level Feynman diagram in Fig.\ref{kpinu} generates an 
effective 
Hamiltonian which contributes to the branching
ratio for
$K^{+}\rightarrow \pi^{+} \nu \bar{\nu}$. Using a Fierz 
rearrangement, 
a straightforward evaluation of this diagram
gives
\begin{equation}
{\cal H}_{\not\!R_p}={\frac{1}{2}}
{\frac{{\left|{\lambda}_{ijk}^{\prime}\right|}^2}
{m^2_{\tilde{d}_{Rk}}}(V_{KMj1}V_{KMj2}^{\ast})
(\bar{s}_L\gamma^{\mu}d_L)(\bar{\nu}_{Li}\gamma_{\mu}\nu_{Li})}.
\end{equation}

There is also a Standard Model contribution to 
this decay \cite{inami}. 
This is an order of magnitude lower than the existing
experimental limit. To obtain a bound on the $\not \!\! R_p\,$
 coupling, we shall 
assume that the $\not \!\! R_p\,$ effects dominate the decay
rate.

As the matrix element for this semi-leptonic decay factors 
into a leptonic 
and a hadronic element, the isospin relation
\begin{equation}
\langle \pi^{+}(\bf{p}\rm)|
\bar{s}\gamma_{\mu}d|K^{+}(\bf{k}\rm)\rangle 
=\sqrt{2}\langle\pi^0(\bf{p}\rm)
|\bar{s}\gamma_{\mu}u|K^{+}(\bf{k}\rm)\rangle
\end{equation}
can be used to relate $\Gamma[K^{+}\rightarrow \pi^{+} 
\nu \bar{\nu}]$ 
to $\Gamma[K^{+}\rightarrow \pi^0 {\nu}e^+]$.
The effective Hamiltonian for the neutral pion decay channel 
arises from
the spectator decay of the strange quark. It is
\begin{equation}
{\cal H}_{eff}=\frac{4G_F}{\sqrt{2}}V_{KM12}^{\ast}
(\bar{s}_L\gamma^{\mu}u_L)(\bar{\nu}_{Li}\gamma_{\mu}e_{Li}).
\end{equation}
So in the limit where the lepton masses can be neglected,
\begin{equation}
\frac{\Gamma[K^{+}\rightarrow \pi^{+} \nu_{i} \bar{\nu}_{i}]}
{\Gamma[K^{+}\rightarrow \pi^0 \nu e^+]}=
{\left(\frac{{\left|{\lambda}_{ijk}^{\prime}\right|}^2}
{4G_Fm_{\tilde{d}_{Rk}}^2}
\right)}^2{\left(\frac{\left|V_{KMj1}V_{KMj2}^{\ast}\right|}
{\left|V_{KM12}^{\ast}\right|}\right)}^2.
\end{equation}
This ratio is valid for $i=1,2$ or 3, since in the massless 
neutrino and 
electron approximation, the integrals over phase space in
the numerator and denominator cancel.
So using $BR[K^{+}\rightarrow \pi^{+} \nu \bar{\nu}]\leq5.2
\times10^{-9}$ 
\cite{atiya} ($90\%CL$) and 
$BR[K^{+}\rightarrow \pi^0 {\nu}e^+]=0.0482$ \cite{pdg1}, 
the constraint is
\begin{equation}
\left|\lambda_{ijk}^{\prime}\right|\leq0.012
\left(\frac{m_{\tilde{d}_{Rk}}}
{100\, \hbox{GeV}}\right) (90\%CL)
\end{equation}
for $j=1$ or $j=2$. 
Using $\left|V_{KM13}\right|\geq0.004$ \cite{pdg1} and 
$\left|V_{KM23}\right|
\geq0.03$ \cite{pdg1}, a conservative upper bound
for $\lambda_{i3k}^{\prime}$ is
\begin{equation}
\left|\lambda_{i3k}^{\prime}
\right|\leq0.52\left(\frac{m_{\tilde{d}_{Rk}}}
{100\, \hbox{GeV}}\right) (90\%CL).
\end{equation}
 
\section{$D^0\!\!-\!\!\bar{D}^0$ Mixing}

If there is only one 
$\tilde{\lambda}_{ijk}$ in the \it mass \rm basis, 
then from Eqn.(\ref{twiddle}) it is clear that flavor
changing
neutral current 
processes will occur in the charge $+2/3$ quark sector. 
Rare processes such as $D^0\!\!-\!\!\bar{D}^0$ mixing,
$D^0\rightarrow\mu^+\mu^-$ 
and $D^+\rightarrow\pi^+l^+l^-$, for example, 
may be used to place tight constraints on
$\tilde{\lambda}_{ijk}$. 
For illustrative purposes, in this section we 
will consider $D^0\!\!-\!\!\bar{D}^0$ mixing.

The interactions 
in Eqn.(\ref{twiddle}) generate box diagrams 
identical to those discussed in the previous
sections if both 
the internal sneutrino (neutrino) propagators are 
replaced with charged slepton (lepton) propagators and the external
quarks lines are 
suitably corrected. Using the same approximations 
that were made 
earlier, the $\not \!\! R_p\,$ effects generate the
following effective Hamiltonian 
\begin{eqnarray}
{\cal H}_{\not\!R_p}&=&\frac{1}{128{\pi}^2}
{\left|\tilde{\lambda}_{ijk}\right|}^4
\left(\frac{1}{m_{{\tilde{l}}_i}^2}+\frac{1}{m_{{\tilde{d}}
_{Rk}}^2}\right)\big((V_{KM})_{2j}(V_{KM})_{1j}^*\big)^2 
(\bar{c}_L \gamma^{\mu} u_L)^2. 
\end{eqnarray}
With $f_D=200\, \hbox{MeV}$ \cite{lattice}, 
$m_D=1864\, \hbox{MeV}$ \cite{pdg1}, and 
$\left|(\Delta m)_{exp}\right|\leq1.32\times10^{-10}\,\hbox{MeV}$ 
\cite{pdg1} ($90\%CL$), 
the constraint on $\tilde{\lambda}_{ijk}$ for $j=1$ or $j=2$ is
\begin{equation}
\left|\tilde{\lambda}_{ijk}
\right|\leq 0.16{\left(\left(\frac{100\, \hbox{GeV}}
{{m_{\tilde{l}_i}}}\right)^2 + \left(\frac{100\, \hbox{GeV}}
{{m_{\tilde{d}_{Rk}}}}\right)
^2\right)}
^{-\frac{1}{4}} (90\%CL). 
\end{equation}

\section{Summary}

In this chapter we have argued 
that $R$-parity breaking interactions 
always lead to flavor changing neutral current processes.
It is possible that there is a single $\not \!\! R_p\,$ 
coupling in the charge 
$+2/3$ quark sector. 
But requiring consistency with electroweak symmetry breaking 
demands that 
$\not \!\! R_p\,$ couplings involving all the 
charge $-1/3$ quarks exist. That is, 
a single coupling scheme may only be possible in either the 
charge $2/3$ or the charge $-1/3$ quark
sector, but not
both. As a result, flavor 
changing neutral current processes always 
exist in one of these sectors. 
We have used $K^+ \rightarrow {\pi}^+ \nu \bar{\nu}$, 
$K^0\!\!-\bar{K}^0$ mixing, $B^0\!\!-\bar{B}^0$ mixing and 
$D^0\!\!-\bar{D}^0$ mixing to constrain the $\not \!\! R_p\,$ 
couplings. If there is
CKM-like mixing in the charged $-1/3$ quark 
sector, then the constraints are
quite stringent; see Table \ref{constraint}. The tightest 
constraint
is on $\left|\lambda_{ijk}^{\prime}\right|$ for
$j=1,2$ and any $i$ and $k$. This comes from the rare decay
$K^+ \rightarrow \pi ^+\nu \bar{\nu}$. The 
constraints we obtain 
for the first two generation couplings are 
more stringent than those presently 
existing in the literature. 

\renewcommand{\arraystretch}{1.5}
\begin{table}
\begin{center}
\begin{tabular}{||l|l||l|l||l|l||}\hline
$\left|\lambda_{1jk}^{\prime}\right|$ 
& &$\left|\lambda_{2jk}^{\prime}\right|$
& &$\left|\lambda_{3jk}^{\prime}\right|$ &  \\ \hline
111 &0.012$^a$ &211 &0.012$^a$ &311 &0.012$^a$ \\ \hline
112 &0.012$^a$ &212 &0.012$^a$ &312 &0.012$^a$ \\ \hline
113 &0.012$^a$ &213 &0.012$^a$ &313 &0.012$^a$ \\ \hline
121 &0.012$^a$ &221 &0.012$^a$ &321 &0.012$^a$ \\ \hline
122 &0.012$^a$ &222 &0.012$^a$ &322 &0.012$^a$ \\ \hline
123 &0.012$^a$ &223 &0.012$^a$ &323 &0.012$^a$ \\ \hline
131 &0.19$^b$  &231 &0.19$^b$ &331 &0.19$^b$  \\ \hline
132 &0.19$^b$ &232 &0.19$^b$ &332 &0.19$^b$ \\ \hline
133 &0.001$^c$ &233 &0.19$^b$ &333 &0.19$^b$ \\ \hline
\end{tabular}
\end{center}
\caption{Constraints on 
$\left|\lambda_{ijk}^{\prime}\right|$ from: (a)
$K^+ \rightarrow \pi ^+ \nu \bar{\nu}$ $(90\% CL)$; (b) 
$b \rightarrow s
\nu \bar{\nu}$ $(90\% CL)$ 
\protect\cite{grossman}; (c) $\nu_e$ mass
$(90\% CL)$ 
\protect\cite{godbole}. These constraints were obtained
assuming $CKM$-like 
mixing in the charged $-1/3$ quark sector. All limits 
are for
$100\, \hbox{GeV}$ sparticle masses.  }
\label{constraint}
\end{table}

\chapter{
Improving the Fine Tuning in 
Models of Low Energy Gauge Mediated Supersymmetry Breaking 
}
In this chapter, the
fine tuning
in models of low energy
gauge mediated supersymmetry breaking
required
to obtain the correct $Z$ mass is quantified.  To
alleviate the fine tuning problem,
a model with 
a non-minimal messenger sector
is presented.  
This chapter is organized as follows. 
In section \ref{mess}, we briefly review
both the ``messenger sector'' in low energy gauge 
mediated SUSY breaking 
models that communicates SUSY 
breaking to the Standard Model and the pattern of the
sfermion and gaugino masses that follows. 
Section \ref{finetune} quantifies
the fine tuning in the minimal model using the 
Barbieri-Giudice 
criterion \cite{barbieri1}. 

In the minimal model,
the messenger fields form complete $SU(5)$ representations. 
Section \ref{toymodel} describes 
a toy model 
with split $({\bf 5+\bar{5}})$ messenger representations 
that 
improves the fine tuning. To maintain gauge coupling 
unification,
additional color triplets are added to the 
low energy theory. They
acquire a mass of $O(500)$ GeV
by a coupling to a gauge singlet. 
The fine tuning in this model 
is 
improved to $\sim 40 \%$.
The sparticle phenomenology of this model is also 
discussed.

In section \ref{NMSSM}, 
we discuss a version of the
toy model where the above mentioned
singlet 
generates the $\mu$ and $\mu^2_3$ terms. 
This is identical to the Next-to-Minimal 
Supersymmetric Standard Model (NMSSM) \cite{fayet} 
with a particular pattern for the 
soft SUSY breaking operators 
that follows from gauge mediated 
SUSY breaking and our solution to
the fine tuning problem.
We show that this model is tuned to $\sim 20 \%$, even if
LEP does not discover SUSY/light Higgs.
We also show that the NMSSM with one complete messenger 
$({\bf 5 + \bar{5}})$ (and extra vector-like quintets)
is fine tuned to $\sim 2 \%$. 

We  
discuss, in section \ref{GUT}, how 
it is possible to make our toy model compatible with
a Grand Unified Theory (GUT) \cite{splitting}
based upon the gauge group $SU(5) \times SU(5)$.
The doublet-triplet splitting mechanism 
of Barbieri, Dvali and Strumia
\cite{barbieri2} is used to 
split both the messenger representations and the Higgs 
multiplets. 
In 
section \ref{complete}, we present a model in which all 
operators consistent with symmetries are present and 
demonstrate that the low energy theory is the 
model of section \ref{NMSSM}. 
In this model $R$-parity $(R_p)$ is the unbroken 
subgroup of a $Z_4$ global discrete symmetry that is
required to solve the doublet-triplet splitting problem.
Our model has some 
metastable particles which might cause a cosmological
problem. In appendix \ref{appA}, we give the expressions 
for the Barbieri-Giudice parameters
(for the fine tuning) for the MSSM and the NMSSM.

\section{Messenger Sector}
\label{mess}
In the models of low energy gauge mediated SUSY breaking
\cite{nirshirman,dns} (henceforth called LEGM models), SUSY
breaking occurs dynamically in a ``hidden'' sector of 
the theory
at a scale $\Lambda_{dyn}$ that is generated through
dimensional transmutation. SUSY breaking is
communicated to the Standard Model fields in 
two stages.
First, a non-anomalous 
$U(1)$ global symmetry of the hidden sector
is weakly gauged. This $U(1)_X$ gauge interaction
communicates SUSY breaking from the original 
SUSY
breaking sector to a messenger sector at a scale
$\Lambda_{mess}\sim \alpha_X \Lambda_{dyn}/(4\pi)$ as follows.
The particle content
in the messenger sector consists of fields $\Phi_{+}$, 
$\Phi_{-}$
charged under this $U(1)_X$, a gauge singlet field $S$,
and vector-like
fields that carry Standard Model quantum numbers 
(henceforth called 
messenger quarks and leptons).
In the minimal LEGM model,
there is one set of vector-like fields, $\bar{q}$, $l$,
and $q$, $\bar{l}$  that together
form a $({\bf \bar{5} + 5})$ of $SU(5)$.\footnote{In this chapter,
to avoid confusion with the SSM fields, we use the notation
$q$ and $l$ for the messenger superfields and their
fermionic components (with tildes for scalar components),
and $\tilde{Q}$ and $\tilde{L}$ for the squark and slepton
$SU(2)_w$ doublets of the SSM.}
This is a suffucient condition to maintain unification of
the SM gauge couplings. The superpotential
in the minimal model is
\begin{equation}
W_{mess}=\lambda _{\Phi} \Phi_{+} \Phi_{-}S +
\frac{1}{3} \lambda _S S^3
+\lambda_q S q\bar{q}+
\lambda_l S l\bar{l}.
\label{eq:potential}
\end{equation}
The scalar potential is
\begin{equation}
V=\sum_i|F_i|^2+m^2_+|\phi_+|^2+m^2_-|\phi_-|^2.
\end{equation}
In the models of \cite{nirshirman,dns}, 
the $\Phi _+, \Phi _-$ fields 
communicate (at two loops) with the hidden sector fields 
through the 
$U(1)$ gauge interactions. 
Then, SUSY breaking
in the original
sector generates a negative value 
$ \sim - \left( \alpha_X \Lambda_{dyn}
\right)^2/(4\pi)^2$
for the mass
parameters $m^2_+$,
$m^2_-$ of the
$\phi_+$ and $\phi_-$ fields. This drives vevs
of $O \left( \Lambda _{mess} \right)$
for
the scalar components of both
$\Phi_+$ and $\Phi_-$,
and also for the scalar and $F$-component of $S$
if the couplings
$\lambda_S$, $g_X$ and $\lambda _{\Phi}$ satisfy the 
inequalities
derived in
\cite{arkani,randall}.\footnote{
This point in field space is a local
minimum. There is a deeper minimum
where SM is broken \cite{arkani,randall}. To avoid this 
problem,
we can, for example, add another singlet to the 
messenger sector \cite{arkani}.
This does not change our conclusions about the fine tuning.}
Generating a vev for both the scalar and $F$-component
of $S$ is
crucial, since this generates a non-supersymmetric
spectrum for the
vector-like fields $q$ and $l$.
The spectrum of each vector-like messenger field
consists of
two complex scalars
with masses $M^2 \pm B$ and
two Weyl fermions with mass $M$
where $M=\lambda S$, $B=\lambda F_S$
and $\lambda$ is the coupling of the vector-like fields to
$S$. Since we do not want the SM to be broken at this
stage, $M^2-B\ge$0.
In the second stage,
the messenger fields are
integrated out.
As these messenger fields have
SM gauge interactions,
SM
gauginos acquire masses at one loop
and the sfermions and Higgs acquire soft scalar
masses at two
loops \cite{gm}.
The gaugino masses at the scale at which the
messenger fields are integrated out, $\Lambda_{mess}
\approx M$ are \cite{dns}
\begin{equation}
M_G =\frac{\alpha_G(\Lambda_{mess})}{4\pi}\Lambda_{SUSY}
\sum_m N^G_R(m)f_1\left(\frac{F_S}{\lambda_mS^2}\right).
\label{gaugino}
\end{equation}
The sum in Eqn.(\ref{gaugino}) is over
messenger fields $(m)$ 
with normalization \\$ \hbox{Tr} (T^a T^b) = N^G_R(m)
\delta ^{ab}$ where the $T$'s are the generators 
of the gauge group $G$
in the representation $R$,
$f_1 (x)=1+O(x)$, and
$\Lambda_{SUSY}\equiv B/M=F_S/S=x\Lambda_{mess}$
with $x=B/M^2$. 
If all the dimensionless couplings in the 
superpotential are 
$\sim O(1)$, then $x$ cannot be much smaller than one. 
Henceforth, we will 
set $\Lambda_{SUSY}\approx\Lambda_{mess}$. The exact one loop 
calculation \cite{unpublished} of the gaugino mass shows that
$f_1(x) \leq$ 1.3 for $x \leq$ 1. 
The soft scalar masses at $\Lambda_{mess}$ are \cite{dns}
\begin{equation}
{m_i}^2=2 \Lambda^2_{SUSY}\sum_{m,G}N^G_R(m)C^G_R(s_i)
\left(\frac{\alpha_G(\Lambda_{mess})}{4\pi}\right)^2 
f_2\left(\frac{F_S}
{\lambda_mS^2}\right),
\label{scalarmass}
\end{equation}
where
$C^G_R(s_i)$ is the Casimir of the representation of the 
scalar $i$
in the gauge group $G$ and $f_2 (x) = 1 + O(x)$.
The exact two loop calculation \cite{unpublished} 
which determines $f_2$ shows that for 
$x\leq$0.8 (0.9), $f_2$ differs from one by less 
than 1$\%$(5$\%$). 
Henceforth we shall 
use $f_1(x)=1$ and $f_2(x)=1$.
In the minimal 
LEGM model
\begin{equation}
M_G(\Lambda_{mess})=\frac{\alpha_G(\Lambda_{mess})}{4\pi}
\Lambda_{mess},
\end{equation}
\begin{eqnarray}
m^2(\Lambda_{mess})&=&2\Lambda^2_{mess}\times \\
 & & \left(C_3\left(
\frac{\alpha_3(\Lambda_{mess})}{4\pi}\right)^2  
  +  C_2\left(\frac{\alpha_2(\Lambda_{mess})}{4\pi}\right)^2+
\frac{3}{5}\left(\frac{\alpha_1(\Lambda_{mess})Y}{4\pi}
\right)^2\right),
\nonumber
\end{eqnarray}
where $Q=T_{3L}+Y$ and $\alpha_1$ is the $SU(5)$ normalized 
hypercharge coupling. Further,
$C_3=4/3$ and $C_2=3/4$ for colored 
triplets and electroweak doublets respectively. 

The spectrum in the models is determined by only a few 
unknown parameters. 
As Eqns.(\ref{gaugino}) and (\ref{scalarmass}) 
indicate, 
the SUSY 
breaking mass parameters for the Higgs, sfermions and
gauginos are
\begin{equation}
m_{\tilde{q}},M_{\tilde{g}}:m_{\tilde{L}},m_{H_i},
M_{\tilde{W}}:m_{\tilde{e}_R},M_{\tilde{B}}
\sim \alpha_3:\alpha_2:\alpha_1.
\end{equation}
The scale of $\Lambda_{mess}$ is chosen
to be $\sim$ 100 TeV so that the lightest 
of these particles escapes detection. It follows that the 
intrinsic scale of supersymmetry breaking, $\Lambda_{dyn}$, is
$\sim 10000 $ TeV.
The goldstino decay of the lightest standard model
superpartner
then occurs outside the detector \cite{thomas}. The phenomenology
of the minimal LEGM model is discussed in detail in 
\cite{thomas}.

\section{Fine Tuning in the Minimal LEGM}
\label{finetune}

A desirable feature of gauge mediated SUSY breaking is
the natural suppression of FCNC
processes since the scalars with the same gauge
quantum numbers are degenerate \cite{gm}. 
But, the minimal LEGM model 
introduces a fine tuning in the Higgs sector 
unless the messenger scale is low.
This has been previously discussed in \cite{nirshirman,arkani} 
and quantified 
more recently in \cite{strumia}. 
We outline the discussion in order to introduce some 
notation.

The superpotential for the MSSM is
\begin{equation}
W=\mu H_u H_d+W_{Yukawa}.
\end{equation}
The scalar potential is
\begin{equation}
V=\mu^2_1|H_u|^2+\mu^2_2|H_d|^2-(\mu^2_3H_u H_d + h.c.) 
\hbox{+D-terms}
+V_{1-loop},
\label{potential}
\end{equation}
where $V_{1-loop}$ is the one loop effective potential.
The vev of $H_u$ ($H_d$), denoted by $v_u (v_d)$,
 is responsible for giving mass
to the up (down)-type quarks, $\mu^2_1=m^2_{H_d}+{\mu}^2$,
$\mu^2_2=m^2_{H_u}+{\mu}^2$ and 
$\mu^2_3$,
\footnote{$\mu^2_3$ is often written as
$B\mu$.} $m^2_{H_u}$, $m^2_{H_d}$ are the SUSY 
breaking mass terms for the Higgs fields.
\footnote{The scale dependence
of the parameters appearing in the potential is implicit.} 
Extremizing this
potential determines, with $\tan\beta\equiv v_u/v_d$,
\begin{equation}
\frac{1}{2}{m_Z}^2=\frac{\tilde{\mu}^2_1-\tilde{\mu}^2_2
\tan^2\beta}{\tan^2\beta-1},
\label{mssm1}
\end{equation}
\begin{equation}
\sin 2\beta=2\frac{\mu^2_3}{\tilde{\mu}^2_1+\tilde{\mu}^2_2},
\label{mssm2}
\end{equation} 
where $\tilde{\mu}^2_i=\mu^2_i
+2\partial V_{1-loop}/\partial v^2_i$. For large 
$\tan\beta$, $m^2_Z/2\approx -(m^2_{H_u}+\mu^2)$. This 
indicates that if $|m^2_{H_u}|$ is large 
relative to $m^2_Z$, the $\mu^2$ term must cancel this
large number to reproduce the correct value for $m^2_Z$. 
This
introduces a fine tuning in the Higgs potential, that is
naively of the order
$m^2_Z/(2|m^2_{H_u}|)$. We shall show 
that this occurs in the minimal LEGM model. 

In the minimal LEGM model, a specification of the
messenger particle content and the 
messenger scale $\Lambda_{mess}$
fixes the sfermion and gaugino spectrum 
at that scale. For example, 
the soft scalar masses
for the Higgs fields are
$\approx \alpha_2 (\Lambda_{mess}) \Lambda_{mess}
/(4 \pi)$.
Renormalization
Group (RG) evolution from $\Lambda_{mess}$ to the 
electroweak scale reduces
$m^2_{H_u}$ due to the
large top quark Yukawa coupling, $\lambda _t$, and the 
squark soft 
masses.
The one loop Renormalization Group Equation (RGE)
 for $m^2_{H_u}$ is (neglecting 
gaugino and the trilinear scalar term
$(H_u \tilde{Q}_3 \tilde{u}_3 ^c)$ contributions ) 
\begin{equation}
\frac{dm_{H_u}^2(t)}{dt} \approx \frac{3 \lambda _t^2}
{8 \pi ^2}
(m_{H_u}^2(t)+m_{\tilde{u}_3 ^c}^2(t) +m_{\tilde{Q}_3}^2(t)),
\end{equation}
which gives
\begin{equation}
m_{H_u}^2 (t \approx 
\ln (\frac{ m_{\tilde{t}} }{ \Lambda _{mess} })) \approx 
m_{H_u,0}^2- 
\frac{3 \lambda _t^2}{8 \pi ^2}
( m_{H_u,0}^2+ m_{\tilde{u}_3 ^c,0}^2 + m
_{\tilde{Q}_3,0}^2)
\ln(\frac{\Lambda_{mess}}{m_{\tilde{t}}}),
\label{approx}
\end{equation}
where the subscript $0$ denotes the masses at the scale $\Lambda_{mess}$.
On the right-hand side of Eqn.(\ref{approx}) 
the 
RG scaling of $m_{\tilde{Q}_3}^2$ and $m_{\tilde{u}_3 ^c}^2$ 
has been neglected.
Since the logarithm 
$|t|\approx$$\ln(\Lambda_{mess}/m_{\tilde{t}})$ 
is small, it is
naively expected 
that $m^2_{H_u}$ will not be driven
negative enough and will not trigger electroweak symmetry 
breaking. 
However since the squarks are $\approx$ 500 GeV (1 TeV) 
for a
messenger scale $\Lambda_{mess}=$ 50 TeV (100 TeV), 
the radiative
corrections from virtual top squarks are large since 
the squarks are
heavy. A numerical solution of 
the one loop RGE (including gaugino 
and the trilinear scalar term
$(H_u \tilde{Q}_3 \tilde{u}_3 ^c)$ contributions) 
determines $-m^2_{H_u}=$(275 GeV)$^2$  
((550 GeV)$^2$) for
$\Lambda_{mess}=$50 TeV (100 TeV) and setting $\lambda_t=1$. 
Therefore, $m^2_Z/(2|m^2_{H_u}|)\sim$0.06 (0.01), an 
indication of the fine tuning required.

To reduce the fine tuning in the Higgs sector, it
is necessary to reduce $|m^2_{H_u}|$; ideally
so that $m^2_{H_u}\approx-$0.5$m^2_Z$.
The large value of $|m^2_{H_u}|$ at the weak scale 
is a consequence of
the large hierarchy in
the soft scalar masses at the
messenger scale: $m^2_{\tilde{e} _R} <
m_{H_u}^2 \ll m_{\tilde{Q}_3,\tilde{u}_3 ^c}^2$. Models of
sections \ref{toymodel}, \ref{NMSSM} and \ref{complete} 
attempt to reduce the ratio
$m^2_{\tilde{Q}_3}/m^2_{H_u}$ at the messenger scale 
and hence 
improve the fine tuning in the Higgs sector.

The fine tuning may be
quantified by applying one of the criteria of 
\cite{barbieri1,anderson}. 
The value $O^{\ast}$ of a physical
observable $O$ will depend on the fundamental 
parameters $(\lambda_i)$
of the theory. The fundamental parameters of the 
theory are to be distinguished from the free parameters of the
theory which parameterize the solutions to $O(\lambda_i)=
O^{\ast}$. 
If the value $O^{\ast}$
is unusually sensitive to the
underlying parameters $(\lambda_i)$ of the theory, then a small
change in $\lambda_i$ produces a large change in the value of
$O$. The Barbieri-Giudice function
\begin{equation}
c(O,\lambda_i)=\left.\frac{\lambda_i^{\ast}}{O^{\ast}}
\frac{\partial O}{\partial \lambda_i} \right| _{O=O^{\ast}}
\end{equation}
quantifies this sensitivity \cite{barbieri1}. This particular
value of $O$ is fine tuned if the sensitivity to $\lambda_i$
is larger at $O=O^{\ast}$ 
than at other values of $O$ \cite{anderson}.
If there are values of $O$ 
for which the sensitivity to $\lambda_i$
is small, then it is probably sufficient to
use $c(O,\lambda_i)$ as the measure of fine tuning.

To determine $c(m^2_Z,\lambda_i)$, we
performed the following. 
The sparticle spectrum in the
minimal LEGM model is determined by the four parameters
$\Lambda_{mess}$, $\mu^2_3$, $\mu$, and $\tan\beta$.
\footnote{We allow for an arbitrary 
$\mu^2_3$ at $\Lambda_{mess}$.}
The scale $\Lambda_{mess}$ fixes the boundary condition for
the soft scalar masses, and an implicit dependence on
$\tan\beta$ from $\lambda_t$, $\lambda_b$ and $\lambda_{\tau}$ 
arises in RG scaling\footnote{The RG 
scaling of
$\lambda_{t}$ was neglected.}
from $\mu_{RG}=\Lambda_{mess}$ 
to the weak
scale, that is chosen to be 
$\mu^2_{RG}=m^2_t+\frac{1}{2}(\tilde{m}^2_{t}+\tilde{m}^2_{t^c})$. 
The extremization conditions 
of the scalar potential (Eqns.(\ref{mssm1}) and (\ref{mssm2}))
together with $m_Z$ and $m_t$ 
leave two free parameters that we choose to be 
$\Lambda_{mess}$ and $\tan\beta$ (see appendix for the expressions 
for the fine tuning functions). 

\begin{figure}
\centerline{\epsfxsize=1\textwidth \epsfbox{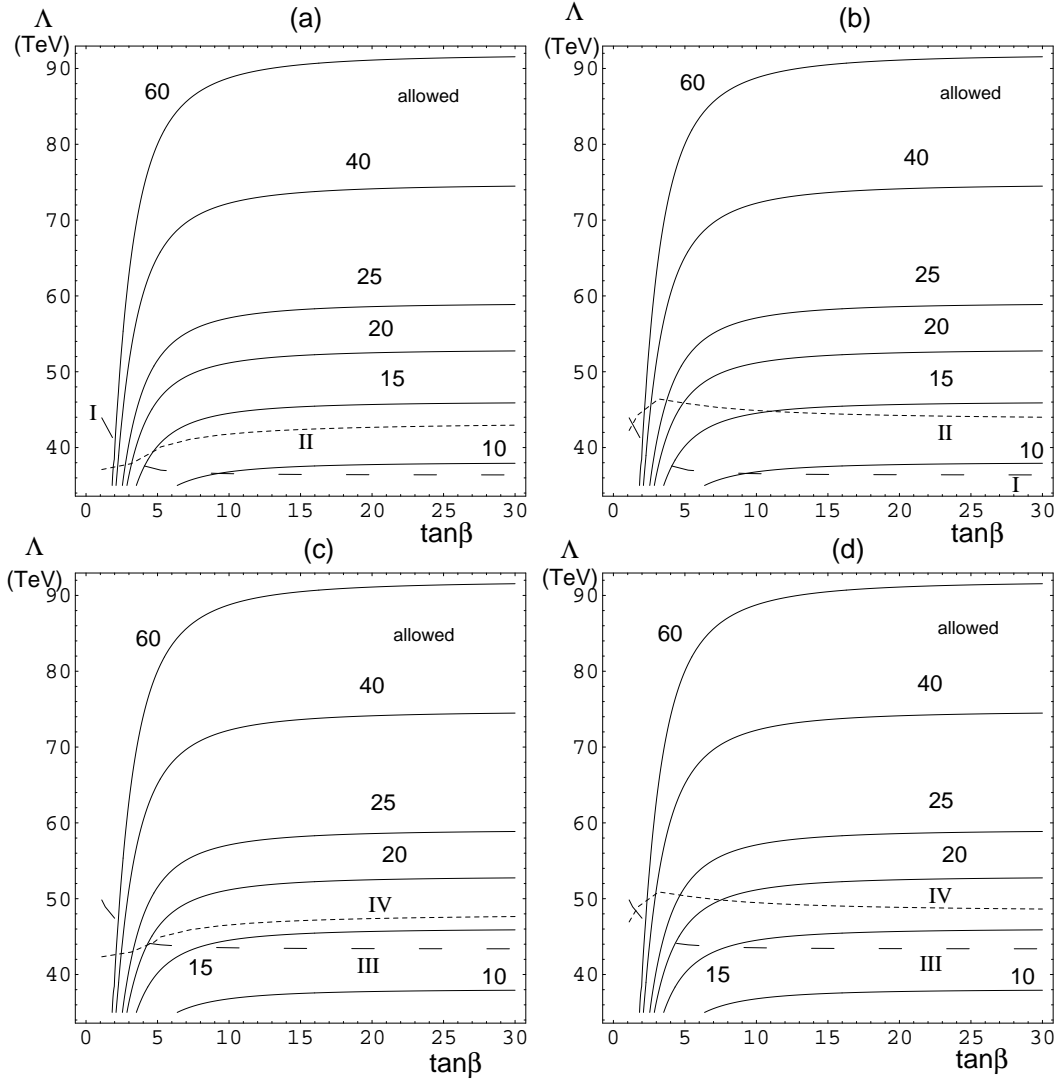}}
\vspace{-1.3in}
\caption{Contours of $c(m^2_Z; \mu^2)=$(10,
15, 20, 25, 40, 60) for a MSSM with a
messenger particle
content of one $({\bf 5+\bar{5}})$. In Figs.$(a)$ and
$(c)$ $sgn(\mu)=-$1 and in
Figs.$(b)$ and $(d)$ $sgn(\mu)=+$1. The
constraints considered are:
(I) $m_{\tilde{e}_R}=$75 GeV , (II)
$m_{{\tilde{\chi}}^{0}_{1}}+m_{{\tilde{\chi}}^{0}_{2}}=$
160 GeV, (III) $m_{\tilde{e}_R}=$85 GeV, and
(IV)
$m_{{\tilde{\chi}}^{0}_{1}}+m_{{\tilde{\chi}}^{0}_{2}}=$
180 GeV.
A central value of
$m_{top}=$175 GeV is assumed.}
\protect\label{ftdns1}
\end{figure}

A numerical analysis yields the value of 
$c(m^2_Z,\mu^2)$ that is displayed in Fig.\ref{ftdns1}
in the 
$(\tan\beta,\Lambda_{mess})$ plane. 

We note that $c(m^2_Z,\mu^2)$ 
is large throughout 
most of the parameter space, except for the region where
$\tan\beta \; \gtap \;$5 and the messenger scale is
 low. A strong
constraint on a lower limit for $\Lambda_{mess}$
comes from the right-handed selectron mass. Contours 
$m_{\tilde{e}_R}=$ 75 GeV ($\sim$ the LEP limit from 
the run at
$\sqrt{s} \approx 170$ GeV 
\cite{aleph}) and 85 GeV ($\sim$ the ultimate LEP2 limit 
\cite{cerngroup2})
are also 
plotted. 
The (approximate) limit on the neutralino masses from
the LEP run at $\sqrt{s} \approx 170$ GeV, 
$m_{\chi ^0_1} + m_{\chi ^0_2} = 160$ GeV and the ultimate 
LEP2 limit, 
$m_{\chi ^0_1} + m_{\chi ^0_2} \sim 180$ GeV are also
shown in Figs.\ref{ftdns1}a and \ref{ftdns1}c for $sgn(\mu)=-1$ and 
Figs.\ref{ftdns1}b and \ref{ftdns1}d for $sgn(\mu)=+1$.
The constraints
from the present and the ultimate LEP2 limits
on the chargino mass are weaker than 
or comparable to those from
the selectron and the neutralino 
masses and are therefore not shown.
If $m_Z$ were much larger, then $c\sim$ 1.
For example,
with $m_Z=$ 275 GeV
(550 GeV) and $\Lambda_{mess}$= 50 (100) TeV, 
$c(m^2_Z;\mu^2)$ varies between 1 and 5  
for $1.4 \; \ltap \; \tan\beta \; \ltap \; 2$, and is 
$\approx 1$ for
$\tan\beta >2$. This suggests that the interpretation 
that a large value for $c(m^2_Z;\mu^2)$ implies that $m_Z$ is 
fine tuned is probably correct.

From Fig.\ref{ftdns1} we conclude 
that in the minimal LEGM model a fine tuning of approximately
$7\%$ in the Higgs potential is required to produce the 
correct value for $m_Z$. Further, for this fine tuning 
the parameters of the model are restricted to 
the region $\tan \beta \; \gtap$ 5 
and $\Lambda_{mess}\approx$ 45 TeV, corresponding to 
$m_{\tilde{e}_R}\approx$ 85 GeV.
We have also checked that adding more complete 
$({\bf 5+\bar{5}})$'s does not
reduce the fine tuning.



\section{A Toy Model to Reduce Fine Tuning}
\label{toymodel}

\subsection{\it Model}

In this section the particle content 
and couplings in the messenger sector 
that are suffucient to reduce $|m_{H_u}^2|$ 
is discussed. The aim is
to reduce $m_ {\tilde{Q}_3} ^2/m_{H_u}^2$ at the scale 
$\Lambda _{mess}$.

The idea is to increase the number of messenger leptons
($SU(2)$ doublets) relative
to 
the number of messenger quarks ($SU(3)$ triplets). 
This reduces both $m_{\tilde{Q}_3}^2/m_{H_u}^2$ and 
$m_{\tilde{Q}_3}^2/m^2_{\tilde{e}_R}$
at 
the scale $\Lambda _{mess}$ (see Eqn.(\ref
{scalarmass})).
This leads to a smaller value of $|m_{H_u}^2|$ in the 
RG scaling
(see Eqn.(\ref{approx})) and the scale $\Lambda_{mess}$
can be lowered since $m_{\tilde{e}_R}$ is larger. 
For example, 
with three doublets and one triplet at a scale 
$\Lambda_{mess} = 30$ TeV, 
so that $m_{\tilde{e} _R} \approx 85$ GeV, we 
find $ |m_{H_u}^2(m_{\tilde{Q}_3})| \approx (100 \hbox{GeV})^2$ for 
$\lambda_t=1$. 
This may be achieved by the following superpotential 
in the messenger sector
\begin{eqnarray}
W & = & \lambda_{q_1} S q_1 \bar{q_1} + 
\lambda_{l_1}S l_1 \bar{l_1} 
+ \lambda_{l_2}Sl_2 \bar{l_2} 
+ \lambda_{l_3}Sl_3 \bar{l_3}
+\frac{1}{3}\lambda_SS^3 \nonumber \\
 & &  + \lambda_{\Phi}S \Phi _- \Phi _+ 
 + \frac{1}{3}\lambda_NN ^3 
+ \lambda_{q_2} N q_2 \bar{q_2} + 
\lambda_{q_3} N q_3 \bar{q_3},
\label{3doublets}
\end{eqnarray}
where $N$ is a gauge singlet.
The two pairs of triplets $q_2, \bar{q} _2$ and 
$q_3, \bar{q} _3$
are required at low 
energies to maintain
gauge coupling unification.
In this model the additional leptons $l_2,\bar{l}_2$ and
$l_3,\bar{l}_3$
 couple to the singlet $S$, whereas the additional
quarks couple to a different singlet $N$ that does not
couple to the messenger fields $\Phi_+$, $\Phi_-$.
This can be enforced by 
discrete symmetries (we discuss such a model in
section \ref{complete}).
Further, we assume the discrete charges also forbid
any couplings between $N$ and $S$ at the renormalizable
 level (this is true of the model
in section \ref{complete}) so that SUSY 
breaking is communicated first to $S$ and to $N$ 
only at a higher loop level. 

\subsection{\it Mass Spectrum}

Before quantifying the fine tuning in this model, 
the mass spectrum of the 
additional states is briefly discussed. 
While these fields form complete representations of
$SU(5)$, they are not degenerate in mass.
The vev and $F$-component of 
the singlet $S$
gives a mass $\Lambda_{mess}$
to the messenger lepton multiplets if the $F$-term
splitting between the scalars is neglected. 
As the
squarks in $q_i+\bar{q_i}$ ($i$=2,3) do not couple to 
$S$, they acquire a soft scalar mass
from the same two loop diagrams that are
responsible for the masses of the
MSSM squarks, yielding
$m_{\tilde{q}}\approx
\alpha_3(\Lambda_{mess}) \; \Lambda_{SUSY}/(\sqrt{6} \pi)$.
The fermions in $q+\bar{q}$ also acquire mass at this
scale since, if either $\lambda _{q_2}$ or 
$\lambda _{q_3} \sim \;O(1)$, a 
negative value for $m^2_N$ (the soft scalar (mass)$^2$
of $N$) is
 generated from the $\lambda _q N q \bar{q}$ coupling 
at one loop
and thus a
vev for $N \sim$ $m_{\tilde{q}}$ is generated. The 
result
is $m_l/m_q\approx\sqrt{6}\pi/\alpha_3
(\Lambda_{mess})(\Lambda_{mess}/\Lambda_{SUSY})
\approx85$.

The mass splitting in the extra fields introduces a
threshold correction to $\sin^2\theta_W$ if it is 
assumed that
the gauge couplings unify at some high scale
$M_{GUT}\approx$10$^{16}$ GeV. 
We
estimate that the splitting shifts the prediction for
 $\sin^2\theta_W$ by an amount
$\approx-$7$\times$ 10$^{-4} \ln(m_l/m_q) n$, 
where $n$ is
the number of split $({\bf 5+\bar{5}})$.\footnote{The
complete $({\bf 5+\bar{5}})$, {\it i.e.}, $l_1, 
\bar{l} _1 \; \hbox{and} \;
q_1, \bar{q} _1$, that couples to $S$
is also split because $\lambda_l\neq\lambda_q$ at 
the messenger scale
due to RG
scaling from $M_{GUT}$ to $\Lambda_{mess}$.
This splitting
is small and neglected.} In this case
$n=$2 and $m_l/m_q \sim$ 85, 
so $\delta$$\sin^2\theta_W\sim-6\times 10^{-3}$.
If
$\alpha_3(M_Z)$ and $\alpha_{em}(M_Z)$ are used
as input, then using the two loop RG equations
$\sin^2\theta_W(\overline{MS})
=0.233 \; \pm \;O(10^{-3})$
is predicted in a minimal SUSY-GUT \cite{langacker}.
The error is a combination of 
weak scale SUSY and GUT 
threshold corrections \cite{langacker}. 
The central value of the
theoretical prediction is a few percent higher
than the measured value of
$\sin^2\theta_W(\overline{MS})=0.231 \pm 0.0003$ \cite{pdg}.
The split extra fields shift the prediction 
of $\sin^2\theta_W$ to $\sim 0.227 \pm\;O(10^{-3})$ which is a few 
percent lower than the experimental value. In sections 
\ref{GUT} and \ref{complete} we show that this spectrum is derivable from 
a $SU(5)\times SU(5)$ GUT in which 
the GUT threshold 
corrections to $\sin^2\theta_W$ 
could be $\sim O(10^{-3})-O(10^{-2})$ \cite{barr}. It is possible that 
the combination of these GUT threshold corrections and the split
extra field threshold corrections make the prediction of 
$\sin^2\theta_W$ more 
consistent with the observed value.

\subsection{\it Fine Tuning}
\label{Fine tuning}
\begin{figure}
\centerline{\epsfxsize=1\textwidth \epsfbox{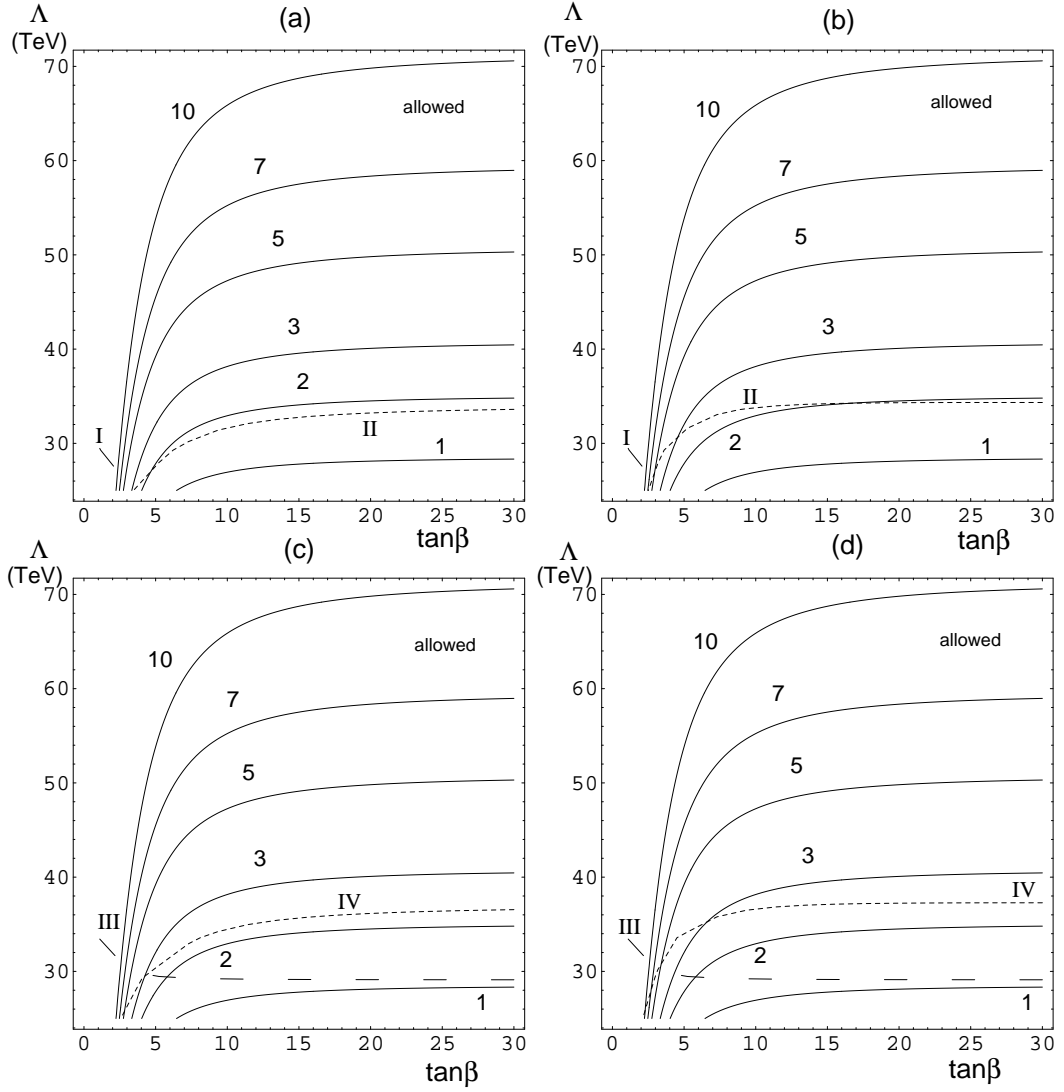}}
\vspace{-1.3in}
\caption{Contours of $c(m^2_Z;\mu^2)=$(1,
2, 3, 5, 7, 10) for a MSSM with
a messenger particle
content of three $(l+\bar{l})$'s and one
$(q+\bar{q})$.
In Figs.$(a)$ and
$(c)$ $sgn(\mu)=-$1 and in
Figs.$(b)$ and $(d)$ $sgn(\mu)=+$1.
The constraints considered are:
(I) $m_{\tilde{e}_R}=$75 GeV , (II)
$m_{{\tilde{\chi}}^{0}_{1}}+m_{{\tilde{\chi}}^{0}_{2}}=$
160 GeV, (III) $m_{\tilde{e}_R}=$85 GeV, and
(IV)
$m_{{\tilde{\chi}}^{0}_{1}}+m_{{\tilde{\chi}}^{0}_{2}}=$
180 GeV.
A central value of
$m_{top}=$175 GeV is assumed.}
\protect\label{ft1}
\end{figure}

To quantify the fine tuning in these class of models the 
analysis of section \ref{finetune} is applied. In our RG analysis the 
RG scaling of $\lambda_t$, the effect of the extra 
vector-like 
triplets on the RG scaling of the gauge couplings, and weak 
scale SUSY threshold corrections were neglected. We have 
checked 
{\it a posteriori} that this approximation is consistent. 
As in section 
\ref{finetune}, the two free parameters are 
chosen to be $\Lambda_{mess}$ and $\tan\beta$.
Contours of constant $c(m^2_Z,\mu^2)$ are 
presented in Fig.\ref{ft1}.  

We show contours of $m_{\chi^0_1} + 
m_{\chi^0_2}= 160$ GeV, and $m_{\tilde{e}_R}= 75$ GeV in Fig.
\ref{ft1}a for $sgn(\mu)=-1$ and in Fig.\ref{ft1}b for $sgn(\mu)=+1$.
These are roughly the
limits from the LEP run at
$\sqrt{s} \approx 170$ GeV \cite{aleph}).
The (approximate) 
ultimate LEP2 reaches \cite
{cerngroup2}: 
$m_{\chi^0_1} + m_{\chi^0_2} = 180$ GeV 
and
$m_{\tilde{e}_R}= 85$ GeV are shown in Fig.\ref{ft1}c for
$sgn(\mu)=-1$ and Fig.\ref{ft1}d for $sgn(\mu)=+1$.
Since $\mu^2 (\approx$ (100 GeV)$^2)$ is much smaller 
in these models than in the minimal LEGM model, 
the neutralinos ($\chi^0_1 \; \hbox{and} \;
\chi^0_2$) are lighter 
so that the neutralino masses
provide a stronger constraint on 
$\Lambda_{mess}$ than does the slepton mass 
limit. The chargino constraints are comparable to the 
neutralino constraints and are thus not shown. It is 
clear that there 
are areas of parameter space in which 
the fine tuning is improved to $\sim$ 40$\%$
(see Fig.\ref{ft1}). 



While this model improves the fine tuning required 
of the $\mu$ parameter, it would be unsatisfactory 
if further fine tunings were required in other sectors 
of the model, for example, the sensitivity of
$m^2_Z$ to $\mu^2_3$, $\Lambda_{mess}$ and 
$\lambda_t$ and the 
sensitivity of $m_t$ to $\mu^2$, $\mu^2_3$,
$\Lambda_{mess}$ and $\lambda_t$. 
We have checked that all these are less than 
or comparable to $c(m^2_Z;\mu^2)$. We
now discuss the other fine tunings in detail.

For large $\tan\beta$, the sensitivity of $m^2_Z$
to $\mu _3 ^2$,
$c(m^2_Z;\mu^2_3)$ $\propto$ $1/\tan^2\beta$, and 
is therefore smaller than $c(m^2_Z;\mu^2)$.
Our numerical analysis shows that for all $\tan \beta$ 
$c(m^2_Z;\mu^2_3) \; \ltap \; c(m^2_Z;\mu^2)$.

In the one loop 
approximation $m^2_{H_u}$ and $m^2_{H_d}$
at the weak scale are proportional 
to  $\Lambda^2 _{mess}$ since 
all the soft masses 
scale with $\Lambda _{mess}$ 
and there is only a weak logarithmic dependence
on $\Lambda _{mess}$ through the gauge couplings. 
We have checked 
numerically 
that $(\Lambda _{mess} ^2/m^2_{H_u}) 
(\partial m^2_{H_u}/ \partial \Lambda _{mess} ^2) \sim 1$. 
Then, $c(m_Z^2;\Lambda _{mess} ^2) \approx
c(m_Z^2;m^2_{H_d}) + c(m_Z^2;m^2_{H_u})$. We find that 
$c(m_Z^2;\Lambda _{mess} ^
2) \approx c(m_Z^2; \mu ^2)+$1 over most of the 
parameter space.

In the one loop approximation, $m^2_{H_u}(t)$ is 
\begin{equation}
m^2_{H_u}(t) \approx m^2_{H_u,0}
+(m^2_{ \tilde{Q} _3,0}+m^2_{\tilde{u} ^c_3,0}
+m^2_{H_u,0})(e^{-\frac{3 \lambda^2_t}{8 \pi^2}t}-1).
\end{equation} \\
Then, using
$t\approx \ln(\Lambda_{mess}/m_{ \tilde{Q} _3})\approx
\ln(\sqrt{6} \pi/\alpha_3)\approx$ 4.5 and  
$\lambda_t \approx$ 1, 
$c(m^2_Z; \lambda_t)$ is (see appendix)
\begin{equation}
c(m^2_Z; \lambda_t) \approx \frac{4}{m^2_Z}
\frac{\partial m^2_{H_u}(t)}{\partial \lambda ^2_t} 
\approx 50 \frac{m^2_{ \tilde{Q} _3}}{(\hbox{600 GeV})^2}.
\end{equation}
This result measures the sensitivity of $m^2_Z$ to 
the value of $\lambda_t$ at the electroweak scale. 
While this 
sensitivity is large, it does not reflect the fact that 
$\lambda_t(M_{Pl})$ is the fundamental parameter of 
the theory, 
rather than $\lambda_t(m_{weak})$.
We find by
 both numerical and 
analytic computations that, for this model 
with three $({\bf 5+ \bar{5}})$'s 
in addition to the MSSM particle content, 
$\delta \lambda_t(m_{weak}) 
\approx 0.1 \times \delta \lambda_t(M_{Pl})$, and 
therefore  
\begin{equation}
c(m^2_Z; \lambda_t(M_{Pl})) \approx 5
\frac{m^2_{ \tilde{Q} _3}}
{(\hbox{600 GeV})^2}.
\end{equation} 
For a scale of $\Lambda_{mess}$
= 50 TeV ($m_{\tilde{Q}_3} \approx$ 600 GeV),
$c(m^2_Z; \lambda_t(M_{Pl}))$ is comparable to 
$c(m^2_Z;\mu^2)$ which is
$\approx$ 4 to 5.
At a lower messenger scale, $\Lambda_{mess} \approx$ 35 TeV, 
corresponding to
 squark masses 
of $\approx$ 450 GeV, the sensitivity of $m^2_Z$ to 
$\lambda_t(M_{Pl})$ is
$\approx$ 2.8. This is comparable to $c(m^2_Z; \mu^2)$ 
evaluated 
at the same scale.

We now discuss the sensitivity of $m_t$ to the fundamental 
parameters.
Since, $m_t^2  = \frac{1}{2} v^2 \sin ^2 \beta \lambda _t ^2$, 
we get
\begin{equation}
c(m_t; \lambda _i) = \delta _{\lambda _t \lambda _i} +
\frac{1}{2}c(m_Z^2; \lambda _i) 
+ \frac{\cos ^3 \beta}{\sin \beta} \frac{\partial \tan \beta}
{\partial \lambda _i}\lambda_i.
\end{equation}
Numerically
we find that the last term in $c(m_t; \lambda _i)$ 
is small compared to 
$c(m_Z^2; \lambda _i)$
and thus over most of 
parameter space $c(m_t; \lambda _i) \approx
\frac{1}{2}c(m_Z^2; \lambda _i)$. 
As before, the sensitivity of $m_t$ 
to the value of $\lambda _t$ at the GUT/Planck scale 
is much smaller than
the sensitivity to the value of $\lambda _t$ at the weak scale.

\subsection{\it Sparticle Spectrum}

The sparticle spectrum is now briefly discussed to
highlight deviations from the mass relations
predicted in the minimal LEGM model. For example, with three
doublets and one triplet at a scale of
$\Lambda=$ 50 TeV, the soft scalar masses (in GeV) at a
renormalization scale $\mu^2_{RG}=m^2_t
+\frac{1}{2}(m^2_{ \tilde{Q}_3}+m^2_{\tilde{u}^c_3})$
$\approx($630 GeV$)^2$, for $\lambda_t=$ 1, are
shown in Table \ref{spectrum}.

\renewcommand{\arraystretch}{0.6}
\begin{table}
\begin{center}
\vspace{0.2in}
\begin{tabular}{lllll} \hline
$m_{ \tilde{Q} _{1,2}}$ & $m_{\tilde{u}^c_{1,2}}$ &
$m_{\tilde{d}^c_{i}}$ & $m_{\tilde{L}_i,H_d}$ &
$m_{\tilde{e}^c_i}$ \\ \hline
687 & 616 & 612 &319 & 125 \\ \hline
\end{tabular} 
\end{center}  

\begin{center}
\begin{tabular}{ll} \hline
$m_{\tilde{Q}_3}$ & $m_{\tilde{u}^c_3}$ \\ \hline
656 & 546 \\ \hline
\end{tabular} 
\end{center} 
\vspace{-0.2in}
\caption{Soft scalar masses in GeV
for messenger particle content
of three $(l+\bar{l})$'s and one $q+\bar{q}$ and a scale 
$\Lambda _{mess} = 50$ TeV.}
\label{spectrum}
\vspace{0.4in}
\end{table}

Two observations that are generic to this type of
model are:
(i) By construction, the spread in the soft
scalar masses is less than in the minimal LEGM model.
(ii) The gaugino masses do not satisfy the one loop
SUSY-GUT relation $M_i/\alpha_i$ = constant. In this case,
for example, $M_3/\alpha_3:M_2/\alpha_2 \approx$ 1$:$3
and $M_3/\alpha_3:M_1/\alpha_1 \approx$ 5$:$11
to one loop.

We have also found that for $\tan\beta$ $\gtap$ 3, the Next
Lightest Supersymmetric Particle (NLSP) is one of the 
neutralinos, whereas for $\tan\beta$ $\ltap$ 3, the NLSP 
is the right-handed stau. Further, for 
these small values of $\tan\beta$,
the three right-handed sleptons are degenerate within 
$\approx$ 200 MeV. 

\section{NMSSM}
\label{NMSSM}
In section \ref{finetune}, the
$\mu$ term and the SUSY breaking mass $\mu^2_3$ were
put in by hand. There it was found that these parameters 
had to be fine tuned in order to correctly reproduce the 
observed $Z$ mass. 
The extent to which this is 
a ``problem'' may only be evaluated within a specific model
that generates 
both the $\mu$ and $\mu^2_3$ terms. 

For this reason, in this section a possible way to 
generate both the $\mu$ term
and $\mu^2_3$ term in a manner that requires a minimal  
modification to the model of either section
\ref{mess} or section \ref{toymodel} is discussed. 
The easiest way to generate these 
mass terms is to introduce a singlet $N$ and add 
the interaction
 $N H_u H_d$ to the superpotential (the NMSSM) \cite{fayet}.
The vev of the scalar component of $N$ 
generates $\mu$ and the vev of the $F$-component of
$N$ generates $\mu_3^2$. 

We note that for the
``toy model'' solution to the fine tuning problem 
(section \ref{toymodel}),
the introduction of the singlet occurs at no additional cost.
Recall that in that model  
it was necessary to introduce a singlet $N$, 
distinct from $S$, 
such that the vev of $N$ gives mass 
to the
extra light vector-like triplets, $q_i,\bar{q} _i \; 
(i=2,3)$ (see
Eqn.(\ref{3doublets})).
Further, discrete 
symmetries (see section \ref{complete})
are imposed to isolate $N$ from SUSY breaking
in the messenger sector.
This last requirement 
is necessary to solve the 
fine tuning problem: if both the 
scalar and $F$-component of $N$ acquired a vev  
at the same scale as $S$, then the extra triplets that 
couple to
$N$ would also act as messenger fields. 
In this case the messenger fields would form 
complete $({\bf 5+\bar{5}})$'s and the fine tuning problem 
would be 
reintroduced. With $N$ isolated from the messenger sector 
at tree level, a vev 
for $N$ at the electroweak scale is naturally generated,
as discussed in section \ref{toymodel}.

We also comment on the necessity and origin of these extra
triplets.
Recall that in the toy model of section \ref{toymodel}
these triplets were required to maintain the SUSY-GUT
prediction for $\sin^2\theta_W$. Further, we shall also 
see that
they are required in order to generate a large enough
$-m^2_N$ (the soft scalar (mass)$^2$ of the singlet $N$). 
Finally, in the GUT model
of section \ref{complete}, the lightness of these triplets
(as compared to the missing doublets) is the consequence
of a doublet-triplet splitting mechanism.

The superpotential in the electroweak symmetry 
breaking sector is  
\begin{equation}
W = \frac{\lambda _N}{3} N^3 
+ \lambda _q N q \bar{q} - \lambda _H N H_u H_d,
\label{WNMSSM}
\end{equation}
which is similar to an  
NMSSM except for the coupling of $N$ to the triplets. 
The superpotential in the messenger sector is 
given by Eqn.(\ref{3doublets}).

The scalar potential is
\footnote{In models of gauge mediated 
SUSY breaking, $A_H$=0 at
tree level and a non-zero value
of $A_H$ is generated at one loop. 
The trilinear scalar term $A_N N^3$ is generated at 
two loops
and is neglected.} 
\begin{eqnarray}
V &=& \sum_{i} | F_i | ^2 + m_N^2 | N | ^2 + 
m_{H_u}^2 | H_u | ^2 + 
m_{H_d}^2 | H_d | ^2 +\hbox{D-terms} \nonumber \\
 & & -(A_H NH_uH_d+h.c.)
+V_{1-loop}.
\label{Vscalar}
\end{eqnarray}
The extremization conditions for 
the vevs of the real components of $N$, $H_u$ 
and $H_d$, denoted by
$v_N$, $v_u$ and $v_d$ respectively (with 
$v = \sqrt{v_u^2 +v_d^2} \approx 250$
GeV), are  
\begin{equation}
v_N (\tilde{m}^2_N + \lambda ^2_H \frac{v^2}{2} + 
\lambda ^2_N v^2_N -
\lambda _H \lambda _N v_u v_d) -
\frac{1}{\sqrt{2}}A_Hv_uv_d= 0,
\label{vn}
\end{equation}
\begin{eqnarray}
\frac{1}{2} m_Z^2 & = & 
\frac{ \tilde{\mu} _1 ^2 - \tilde{\mu} _2 ^2 \tan ^2 \beta }
{\tan ^2 \beta - 1 }, 
\label{NMSSM1}
\\
\sin 2 \beta & = &  2 \frac{\mu ^2 _3}
{\tilde{\mu} _2 ^2 + \tilde{\mu} _1 ^2},
\label{NMSSM2}
\end{eqnarray}
with
\begin{eqnarray}
\mu ^2 & = & \frac{1}{2} \lambda _H ^2 v_N^2 ,    \\
\mu _3 ^2 & = & -\frac{1}{2} \lambda _H ^ 2 v_u v_d 
+ \frac{1}{2} 
\lambda _H \lambda _N v_N^2+A_H\frac{1}{\sqrt{2}}v_N, 
\label{Bmu}
\\
\tilde{m}^2_i & = & m^2_i
+2\frac{\partial V_{1-loop}}{\partial v^2_i}
\hbox{}; \; \; i=(u,d,N).
\end{eqnarray}

We now comment on the expected size of the Yukawa couplings 
$\lambda_q$, $\lambda_N$ and $\lambda_H$.
We must use the RGE's to evolve these couplings from their 
values at
$M_{GUT}$ or $M_{Pl}$ to the weak scale. The quarks and the
Higgs doublets receive 
wavefunction renormalization from $SU(3)$ and 
$SU(2)$ gauge interactions respectively, whereas the 
singlet $N$
does not receive any
 wavefunction renormalization from gauge interactions at 
one loop.
So, the couplings at the weak scale are in the order:
$\lambda _q \sim O(1) > \lambda _H > \lambda _N$ if 
they all 
are $O(1)$
at the GUT/Planck scale. 

We remark that 
without the $N q \bar{q}$ coupling, it is difficult to 
drive a vev for
$N$ as we now show below.
The one loop RGE for $m_N^2$ is
\begin{equation}
\frac{dm_N^2}{dt} 
\approx \frac{6 \lambda _N^2}{8 \pi ^2} m_N^2(t) +
\frac{2 \lambda _H^2}{8 \pi ^2} (m_{H_u}^2(t)+m_{H_d}^2(t) 
+m_N^2(t)) +
\frac{3 \lambda _q^2}{8 \pi ^2} (m_{\tilde{q}}^2(t) + 
m_{\tilde{
\bar{q}}}^2(t)).
\end{equation}
Since $N$ is a gauge-singlet, 
$m_N^2 =0 $ at $\Lambda _{mess}$. 
Further, if $\lambda _{q} = 0$, an estimate 
for $m_{N}^2$ at the weak scale is then
\begin{equation}
m_N^2 \approx - \frac{2 \lambda ^2_H}{8 \pi ^2} 
(m_{H_u,0}^2 + 
m_{H_d,0}^2) \ln \left(
\frac{\Lambda _{mess}}{m_{H_d}} \right),
\label{mn1}
\end{equation}
{\it i.e.}, $\lambda _H$ drives $m_N^2$ negative.
The extremization condition for $v_N$, Eqn.(\ref{vn}), and 
using Eqns.(\ref{NMSSM2}) and (\ref{Bmu}) (neglecting $A_H$)
shows that 
\begin{equation}
m^2_N + \lambda ^2_H \frac{v^2}{2}
\approx \lambda ^2 _H \left(\frac{v^2}{2} - 
\frac{2}{8 \pi ^2}
(m_{H_u,0}^2 +m_{H_d,0}^2) \ln \left(
\frac{\Lambda _{mess}}{m_{H_d}} \right) \right)
\end{equation}
has to be negative for $N$ to acquire a vev. This implies 
that 
$m_{H_u}^2$ and $m_{H_d}^2$ at $\Lambda _{mess}$ 
have to be greater
than $\sim (350 \; \hbox{GeV})^2$ which implies that 
a fine tuning of a few percent is required 
in the electroweak symmetry
breaking sector.
With $\lambda _q \sim O(1)$, however,
there is an additional negative contribution to 
$m_N^2$ given approximately by
\begin{equation}
- \frac{3 \lambda ^2_q}{8 \pi ^2} (m_{\tilde{q},0})^2 +
m_{\tilde{\bar{q}},0}^2) \ln \left( 
\frac{ \Lambda _{mess} }{m_{
\tilde{q} } } \right).
\end {equation}
This contribution dominates the one in Eqn.(\ref{mn1})
since
$\lambda _q > \lambda _H$ and the squarks $\tilde{q}$, 
$\tilde{\bar{q}}$ have soft 
masses larger than the
Higgs. 
Thus, with $\lambda _q \neq 0$, 
$m^2_N + \lambda ^2_H v^2/2$ is naturally negative.

Fixing $m_Z$ and $m_t$,
we have the following 
parameters : $\Lambda _{mess}$, 
 $\lambda_q$, $\lambda_{H}$,
 $\lambda_N$, $\tan\beta$,
and $v_N$.
Three of the parameters are fixed by the 
three extremization conditions, leaving three 
free parameters that for convenience are
chosen
to be $\Lambda_{mess}$, $\tan\beta\geq$0,
and $\lambda_H$. The signs of the vevs are fixed to be 
positive by requiring a stable vacuum and no  
spontaneous CP violation.
The three extremization equations determine the following
relations
\begin{eqnarray}
\lambda_N &=&\frac{2}{\lambda_H v^2_N}(\mu^2_3
+\frac{1}{4}\lambda^2_H 
\sin2\beta v^2-\frac{1}{\sqrt{2}}A_Hv_N) ,\\ 
v_N & = & \sqrt {2} \frac{\mu}{\lambda _H}, \\
\tilde{m}^2_N &=& \lambda_N \lambda_H \frac{1}{2}\sin2\beta v^2
-\lambda^2_N v^2_N-\frac{1}{2}\lambda^2_H v^2
+\frac{1}{2 \sqrt{2}}A_H\sin2\beta \frac{v^2}{v_N},
\label{sol}
\end{eqnarray}
where
\begin{eqnarray}
\mu ^2 & = & - \frac{1}{2} m^2_Z 
+ \frac{ \tilde{m}_{H_u}^2 \tan ^2 \beta - \tilde{m}^2_{H_d} }
{ 1 - \tan ^2 \beta } ,\\
2 \mu ^2_3 & = &\sin 2 \beta (2 \mu ^2 + 
\tilde{m}_{H_u}^2 + \tilde{m}^2_{H_d}).
\end{eqnarray}
The superpotential term $NH_uH_d$ couples the RGE's for 
$m^2_{H_u}$, $m^2_{H_d}$ and $m^2_N$. Thus the values of 
these masses
at the electroweak scale are, in general, complicated 
functions of 
the Yukawa parameters $\lambda_t$, $\lambda_H$, 
$\lambda_N$ and 
$\lambda_q$. 
In our case, two of these Yukawa parameters 
($\lambda _q$ and 
$\lambda _N$) are determined by the
extremization equations and a closed form expression for 
the derived 
quantities cannot be found.
To simplify the analysis,
we neglect the dependence of
$m^2_{H_u}$ and $m^2_{H_d}$ on $\lambda_H$ induced in
RG scaling from $\Lambda_{mess}$ to
the weak scale. Then
$m_{H_u}^2$ and $m_{H_d}^2$ depend only on $\Lambda _{mess}$ 
and $\tan \beta$
and thus closed form solutions for
$\lambda_N$, $v_N$ and
$\tilde{m}^2_N$ can be obtained using the above equations.
Once $\tilde{m}_N^2$ at the weak scale is obtained, 
the value of $\lambda _q$ is obtained by using an 
approximate analytic solution. 
An exact numerical solution of the
RGE's then shows that the above approximation is 
consistent.

\subsection{\it Fine Tuning and Phenomenology}

\begin{figure}
\vspace{-0.875in}
\centerline{\epsfxsize=1\textwidth \epsfbox{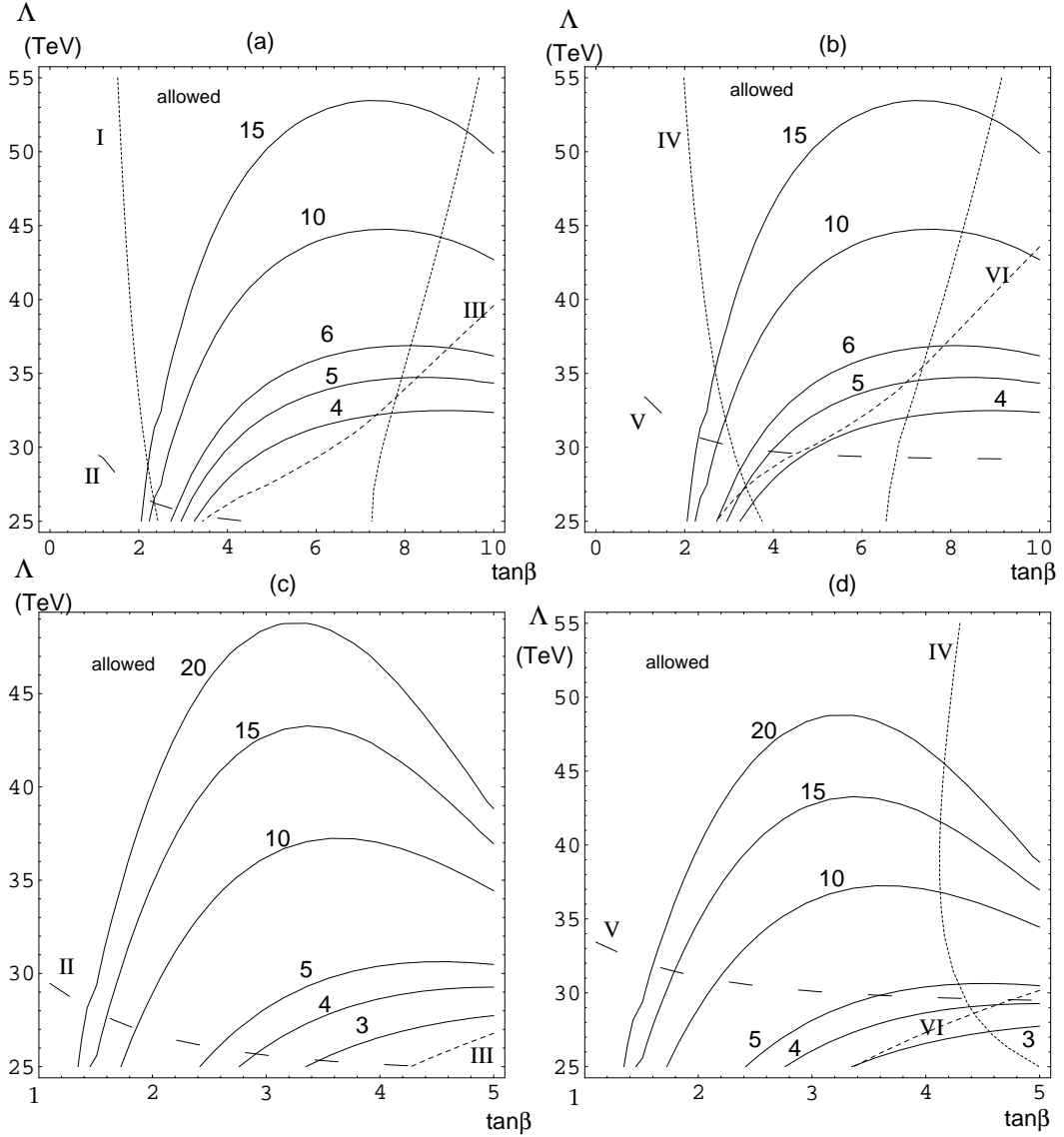}}
\vspace{-1.45in}
\caption{
Contours
of $c(m^2_Z;\lambda_H)$ for the NMSSM of section \ref{NMSSM} and
a messenger particle content of
three $(l+\bar{l})$'s and one
$(q+\bar{q})$. In Figs.$(a)$ and $(b)$,
$c(m^2_Z;\lambda_H)$=(4, 5, 6,
10, 15) and $\lambda_H=$0.1.
In Figs.$(c)$ and
$(d)$, $c(m^2_Z;\lambda_H)=$(3, 4, 5,
10, 15, 20) and $\lambda_H$=0.5.
The constraints considered are:
5(I) $m_h+m_a=m_Z$, (II) $m_{\tilde{e}_R}=$75 GeV,
(III)
$m_{{\tilde{\chi}}^{0}_{1}}+m_{{\tilde{\chi}}^{0}_{2}}=$
160 GeV,
(IV) $m_h=$ 92 GeV, (V) $m_{\tilde{e}_R}=$85 GeV,
and (VI)
$m_{{\tilde{\chi}}^{0}_{1}}+m_{{\tilde{\chi}}^{0}_{2}}=$
180 GeV. For $\lambda_H=$0.5,
the limit $m_h \stackrel{>}{\sim}$ 70 GeV constrains
$\tan \beta \stackrel{<}{\sim}$ 5 (independent of $\Lambda_{mess}$)
and is thus not shown.
A central value of $m_{top}=$175 GeV
is assumed.}
\protect\label{ftN1}
\end{figure}

The fine tuning functions we consider below are
$c(O;\lambda_H)$, $c(O;\lambda_N)$, $c(O;\lambda_t)$,
$c(O;\lambda_q)$ and $c(O;\Lambda_{mess})$ where
$O$ is either $m^2_Z$ or $m_t$.
The expressions for the fine tuning functions and other 
details are
given in the appendix. In our RG analysis the approximations
 discussed in 
subsection \ref{Fine tuning} and above were used and 
found to 
be consistent.
Fine tuning contours of 
$c(m^2_Z;\lambda_H)$ are displayed in 
Figs.\ref{ftN1}a and Fig.\ref{ftN1}b for
$\lambda_H=0.1$ and Figs.\ref{ftN1}c and \ref{ftN1}d 
for $\lambda_H=0.5$.
We have found by numerical computations that
the other fine tuning functions are either smaller or
comparable to $c(m^2_Z; \lambda_H)$. \footnote{
In computing
these functions
the weak scale value of the couplings $\lambda_N$
and $\lambda_H$ has been used.
But since $\lambda_N$
and
$\lambda_H$ do not have a fixed point behavior, we have found
that
$\lambda _{H}(M_{GUT})/\lambda _{H}(m_Z) \;
\partial \lambda _{H}(m_Z)/\partial \lambda _{H}(M_{GUT})
 \sim 1$ so that, for example,
$c( m^2_Z;\lambda_H(M_{GUT}) )\approx
c( m^2_Z;\lambda_H(m_Z) )$.}

We now discuss the existing phenomenological constraints 
on our model
and also the ultimate
constraints if LEP2 does not discover SUSY/light Higgs($h$).
These are shown in Figs.\ref{ftN1}a, \ref{ftN1}c and Figs.\ref{ftN1}b,
\ref{ftN1}d 
respectively.
We consider the processes $e^+ e^-\rightarrow Z h$, 
$e^+ e^-$$\rightarrow (h + \hbox{pseudoscalar})$,
$e^+ e^-$$\rightarrow$ $\chi^+\chi^-$,
$e^+ e^-$$\rightarrow$ $\chi^0_1\chi^0_2$, and 
$e^+ e^-$$\rightarrow$ $\tilde{e}_R \tilde{e}^{*}_R$ 
observable at LEP.
Since this model also has a light pseudoscalar, 
we also consider upsilon decays
$\Upsilon$$\rightarrow (\gamma +  \hbox{pseudoscalar})$.
We find that the 
model is phenomenologically viable and requires 
a $\sim$ 20$\%$ tuning even if  
no new particles are discovered at LEP2.

We begin with the constraints on the
scalar and pseudoscalar spectra of this model.
There are three neutral scalars, two neutral 
pseudoscalars and one complex charged scalar.
We first consider the mass spectrum of the pseudoscalars.
At the boundary scale $\Lambda_{mess}$, SUSY is 
softly broken 
in the visible sector only by the soft scalar masses and 
the gaugino 
masses. Further, the superpotential of 
Eqn.(\ref{WNMSSM}) has
an $R$-symmetry. Therefore,
at the tree level, {\it i.e.,}
with $A_H=$0, the scalar potential
of the visible sector (Eqn.(\ref{Vscalar})) has  
a global symmetry.
This symmetry is spontaneously broken
by the vevs of $N^R$, $H^R_u$, and $H^R_d$ (the
superscript $R$ denotes the real component of fields), 
so that one 
physical
pseudoscalar
is massless at tree level. 
It is 
\begin{equation}
a=\frac{1}{ \sqrt{ v^2_N + v^2 \sin ^2 2 \beta } } \left(
v_N N^I + v \sin 2 \beta \cos \beta H_u^I + v \sin 2 \beta
\sin \beta H_d^I \right),
\label{Raxion}
\end{equation}
where the superscripts $I$ denote the imaginary components 
of the 
fields.
The second pseudoscalar, 
\begin{equation}
A\sim- \frac{2}{v_N} N^I + \frac{H_u^I}{v \sin \beta} + 
\frac{H_d^I}{v
\cos \beta},
\end{equation}
acquires a mass
\begin{equation}
m^2_A=\frac{1}{2}\lambda_H\lambda_Nv^2_N(\tan\beta+\cot\beta)
+\lambda_H\lambda_Nv^2\sin2\beta
\label{mA}
\end{equation}
 through the $| F_N | ^2$ term in the scalar potential.

The pseudoscalar $a$ acquires a mass once an 
$A_H$-term is generated, at one loop, through interactions 
with the gauginos.
Including only
the wino contribution in
the one loop RGE, $A_H$ is given by
\begin{eqnarray}
A_H & \approx &
6 \frac{\alpha _2 (\Lambda _{mess})}{4 \pi} M_2 \lambda _H \ln 
\left( \frac{\Lambda _{mess}}{M_2} \right), \nonumber \\
 & \approx & 20 \; \lambda _H  
\left( \frac{M_2}{280 \hbox{GeV}} \right) \hbox{GeV},
\label{A}
\end{eqnarray}
where $M_2$ is the wino mass at the weak scale.
Neglecting the mass mixing between the two pseudoscalars,
the mass of the pseudo-Nambu-Goldstone boson 
is computed to be 
\begin{eqnarray} 
m^2_a & = & \frac{9}{\sqrt{2}} A v_N v_u v_d
/ (v_N^2 + v^2 \sin ^2 2 \beta) \nonumber \\
 & \approx &  (40)^2 \left( \frac{\lambda _H}{0.1} \right)
\frac{M_2}{280 \hbox{GeV}}
\sin 2 \beta 
\left( \frac{ \frac{ {\textstyle v_N} }
{ {\textstyle 250} \hbox{GeV} } }
{\sin^2 2\beta+ \left ( \frac{ {\textstyle v_N} }
{ {\textstyle 250} \hbox{GeV} } \right)^2}\right )
(\hbox{GeV})^2.
\label{ma}
\end{eqnarray}
If the mass of $a$ is less than 7.2 GeV, it could 
be detected in the decay  
$\Upsilon \rightarrow a + \gamma$ \cite{pdg}.
Comparing 
the ratio of decay width for
$\Upsilon \rightarrow a + \gamma$ to 
$\Upsilon \rightarrow \mu ^- +\mu ^+$ \cite{pdg,wilczek},
the limit
\begin{equation}
\frac{\sin 2 \beta \tan \beta}
{ \sqrt{ ( \frac{ {\textstyle v_N} }
{ {\textstyle 250} \hbox{GeV} } )^2 
+ \sin ^2 2 \beta } } < 0.43
\label{bb}
\end{equation}
is found.

Further constraints on the spectra are obtained from 
collider searches.
The non-detection of $Z \rightarrow$ scalar + $a$ at 
LEP implies that 
the combined mass of the lightest Higgs scalar and $a$ must 
exceed $\sim$ 92 GeV. 
Also, the process $e^{+}e^{-}$
$\rightarrow$$Zh$ may be observable at LEP2.
For $\lambda _H = 0.1$, 
the constraint $m_h + m_a \; \gtap \;
92$ GeV is stronger than $m_h \; \gtap \; 70$
GeV which is 
the limit 
from LEP at
$\sqrt{s} \approx 170$ GeV \cite{aleph}. 
The contour of $m_h + m_a =
92$ GeV is shown in Fig.\ref{ftN1}a.
In Fig.\ref{ftN1}b, we show
the contour of $m_h = 92$ GeV ($\sim$ 
the ultimate LEP2 reach \cite
{cerngroup}).
For $\lambda _H = 0.5$, we find that the constraint 
$m_h \; \gtap \; 70$ GeV
is stronger than $m_h + m_a \; \gtap \; 92$ GeV and restricts
 $\tan \beta \; \ltap \; 5$ independent of $\Lambda _{mess}$.
The contour $m_{h} = 92 $ GeV is shown
in Fig.\ref{ftN1}d. 
We note that the allowed parameter space is not 
significantly constrained.  
We find that 
these limits make 
the constraint of Eqn.(\ref{bb}) redundant.
The left-right mixing between the two top squarks was 
neglected in 
computing the top squark radiative corrections to the 
Higgs masses.

The pseudo-Nambu-Goldstone boson $a$ might be 
produced along with the lightest scalar $h$ 
at LEP. 
The (tree-level) 
cross section in units of $R=87/s$ nb
is  
\begin{equation}
\sigma (e^+e^- \rightarrow h\;a) \approx 0.15
\frac{s^2}{(s-m_Z^2)^2} \; \lambda ^2 \; 
v\left(1,\frac{m^2_h}{s},\frac{m^2_a}{s}\right)^3,
\end{equation} 
where $g \lambda /\cos\theta_W$ is the 
$Z(a$$\partial$$h-h$$\partial$$a)$ coupling, and \\
$v(x,y,z)=\sqrt{(x-y-z)^2-4yz}$. 
If $h = c_N N^R + c_u H_u^R + c_d H_d^R $, then
\begin{equation}
\lambda =  \sin 2 \beta \frac{\cos \beta \;
c_u - \sin \beta \; c_d}
{ \sqrt{ ( \frac{ {\textstyle v_N} }
{ {\textstyle 250} \hbox{GeV} } )^2
+ \sin ^2 2 \beta }  } .
\end{equation}
We have numerically checked the parameter space allowed 
by $m_h \; \gtap \; 70$ GeV 
and $\lambda_H\leq$0.5 and have found the production 
cross section for 
$h\;a$ to be less than both the current limit set by DELPHI 
\cite{delphi} 
and a 
(possible) exclusion limit of 30 fb \cite{cerngroup}
at $\sqrt{s} \approx $ 192 GeV.
The production cross-section for $h\;A$ is larger 
than for $h\;a$ and $A$ is therefore in principle easier 
to detect.
However, for the 
parameter space allowed by $m_h \; \gtap \; 70$ GeV, 
numerical calculations show that 
$m_A \; \gtap \; $ 125 GeV,
so that this channel is not kinematically accessible.

The charged Higgs mass is
\begin{equation}
m^2_{H ^\pm} = m^2_W + m^2_{H_u} + m^2_{H_d} + 2 \mu ^2
\end{equation}
which is greater than about 200 GeV in this model 
since $m_{H_d}^2 \; \gtap \; (200 \; \hbox{GeV} )^2$
for $\Lambda _{mess} \; \gtap \; 35$ TeV
and as $\mu ^2 \sim - m^2_{H_u}$.

The neutralinos and charginos may be observable at LEP2 at
$\sqrt{s} \approx 192$ GeV
if $m_{\chi^+} \; \ltap \; 95$ GeV and 
$m_{\chi^0_1}+m_{ \chi^0_2} \; \ltap \; 180$ GeV.
These two constraints are comparable, and thus 
only one of these
is displayed in Figs.\ref{ftN1}b and \ref{ftN1}d, for 
$\lambda _H = 0.1$ and
$\lambda _H = 0.5$
repectively. Also, 
contours of $m_{\chi^0_1}+m_{ \chi^0_2} =$ 160 GeV ($\sim$
the LEP kinematic limit at  $\sqrt{s} \approx 170$ GeV)
are shown in Figs.\ref{ftN1}a and \ref{ftN1}c.
Contours of 85 GeV ($\sim$ the ultimate LEP2 limit)
and 75 GeV ($\sim$ the LEP 
limit from $\sqrt{s} \approx 170$ GeV)
for the right-handed selectron mass further constrain the 
parameter space. 

The results presented in all the figures are for a 
central value
of $m_t$=175 GeV. We have varied the top quark 
mass by 10 GeV about the central value of 
$m_t$= 175 GeV and
have found that both the fine tuning measures and the 
LEP2 constraints (the Higgs
mass and the neutralino masses) vary by $\approx$ 30 $\%$, 
but the 
qualitative features are unchanged.

We see from Fig.\ref{ftN1} that there is 
parameter space allowed by the present limits 
in which the tuning is $\approx$
30 $\%$. Even if no new particles are discovered at
LEP2, the tuning required for some region is
$\approx$ 20$\%$. 

\begin{figure}
\centerline{\epsfxsize=1\textwidth \epsfbox{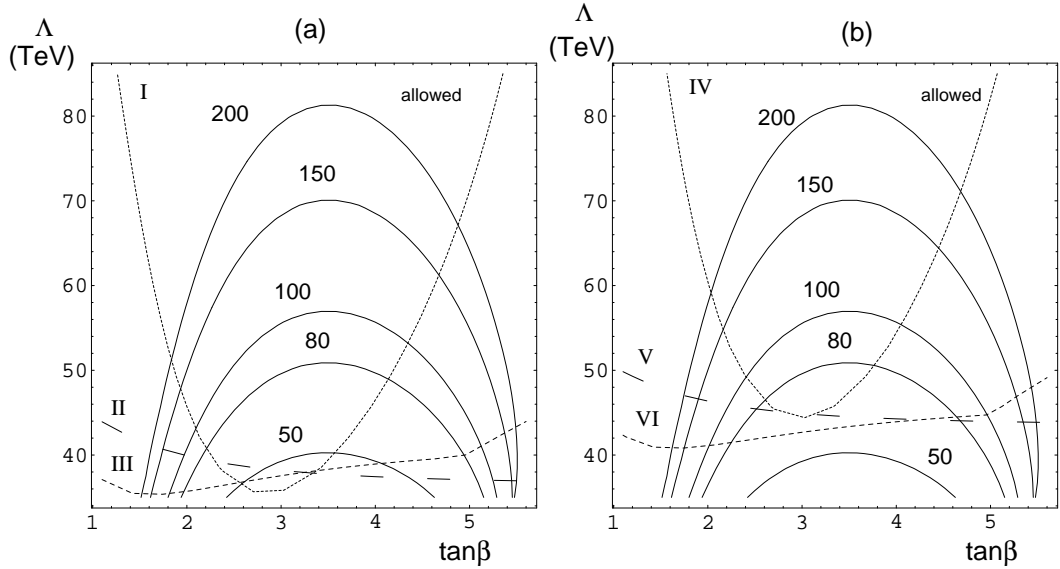}}
\vspace{-2in}
\caption{Contours of
$c(m^2_Z;\lambda_H)=$(50, 80, 100, 150, 200)
 for the NMSSM of section \ref{NMSSM} with
$\lambda_H=$0.1 and a messenger
particle content of one $({\bf 5+\bar{5}})$.
The constraints considered are:
(I) $m_h+m_a=m_Z$,
(II) $m_{\tilde{e}_R}=$75 GeV,
(III)
$m_{{\tilde{\chi}}^{0}_{1}}+m_{{\tilde{\chi}}^{0}_{2}}=$
160 GeV,
(IV) $m_h=$92 GeV,
(V) $m_{\tilde{e}_R}=$85 GeV, and
(VI)
$m_{{\tilde{\chi}}^{0}_{1}}+m_{{\tilde{\chi}}^{0}_{2}}=$
180 GeV.
A central
value of $m_{top}=$175 GeV
is assumed.}
\protect\label{ftdns2}
\end{figure}

It is also interesting to compare the fine tuning 
measures with those found in the minimal LEGM model
 (one messenger $({\bf 5+\bar{5}})$)
with an extra singlet $N$ to generate the $\mu$ and
$\mu^2_3$ terms.\footnote{We assume that the
model contains some mechanism to generate 
$-m^2_N\sim(100 \hbox{GeV})^2 - (200 \hbox{GeV})^2$;
for example, the singlet is coupled to an extra 
$({\bf 5+\bar{5}}$).}
In Fig.\ref{ftdns2} the fine tuning 
contours for $c(m^2_Z;\lambda_H)$ are presented for 
$\lambda_H$=0.1. 

Contours of $m_{\tilde{e} _R} = 75$ GeV and 
$m_{\chi^0_1}+m_{ \chi^0_2} =$ 160 GeV are also shown in 
Fig.\ref{ftdns2}a.  
For $\lambda _H = 0.1$, the
constraint $m_h + m_a \; \gtap \; 92$ GeV is stronger than 
the limit $m_h \; \gtap \; 70$ GeV and is shown in the 
Fig.\ref{ftdns2}a.
In Fig.\ref{ftdns2}b, we show the (approximate)
ultimate LEP2 limits, {\it i.e.,}
$m_h = 92$ GeV, $m_{\chi^0_1}+m_{ \chi^0_2} =$ 180 GeV
and  $m_{\tilde{e} _R} = 85$ GeV. 
Of these constraints, the bound on the lightest Higgs mass
(either $m_h + m_a \; \gtap \; 92$ GeV or $m_h \; 
\gtap \; 92$ GeV) provides
a  strong lower limit on the messenger scale.
We see that in the parameter space allowed by
present limits 
the fine tuning is $\ltap \;2 \%$
and if LEP2 does
not discover new particles, the fine tuning will
be $\ltap \;1 \%$.
The coupling $\lambda_H$ is constrained to 
be not significantly larger than 0.1 if the
constraint 
$m_h + m_a \; \gtap \; 92$ GeV (or $m_h \; \gtap$ 92 GeV) 
is imposed and if the
fine tuning is required to be no worse than 1$\%$.
 


\section{Models Derived from a GUT}
\label{GUT}
In this section, we discuss how the toy model 
of section \ref{toymodel} 
could be derived from a GUT model.

In the toy model of section \ref{toymodel}, the singlets 
$N$ and $S$ do not separately couple to complete $SU(5)$ 
representations (see Eqn.(\ref{3doublets})). 
If the extra fields introduced to solve the fine tuning 
problem
were originally part of $({\bf 5+\bar{5}})$ multiplets,
then the missing triplets (missing doublets)
necessarily couple to the singlet $S(N)$. The triplets
 must
be heavy in order to suppress their contribution to the
soft SUSY breaking mass parameters. If we assume
 the only
other mass scale is $M_{GUT}$, they must acquire a mass
at $M_{GUT}$. This is just the usual problem of 
splitting a $({\bf 5+\bar{5}})$ \cite{splitting}. 
For example, if the 
superpotential in the messenger sector 
contains four $({\bf 5+\bar{5}})$'s,  
\begin{equation}
W=\lambda_1S\bar{5}_{l1} 5_{l1} + 
\lambda_2S\bar{5}_{l2} 5_{l2} 
+ \lambda_3S\bar{5}_{l3} 5_{l3} + \lambda_4S\bar{5}_{q} 5_{q},
\end{equation}
then the $SU(3)$ triplets in the $(\bar{5}_l+ 5_l)$'s 
and the $SU(2)$ doublet in $(\bar{5}_q + 5_q)$ must be 
heavy at $M_{GUT}$
so that in the low energy theory there are three doublets 
and one triplet 
coupling to $S$.
This problem can be solved 
using the method of Barbieri, Dvali and Strumia 
\cite{barbieri2} 
that solves 
the usual Higgs doublet-triplet splitting problem. 
The mechanism in this model is
attractive since 
it is possible to make either the doublets or triplets 
of a 
quintet heavy at the GUT scale. We next describe 
their model.
 
 The gauge group is $SU(5) \times 
SU(5) ^{\prime}$, with the particle content
$\Sigma ({\bf 24,1}), \\
\Sigma ^{\prime}({\bf 1,24}),
\Phi ({\bf 5, \bar{5}}) \, \hbox{and} \,
\bar{\Phi} ({\bf \bar{5}, 5}) $
and the superpotential can be written as 
\begin{eqnarray}
W&=& \bar{\Phi} ^{\beta} _{\alpha ^{\prime}}
(M_{\Phi} \delta ^{\alpha ^{\prime}}
_{\beta ^{\prime}} \delta ^{\alpha}
_{\beta}+ \lambda \Sigma ^{\alpha}_{\beta} \delta 
^{\alpha ^{\prime}}
_{\beta ^{\prime}}
+\lambda ^{\prime} {\Sigma ^{\prime}}^{\alpha ^{\prime}}
_{\beta^{\prime}} \delta ^{\alpha}
_{\beta}) \Phi ^{\beta ^{\prime}}_{\alpha} 
+ \nonumber \\
 & &  + \frac{1}{2} M_{\Sigma} \hbox{Tr} (\Sigma ^2) 
+ \frac{1}{2} 
M_{\Sigma ^{\prime}} 
\hbox{Tr} (\Sigma^{\prime 2}) + \nonumber \\
 & & \frac{1}{3} \lambda _{\Sigma} \hbox{Tr} \Sigma ^3 +
\frac{1}{3} \lambda _{\Sigma ^{\prime}} \hbox{Tr} 
\Sigma ^{\prime3}.
\end{eqnarray}
A supersymmetric minimum of the scalar potential 
satisfies the 
$F$ - flatness
conditions
\begin{eqnarray}
0&=&F_{\bar{\Phi}}=(M_{\Phi} \delta ^{\alpha ^{\prime}}
_{\beta ^{\prime}} \delta ^{\alpha}
_{\beta}+ \lambda \Sigma ^{\alpha}_{\beta} 
\delta ^{\alpha ^{\prime}}
_{\beta ^{\prime}}
+\lambda ^{\prime} \Sigma
 ^{\prime\alpha^{\prime}}_{\beta ^{\prime}} \delta ^{\alpha}
_{\beta} )
\Phi ^{\beta^{\prime}} _{\alpha} ,\nonumber \\
0&=&F_{\Sigma}=\frac{1}{2} M_{\Sigma} 
\Sigma _{\alpha}^{\beta} + 
\frac{1}{2} \left( \lambda \bar{\Phi} 
_{\alpha ^{\prime}} ^{\beta}
\Phi ^{\alpha ^{\prime}} _{\alpha} - \lambda 
\frac{1}{5} \delta _
{\alpha} ^{\beta} \hbox{Tr} 
(\bar{\Phi} \Phi) \right) 
 + \lambda _{\Sigma} ( \Sigma ^2 - \frac{1}{5} \hbox
{Tr} \Sigma ^2 ) ,\nonumber \\
0&=&F_{\Sigma ^{\prime}}=\frac{1}{2} M_{\Sigma ^{\prime}} 
\Sigma^{\prime\beta^{\prime}}  
_{\alpha ^{\prime}} +
\frac{1}{2} \left(  \lambda ^{\prime} 
\bar{\Phi} ^{\alpha} _{\alpha ^{\prime}}
\Phi ^{\beta ^{\prime}} _{\alpha} - \lambda ^{\prime} 
\frac{1}{5} \delta _
{\alpha ^{\prime}} ^{\beta ^{\prime}} \hbox{Tr} 
(\bar{\Phi} \Phi) \right)
+ \lambda _{\Sigma ^{\prime}} ( \Sigma ^{\prime2} 
- \frac{1}{5} \hbox
{Tr} {\Sigma ^{\prime}}^2 ).
\nonumber \\
\end{eqnarray}
With the ansatz 
\footnote{The 
two possible solutions to the $F$-flatness conditions are
$\langle \Sigma \rangle = v_{\Sigma} \, \hbox{diag}(2,2,2,-3,-3)$ and 
$\langle \Sigma \rangle = v_{\Sigma} \, \hbox{diag}(1,1,1,1,-4)$.}
\begin{equation}
\Sigma = v_{\Sigma} \, \hbox{diag}(2,2,2,-3,-3) \, ,
\Sigma ^{\prime} 
= v_{\Sigma ^{\prime}} \, \hbox{diag}(2,2,2,-3,-3),
\end{equation}
the $F_{\bar{\Phi}}=0$ condition is 
\begin{equation}
\hbox{diag} [M_3, M_3, M_3, M_2, M_2]\cdot
\hbox{diag} [v_3,v_3,v_3,v_2,v_2]=0,
\end{equation}
where $M_3 = M_{\Phi} + 2 \lambda v_{\Sigma} + 
2 \lambda ^{\prime} 
v_{\Sigma ^{\prime}}$
and $M_2 = M_{\Phi} - 3 \lambda v_{\Sigma} 
- 3 \lambda ^{\prime} v_{\Sigma ^{\prime}}$ and the second 
matrix is 
the vev of $\Phi$. 
To satisfy
this condition, there is a discrete choice for the pattern 
of vev of 
$\Phi$ :
i) $v_3 \neq 0 \, \hbox{and} \, M_3=0 $ or
ii) $v_2 \neq 0 \, \hbox{and} \, M_2=0 $. 
Substituting either i) or ii) in the $F_{\Sigma}$ and 
$F_{\Sigma ^{
\prime}}$ conditions then 
determines
 $v_3$ (or $v_2$).
With two sets of fields, $\Phi _1, \bar{\Phi} _1$ 
with $v_3 \neq 0$ 
and $\Phi _2, \bar{\Phi} _2$ with $v_2 \neq 0$ , we have the 
following pattern
of symmetry breaking
\begin{eqnarray}
SU(5) \times SU(5) ^{\prime} 
& \stackrel{v_{\Sigma},v_{\Sigma ^{\prime}}}
{\rightarrow} &
 (SU(3) \times SU(2) \times U(1))
\times (SU(3) \times SU(2) \times U(1)) ^{\prime} \nonumber \\
  & \stackrel{v_3,v_2}
{\rightarrow}& SM \, 
\hbox{(the diagonal subgroup)}.
\end{eqnarray}
If the scales of the two stages of symmetry breaking 
are about equal,
{\it i.e.} $v_{\Sigma} , \, v_{\Sigma ^{\prime}}
, \, \sim v_3 \, , v_2 \, \sim M_{GUT}$, then
the SM gauge couplings unify at the scale $M_{GUT}$.
\footnote{See \cite{barbieri2} and \cite{barr} 
for models which give this structure of vevs
for the $\Phi$ fields without using the adjoints.}

The particular structure of the 
vevs of $\Phi _1$ and $\Phi _2$ can be 
used to 
split representations as follows. 

Consider the Higgs doublet-triplet splitting problem. 
With the particle
content
$5_h ({\bf 5,1})$,
 $\bar{5}_h ({\bf {\bf \bar{5},1}})$ and
$X ({\bf 1,5})$, $\bar{X} ({\bf 1,\bar{5}})$
and the superpotential
\begin{equation}
W = 5 _{h \alpha} \bar{X} ^{\alpha ^{\prime}} 
\bar{\Phi}^{\alpha} _
{1\alpha ^{\prime}} 
+ \bar{5}_h ^{\alpha} X _{\alpha ^{\prime}} {\Phi _1}
 ^{\alpha ^{\prime}} _{\alpha} ,
\end{equation}
the $SU(3)$ triplets in $5_h$, $\bar{5}_h$ and $X$, 
$\bar{X}$ acquire a
mass of order $M_{GUT}$ whereas the doublets in $5_h$, 
$\bar{5}_h$ and 
$X$, $\bar{X}$ are massless. We want only one pair of 
doublets in the 
low energy theory
(in addition to the usual matter fields). The doublets 
in $X$, $\bar{X}$
can be made heavy by a bare mass term $M_{GUT} X \bar{X}$. 
Then
the doublets in 
$5_h, \bar{5}_h$ are the standard Higgs doublets.
But if all
terms consistent with symmetries are allowed 
in the superpotential, then allowing
$M_{GUT} \Phi _1 \bar{\Phi} _1$, $M_{GUT} X \bar{X}$, 
$5_h \bar{X} \Phi _1$ 
and 
$\bar{5}_h X \bar{\Phi} _1$ implies that a bare mass 
term for $5_h \bar{5}_h$ is 
allowed. Of course, we can by hand put in 
a $\mu$ term $\mu 5_h \bar{5}_h$ of 
the order of the weak scale as in section \ref{toymodel}. 
However, 
it is theoretically more desirable to relate all electroweak 
mass scales to the original SUSY breaking scale. 
So, we would like to relate the $\mu$ term to the SUSY 
breaking
scale. We showed in section \ref{NMSSM} that the NMSSM is 
phenomenologically viable and ``un-fine tuned''
in these models.

The vev structure of $\Phi _2$, 
$\bar{\Phi} _2$ can be used to 
make the doublets in a $({\bf 5 + \bar{5}})$ heavy. Again, 
we get two pairs of 
light triplets and one of these 
pairs can be given a mass  at the GUT scale.
  
We can use this mechanism of making either doublets or 
triplets in
a $({\bf 5+\bar{5}})$ heavy to show how the model of section 
\ref{toymodel} is derivable from 
a GUT. 
The model with three messenger doublets and
one triplet is obtained from a GUT with the following 
superpotential
\begin{eqnarray}
W & = & S 5 \bar{5} + S 5_l \bar{5}_l 
+ S X_l \bar{X}_l + \nonumber \\
 & &  5_l \bar{X}_l \bar{\Phi} _1 
+ \bar{5}_l X_l \Phi _1 + \nonumber  \\
 & &  5_q \bar{X}_q \bar{\Phi} _2 
+ \bar{5}_q X_q \Phi _2 + \nonumber \\
 & &  M_{GUT} X_h \bar{X}_h 
+ 5_h \bar{X}_h \bar{\Phi} _1 
 + \bar{5}_h X_h \Phi _1 
+ \mu 5_h \bar{5} _h\nonumber \\
 & &  + N^3 + N5_q \bar{5} _q 
+ N X_q \bar{X} _q .
\end{eqnarray}
Here, some of the ``extra'' triplets and doublets 
resulting from 
splitting $({\bf 5+\bar{5}})$'s are massless at the GUT scale. 
For example, the 
``extra'' light doublets are used 
as the additional messenger leptons.
After inserting the vevs and integrating out the heavy 
states, this corresponds to the 
superpotential in Eqn.(\ref{3doublets}) with the 
transcription:
\begin{eqnarray}
5,\bar{5} & \rightarrow & q_1,\bar{q} _1 
+ l_1,\bar{l} _1 \nonumber \\
5_l, \bar{5}_l 
& \rightarrow & l_2,\bar{l} _2 \nonumber \\
X_l, \bar{X}_l & 
\rightarrow & l_3,\bar{l} _3 \nonumber \\
5_q, \bar{5} _q & 
\rightarrow & q_2, \bar{q} _2 \nonumber \\
X_q, \bar{X} _q & 
\rightarrow & q_3, \bar{q} _3 .
\end{eqnarray}                                                                  
                                              
We conclude this section with a remark about light 
singlets in SUSY-GUT's with low energy 
gauge mediated SUSY breaking.
In a SUSY GUT with a singlet $N$
coupled to the Higgs multiplets,
there is a potential problem of destabilising
the $m_{weak} / M_{GUT}$ hierarchy, if the singlet
is light and if the Higgs triplets have a
SUSY invariant mass of $O(M_{GUT})$ \cite{srednicki}.
In the LEGM models,
a
B-type mass for
the Higgs triplets and doublets is generated at one loop
with gauginos and
Higgsinos in the loop, and with
SUSY breaking coming from the gaugino mass.
Since SUSY breaking (the gaugino mass and the soft scalar
masses) becomes soft above
the messenger scale,
$\Lambda _{mess} \sim$ 100 TeV,
the B-type mass term generated for the Higgs triplets
 is suppressed, {\it i.e.},
it is $O( (\alpha/4 \pi) M_2
\Lambda _{mess}^2 / M_{GUT})$.
Similarly the soft (mass)$^2$
for the Higgs triplets are
$O(m_{weak} ^2 \Lambda _{mess}^2 / M_{GUT}
^2)$. Since the triplets couple to the singlet $N$, 
the soft scalar mass and $B$-term generate at one loop
a 
linear term for the scalar and $F$-component of $N$ 
respectively. These tadpoles are harmless since 
the SUSY breaking masses for the triplets are so 
small.
This is to be contrasted with supergravity theories,
where the $B$-term$\sim O(m_{weak}M_{GUT})$ 
and the soft mass $\sim O(m_{weak})$ for the 
triplet Higgs generate a mass for the Higgs doublet
that is at least $\sim O(\sqrt{m_{weak}M_{GUT}}/(4\pi))$.
 
\section{One complete Model}
\label{complete}
The model is based on the gauge group 
$G_{loc}=SU(5) \times SU(5)'$ and 
the global symmetry group 
$G_{glo}=Z_3 \times Z_3' \times Z_4$. The global symmetry acts 
universally on the three
 generations of the SM. The particle 
content and their 
$G_{loc} \times G_{glo}$ quantum numbers are given
in Table \ref{completemodel}. 

\renewcommand{\arraystretch}{0.6}
\begin{table}
\begin{center}
\begin{tabular}{||l||l|l|l|l||}\hline
$\Psi$ &$\bar{5}_i$ &$10_i$ &$5_h$ &$\bar{5}_h$ \\ \hline
$G_{loc}$ &$({\bf \bar{5},1})$ &$({\bf 10,1})$ &$({\bf 5,1})$
&$({\bf \bar{5},1})$ \\ \hline
$Z_3$ &$1$ &$a$ &$a$ &$a^2$ \\ \hline
$Z'_3$ &$b$ &$1$ &$1$ &$b^2$ \\ \hline
$Z_4$ &$c$ &$c$ &$c^2$ &$c^2$ \\ \hline
\end{tabular}
\end{center}  

\begin{center}
\begin{tabular}{||l||l|l|l|l|l|l||}\hline
$\Psi$ &$\Sigma$ &$\Sigma'$ &$\bar{\Phi} _2$ &$\Phi_2$
&$\bar{\Phi} _1$ &$\Phi _1$ \\ \hline
$G_{loc}$ &$({\bf 24,1})$ &$({\bf 1,24})$ &$({\bf \bar{5},5})$
&$({\bf 5,\bar{5}})$ &$({\bf \bar{5},5})$ &$({\bf 5,\bar{5}})$
\\ \hline
$Z_3$ &$1$ &$1$ &$1$ &$1$ &$1$ &$1$ \\ \hline
$Z'_3$ &$1$ &$1$ &$1$ &$1$ &$1$ &$1$ \\ \hline
$Z_4$ &$1$ &$1$ &$1$ &$1$ &$c^2$ &$c^2$ \\ \hline
\end{tabular}
\end{center} 

\begin{center}
\begin{tabular}{||l||l|l|l|l|l|l||}\hline
$\Psi$ &$5_l$ &$\bar{5}_l$ &$X_l$ &$\bar{X}_l$
&$5_q$ &$\bar{5}_q$   \\ \hline
$G_{loc}$ &$({\bf 5,1})$ &$({\bf \bar{5},1})$
&$({\bf 1,5})$ &$({\bf 1,\bar{5}})$ &$({\bf 5,1})$ &$({\bf \bar{5},1})$
 \\ \hline
$Z_3$ &$a^2$ &$1$ &$1$ &$a$ &$1$ &$a^2$
  \\ \hline
$Z'_3$ &$1$ &$1$ &$1$ &$1$ &$b^2$ &$b$  \\ \hline
$Z_4$ &$c^2$ &$c^2$ &$1$ &$1$ &$1$ &$1$  \\ \hline
\end{tabular}
\end{center}

\begin{center}
\begin{tabular}{||l||l|l|l|l|l|l||}\hline
$\Psi$ &$X_q$ &$\bar{X}_q$ &$X_h$ &$\bar{X}_h$ &$X$
&$\bar{X}$ \\ \hline
$G_{loc}$ &$({\bf 1,5})$ &$({\bf 1,\bar{5}})$
 &$({\bf 1,5})$ &$({\bf 1,\bar{5}})$ &$({\bf 1,5})$ &$({\bf 1,\bar{5}})$ 
\\ \hline
$Z_3$ &$a$ &$1$ &$a$ &$a^2$ &$a^2$ &$a$ \\ \hline
$Z'_3$ &$b^2$ &$b$ &$b$ &$1$ &$1$ &$b^2$ \\ \hline
$Z_4$ &$1$ &$1$ &$1$ &$1$ &$1$ &$1$ \\ \hline
\end{tabular}
\end{center} 

\begin{center}
\begin{tabular}{||l||l|l|l|l|l||}   \hline
$\Psi$ &$S$ &$N$ &$N'$ &$\Phi_+$ &$\Phi_-$\\ \hline
$Z_3$ &$a$ &$1$ &$a$ &$a$ &$a$\\ \hline
$Z'_3$ &$1$ &$b$ &$b^2$ &$1$ &$1$\\ \hline
$Z_4$ &$1$ &$1$ &$1$ &$1$ &$1$\\ \hline
\end{tabular}
\end{center}
\caption{$SU(5) \times SU(5)' \times Z_3
\times Z'_3 \times Z_4$ quantum
numbers for the fields of the model
discussed in section 3.6. The generators
of $Z_3 \times Z'_3 \times Z_4$ are labeled by $(a,b,c)$.
The three SM generations are
labeled by the index $i$.}
\label{completemodel}
\end{table}

The most general renormalizable 
superpotential that is consistent with these symmetries 
is
\begin{equation}
W=W_1+W_2+W_3+W_4+W_5+W_6+W_7,
\end{equation}
where,
\begin{eqnarray}
W_1&=&\frac{1}{2}M_{\Sigma}\hbox{Tr}\Sigma^2+
\frac{1}{3}\lambda_{\Sigma}
\hbox{Tr}\Sigma^3
+\frac{1}{2}M_{\Sigma ^{\prime}} \hbox{Tr} \Sigma ^{\prime 2}
+\frac{1}{3}\lambda _{\Sigma ^{\prime}}
\hbox{Tr}\Sigma ^{\prime 3} \nonumber \\
& & +\Phi _2(M_{\Phi _2}+\lambda_{\Phi _2}\Sigma
+\lambda ^{\prime} _{\Phi _2} \Sigma^{\prime}) \bar{\Phi} _2
\nonumber \\
 & & +\Phi _1(M_{\Phi _1}+\lambda_{\Phi _1}\Sigma+
\lambda ^{\prime} _{\Phi _1} \Sigma ^{\prime}) \bar{\Phi} _1,\\
W_2&=&M_1\bar{X}_lX ,\\
W_3&=&\lambda_1\bar{5}_h\Phi _1X_h
+\bar{\lambda}_15_h\bar{\Phi}_1\bar{X}_h
+\lambda_2\bar{5}_l\Phi _1X_l
+\bar{\lambda}_25_l\bar{\Phi}_1\bar{X}_l, \\
W_4&=&\lambda_3\bar{5}_q\Phi _2X_q
+\bar{\lambda}_35_q\bar{\Phi}_2\bar{X}_q ,\\
W_5&=&\lambda_6S5_l\bar{5}_l+\lambda_7S5_q\bar{5}_q
+\lambda_8S\bar{X}_hX_l+\lambda_9S\bar{X}X_h
+\frac{1}{3}\lambda_{S}S^3 ,\\
W_6&=&-\lambda_H 5_h\bar{5}_hN
+\frac{1}{3}\lambda_{N}N^3+
\bar{\lambda} _q N X \bar{X} \nonumber \\
& &+\lambda_{10}N'\bar{X}X_q+\lambda_{11}N'\bar{X}_qX
+\frac{1}{3}\lambda_{N'}N'^3, \\
W_7&=&\lambda^D_{ij}\bar{5}_i10_j\bar{5}_h
+\lambda^U_{ij}10_i10_j5_h.
\end{eqnarray}
The origin of each of the $W_i$'s appearing in the 
superpotential is easy to understand. In computing the
$F$=0 equations at the GUT scale, the only non-trivial 
contributions come 
from fields appearing in $W_1$, since all other $W_i$'s 
are bilinear in fields that do not acquire vevs at the 
GUT scale. The function of $W_1$ is to generate the vevs 
$\Sigma,\Sigma'\sim$ diag $[2,2,2,-3,-3]$, 
$\bar{\Phi}^T_2=\Phi _2\sim$ diag $[0,0,0,1,1]$ 
and $\bar{\Phi}^T_1=\Phi _1\sim$ diag $[1,1,1,0,0]$.
These vevs are necessary to break 
$G_{loc}\rightarrow$$SU(3)_c \times SU(2) \times U(1)_Y$
(this was explained in section \ref{GUT}). 
The role of
$W_3$ and $W_4$ 
is to generate the necessary splitting within the 
many $({\bf 5+\bar{5}})$'s of $G_{loc}$ that is necessary
to solve the usual doublet-triplet splitting problem, as
well as to solve the fine tuning problem that is discussed 
in sections \ref{finetune}, \ref{toymodel} and \ref{NMSSM}. 
The messenger sector 
is given by $W_5$. It will shortly be 
demonstrated that at low energies this sector contains 
three vector-like doublets and one vector-like triplet. The 
couplings in $W_6$ and $W_7$ at low energies contain the 
electroweak symmetry breaking sector of the NMSSM, the 
Yukawa couplings of the SM fields, and the 
two light vector-like triplets necessary to maintain 
the few percent prediction for $\sin^2\theta_W$ as well as
to generate a vev for $N$. 

We now show that the low energy theory of this model is the 
model that is discussed in section \ref{NMSSM}.

Inserting the 
vevs for $\Phi _1$ and $\bar{\Phi} _1$ into $W_3$
and from $W_2$, the 
following 
mass matrix for the colored triplet chiral 
multiplets is obtained:
\begin{equation}
(\bar{5}_h,\bar{X}_h,\bar{5}_l,\bar{X}_l)
\left(\begin{array}{ccccc}
0& \lambda_1 v_{\Phi _1}& 0 &0 &0 \\
\bar{\lambda}_1 v_{\Phi _1} &0 &0 &0 &0  \\
0& 0& 0& \lambda_2 v_{\Phi _1}& 0 \\
0 &0 &\bar{\lambda}_2 v_{\Phi _1} &0 &M_1  \\
\end{array} \right)
\left(\begin{array}{c}
5_h \\
X_h \\
5_l \\
X_l \\
X \\
\end{array} \right)
\end{equation}
and all other masses are zero.
There are a total of four  
vector-like colored triplet fields that 
are massive at $M_{GUT}$.
These are
the triplet components of 
$(5_h,\bar{X}_h)$, $(\bar{5}_h,X_h)$, 
$(\bar{5}_l,X_l)$ and $(\bar{X}_l,T_H)$,
where
$T_H$ is that linear combination of triplets in $5_l$ and 
$X$ that marries the triplet component of $\bar{X}_l$. 
The 
orthogonal combination to $T_H$, $T_L$, is massless at 
this scale.
The massless triplets at $M_{GUT}$ are $(5_q,\bar{5}_q)$, 
$(X_q,\bar{X}_q)$ and $(\bar{X},T_L)$, for a total of 
three vector-like
triplets. By inspection of $W_5$, the only light triplets that 
couple to $S$
at a renormalizable level 
are $5_q$ and $\bar{5}_q$, which was desirable in 
order to solve 
the fine tuning problem. Further, since $X$ contains a 
component of 
$T_L$, the couplings of the other light triplets to the 
singlets 
$N$ and $N'$ are 
\begin{equation}
W_{eff}=\lambda_{10}N'\bar{X}X_q
+\bar{\lambda}_{11}N'\bar{X}_qT_L
+\lambda _q NT_L\bar{X}     ,
\end{equation}
where $\lambda _q = \bar{\lambda} _q \cos \alpha ^{\prime}$,
$\bar{\lambda}_{11}=\lambda_{11} \cos \alpha^{\prime}$ 
and $\alpha^{\prime}$ is the mixing angle 
between
the triplets in 
$5_l$ and $X$, {\it i.e.},
$T_L = \cos \alpha^{\prime} X - \sin \alpha^{\prime} 5_l$. 
The $\lambda _q N T_L \bar{X}$ coupling is also desirable 
to generate
acceptable $\mu$ and $\mu^2_3$ terms (see section \ref{NMSSM}). 

In sections \ref{toymodel} and \ref{NMSSM} it was also 
demonstrated that with a total of three messenger doublets 
the fine tuning required in 
electroweak symmetry breaking could be alleviated.
By inserting the vev for $\Phi _2$ into $W_4$ and from $W_2$, the doublet 
mass matrix is
given as
\begin{equation}
(\bar{X}_l,\bar{5}_q,\bar{X}_q)
\left(\begin{array}{ccc}
M_1 &0 &0 \\
0 &0 &\lambda_3 v_{\Phi _2} \\
0 &\bar{\lambda}_3 v_{\Phi _2} &0 \\
\end{array} \right)
\left(\begin{array}{c}
X \\
5_q \\
X_q \\
\end{array} \right) 
\end{equation}
and all other masses are zero. At $M_{GUT}$ the heavy 
doublets are $(\bar{X}_l,X)$, $(5_q,\bar{X}_q)$ and 
$(\bar{5}_q,X_q)$, leaving the four vector-like doublets in 
$(5_h,\bar{5}_h)$, $(5_l,\bar{5}_l)$, $(\bar{X},X_l)$ and
$(X_h,\bar{X}_h)$ massless at this scale. Of these four pairs, 
$(5_h,\bar{5}_h)$ are the usual Higgs doublets and 
the other three pairs couple to $S$ (see $W_5$).

The (renormalizable) superpotential at 
scales below $M_{GUT}$ is then
\begin{eqnarray}
W&=&\lambda _q N \bar{q}_2 q_2
+\frac{1}{3}\lambda_{N}N^3+\lambda_{10} N^{\prime} q_3 \bar{q}_2
\nonumber \\
& & +\lambda_{11}N^{\prime} q_2 \bar{q}_3 - \lambda_H N H_u H_d
+\frac{1}{3}\lambda_{N^{\prime}}{N ^{\prime}} ^3 \nonumber \\
& &+\lambda_6S\bar{l}_1l_1
+\lambda_7S\bar{q}_1 q_1 + \lambda_8S\bar{l}_2l_2 
\nonumber \\
& &+\lambda_9 S \bar{l}_3l_3+\frac{1}{3}\lambda_{S}S^3+W_7,
\end{eqnarray}
where the fields have been relabeled to make, in an obvious 
notation,
their $SU(3) \times SU(2) \times U(1)$ quantum numbers apparent.
 
We conclude this section with comments about both the 
choice of 
$Z_4$ as a discrete symmetry and 
about non-renormalizable operators in our model.

The usual $R$-parity violating operators 
$10_{SM}\bar{5}_{SM}\bar{5}_{SM}$ are
not allowed by the discrete symmetries, even at the 
non-renormalizable level. In fact, $R$-parity
 is a good symmetry of the effective theory 
below $M_{GUT}$. By inspection, the fields that acquire 
vevs 
at $M_{GUT}$ are either
invariant under $Z_4$ or have a $Z_4$
charge of $2$ (for example, $\Phi _1$), so that a 
$Z_2$ symmetry is left
unbroken. In fact, the vevs of 
the other fields $S$, $N$, 
$N'$ and the Higgs doublets do not break 
this $Z_2$ either.
By inspecting the $Z_4$ charges of the
SM fields, we see that the unbroken $Z_2$ 
is
none other than the usual $R$-parity. So at $M_{GUT}$, 
the discrete symmetry $Z_4$ is broken 
to $R_p$.
We also note that the $Z_4$ symmetry 
is sufficient to maintain, to 
all orders in $1/M_{Pl}$ operators, the 
vev structure of $\Phi _1$ and $\Phi _2$, 
{\it i.e.}, to forbid
unwanted couplings between 
$\Phi _1$ and $\Phi _2$ that might destabilize 
the vev structure \cite{barr}. This pattern of
vevs  
was essential to solve the 
doublet-triplet splitting problem. 
It is interesting that both 
$R$-parity 
and requiring a viable solution 
to the doublet-triplet splitting problem can be 
accommodated by the same $Z_4$ symmetry.

The non-SM matter fields ({\it i.e.}, the messenger
$5$'s and $X$'s and the light triplets) have the opposite 
charge to the 
SM matter fields under the unbroken $Z_2$. Thus, there is 
no mixing 
between the SM and the non-SM matter fields. 

Dangerous proton decay operators are forbidden 
in this model by the discrete symmetries. Some
higher dimension operators that lead to 
proton decay are allowed, but are sufficiently 
suppressed. We discuss these below.

Renormalizable operators such as
$ 10_{SM} 10_{SM} 5_q$ and 
$ 10_{SM} \bar{5} _{SM} \bar{5} _q$
are 
forbidden by the $Z_3$ symmetries. This is 
necessary to avoid a large proton decay rate.  
A dimension-6 proton decay operator is
obtained by integrating out the colored triplet scalar
components of $5_q$ or $\bar{5}_q$.
Since the colored scalars in $5_q$ and $\bar{5}_q$ 
have a mass $\sim$$O($50 TeV$)$, the presence of these
operators would have led to an unacceptably large 
proton decay rate.

The operators 
$10_{SM} 10_{SM} 10_{SM} \bar{5}_{SM}/M_{Pl}$
and
$10_{SM} 10_{SM} 10_{SM} \bar{5}_{SM} \\ 
(\Phi \bar{\Phi}/M_{Pl}^2)^n /M_{Pl}$,
which give dimension-5 proton decay operators,
are also forbidden by the two $Z_3$ symmetries.
The allowed non-renormalizable operators
that generate dimension-5 proton decay 
operators 
are suffuciently suppressed. 
The operator
$10_{SM} 10_{SM} 10_{SM} \bar{5}_{SM} 
N'/(M_{Pl})^2$,
for example,
is allowed by the discrete symmetries,
but the proton decay rate is safe since 
$v_{N^{\prime}} \sim $ 1 TeV.

The operators 
$10_i 
\bar{5} _j \bar{\Phi} _1 (\bar{X} \; \hbox{or} \; \bar{X}_q) /M_{Pl}$ 
could, in principle, also lead to a large proton
decay rate. Setting $\bar{\Phi} _1$ to its vev, 
the superpotential couplings,
for example,
$\lambda_{ij}(U^c_i D^c_j \bar{X}(\bar{3})
+Q_i L_j \bar{X}(\bar{3}))$ are generated with 
$\lambda_{ij}$ suppressed only by 
$v_{\Phi _1}/M_{Pl}$.
In this model the colored triplet (scalar) components of
$\bar{X}$ and $\bar{X}_q$ have a mass $m_{\tilde{q}} \sim $
500 GeV, giving a  
potentially large proton decay rate. But, in this model 
these operators are forbidden by the discrete symmetries. 
The operator $10_i
\bar{5} _j \bar{\Phi} _1 \bar{X} S/M_{Pl}^2$ is allowed giving a 
four SM fermion proton decay operator with coefficient 
$\sim(v_{\Phi _1} \; v_S /M^2_{Pl})^2 /m_{\tilde{q}}^2 \sim
10^{-34} \hbox{GeV} ^{-2}$. 
This is smaller than the coefficient
generated by exchange of the heavy gauge bosons of mass 
$M_{GUT}$,
which is $\sim g^2_{GUT} / M^2_{GUT} \sim 1/2 \times \; 10^{-32} 
\hbox{GeV} ^{-2}$ and so this operator leads to
proton decay at a tolerable rate.

With
our set of discrete symmetries, some of the 
messenger states and the light color triplets 
are stable at the renormalizable level.
Non-renormalizable
operators lead to decay lifetime for 
some of these particles of more than about 100
seconds. This is a problem 
from the viewpoint of cosmology, since these
particles decay after Big-Bang Nucleosynthesis (BBN).
With a non-universal choice of discrete symmetries, it 
might be possible
to make these
particles decay before BBN through either 
small renormalizable 
couplings
to the third generation 
(so that the constraints from 
proton decay and FCNC are avoided)
or non-renormalizable operators. This is, however,
 beyond the 
scope of this chapter.

\section{Conclusions}
\label{conclude}
We have quantified the fine tuning 
required in
models of low energy gauge-mediated SUSY breaking to obtain
the correct $Z$ mass. We showed 
that the minimal model requires a fine tuning 
of order $\sim$
7$\%$  
if LEP2 does not discover a
right-handed slepton. We discussed 
how models with more messenger doublets than triplets
can improve the fine tuning.
In particular, a model
with a messenger field particle content of
three $(l+\bar{l})$'s and only one
$(q+\bar{q})$ was tuned to $\sim 40 \%$.
We found that it was necessary to introduce 
an extra singlet to give mass to some color triplets
(close to the weak scale) which are required to 
maintain gauge coupling 
unification. We also discussed 
how the vev and $F$-component of this singlet 
could be used to generate the $\mu$ and 
$B\mu$ terms. We 
found that
for some region of the parameter space this model
requires $\sim \; 25 \%$ tuning and have shown that
limits from LEP do not constrain the parameter space.
This is in contrast to an NMSSM with extra vector-like quintets
and with one
$({\bf 5+\bar{5}})$ messenger field, for which we found that
a fine tuning of $\sim \; 1\%$ is required and that
limits from LEP do significantly constrain the
parameter space.

We further discussed how the model with split messenger 
field representations
could be the
low energy theory of a $SU(5) \times SU(5)$ GUT.
A mechanism similar to
the one used to solve the usual
Higgs doublet-triplet splitting problem was used to split
the messenger field representations.
All operators
consistent with gauge and discrete symmetries
were allowed.
In this model $R$-parity is the unbroken subgroup of one
of
the discrete symmetry groups.
Non-renormalizable operators
involving non-SM fields lead to proton decay, but at a safe 
level.

\chapter{
Supersymmetry Breaking 
and the Supersymmetric Flavour Problem: An Analysis 
of Decoupling the First Two Generation Scalars 
} 
The supersymmetric contributions to
the Flavor Changing
Neutral Current processes may be
suppressed by decoupling the
scalars of the first and second generations.
It is known, however, that the heavy scalars drive the stop
(mass)$^2$ negative through the two loop
Renormalization Group
evolution.
To avoid negative stop (mass)$^2$ at the weak scale,
the boundary value of the stop mass has to be large leading
to fine tuning in EWSB.
This tension is studied in detail in this chapter. 

The chapter is organised as follows.
In section \ref{setup}, an overview of the 
ingredients of our analysis is presented. Some philosophy and notation are
discussed. Section \ref{mk} discusses the constraints on the masses and 
mixings of the first two generation scalars obtained from $\Delta m_K$
after including QCD corrections. It is found, in particular, 
that 
a mixing among both left-handed and right-handed 
first two generation squarks
of the order of the Cabibbo angle ($\lambda$), {\it i.e.,} $\sim 0.22$
requires them to be heavier than 40 TeV. Section \ref{rgeanal} discusses the 
logic of our RG analysis, and some formulae are presented. 
This analysis is independent of the $\Delta m_K$ analysis. 
Sections \ref{lowsusy} and 
\ref{highsusy} apply this machinery to the cases of low energy and high 
energy supersymmetry
breaking, respectively.  

Section \ref{lowsusy} deals with the case in which the scale at 
which SUSY
breaking is communicated to the SM sparticles 
is close to
the mass of the heavy scalars. We use the finite parts 
of the two loop
diagrams to estimate the negative contribution of the 
heavy scalars.
We find that 
a mixing among both left-handed and right-handed
first two generation squarks 
of the order of $\lambda$, {\it i.e.,} $\sim 0.22$,
implies that the boundary value of the stop masses 
has to greater
than $\sim 2$ TeV to keep
the stop (mass)$^2$ positive at the
weak scale. 
This results in a fine tuning of
naively $1\%$ in electroweak symmetry breaking \cite{barbieri1}. 
We also discuss the cases where 
there is $O(1)$ mixing among only the right or left
squarks of the first two generations, 
and find that requiring 
positivity of the
slepton (mass)$^2$ implies a constraint 
on the stop masses of $\sim 1$ TeV 
if gauge mediated boundary conditions 
are used to relate the two masses. This is comparable to the 
direct constraint on the initial stop masses. 

In section \ref{highsusy}, we consider the case where the SUSY 
breaking masses
for the SM sparticles are generated at a high scale 
($\sim 10^{16}$ GeV).
In this case, the negative contribution of the 
heavy scalars
is RG log enhanced. 
We consider various boundary conditions for the
stop and Higgs masses and find that 
with a degeneracy
between the first two generation squarks
of the order of the Cabibbo angle, the 
boundary value
of the stop mass needs to be larger than $\sim 7$ TeV.
This gives a fine tuning of naively 
$0.02 \%$ \cite{barbieri1}. 
For $O(1)$ mixing
between the left (right) squarks only, the minimum 
initial value
of the stop mass is $\sim 4 (2)$ TeV.
We conclude in section \ref{end}. In appendix \ref{appB}, we discuss 
the computation
of the two loop diagrams which give the negative 
contribution of 
the heavy scalars to the light scalar (mass)$^2$.

\section{Preliminaries}
\label{setup}
The chiral 
particle content of the Minimal Supersymmetric Standard
Model (MSSM) contains 3 generations of ${\bf \bar{5}+10}$
representations of $SU(5)$. The supersymmetry must be softly
broken to not be excluded by experiment. 
Thus the theory must also be supplemented by some `bare' soft
supersymmetry breaking parameters, as well as a physical 
cutoff, $M_{SUSY}$. 
The
`bare' soft supersymmetry breaking parameters are then the
coefficients appearing in the Lagrangian, defined with a
cutoff $M_{SUSY}$. It will be assumed for simplicity 
that the bare soft masses,
$m^2_{\tilde{f},0}$, the bare gaugino masses $M_{A,0}$, and a bare
trilinear term for the stops, $\lambda_t A_{t,0}$, are all 
generated close to this scale. 
The MSSM is then a good effective theory at energies 
below the scale $M_{SUSY}$,
but above the mass 
of the heaviest superpartner.

The physical observables at low energies will depend on these
parameters.
If an unnatural degree of cancellation is required between the bare 
parameters of the theory to produce a measured observable, the theory
may be considered 
to be fine tuned. 
Of course, it is possible that a more fundamental theory
may resolve in a natural manner 
the apparent 
fine tuning.  
The gauge-hierarchy problem is a well-known example of this. 
The Higgs boson mass
of the SM
is fine tuned if the SM is valid at energies above a few TeV.
This fine tuning is removed if at energies close to the weak scale 
the SM is replaced by a more fundamental 
theory that is supersymmetric as discussed in chapter 1. 
 
One quantification of the fine tuning of an observable ${\cal O}$
with respect to a bare parameter $\lambda_0$  
is given by Barbieri and Giudice \cite{barbieri1} to be 
\begin{equation}
c({\cal O};\lambda_0)
=(\delta {\cal O}/ {\cal O})/(\delta \lambda_0 / \lambda_0)
=\frac{\lambda_0}{{\cal O}}\frac{\partial}{\partial \lambda_0} {\cal O}.
\label{ft}
\end{equation}
It is argued that this only measures the sensitivity of ${\cal O}$ to 
$\lambda_0$, and care should be taken when interpreting whether 
a large value of $c$ necessarily implies that ${\cal O}$ is fine tuned
\cite{anderson}. It is not the intent of this chapter to quantify fine tuning; 
rather, an estimate of the fine tuning is sufficient and 
Eqn.(\ref{ft}) will be used. In this chapter the value of 
${\cal O}$ is considered extremely unnatural if $c({\cal O}; 
\lambda_0) >100$.

The theoretical prediction for $\Delta m_K$ (within the MSSM) 
and its measured value 
are an example of such a fine tuning: 
Why should the masses of the first two generation scalars be degenerate 
to within 1 GeV, when their masses are $O(\hbox{500 GeV})$? Phrased 
differently, the first two generation scalars must be extremely 
degenerate for the MSSM to not be excluded by the measured value 
of $\Delta m_K$. 
An important direction in supersymmetry model building is aimed 
at attempting to explain the origin of this degeneracy.

One proposed solution to avoid this fine tuning 
is to decouple the first two generation scalars 
since they are the ones most stringently constrained by the 
flavor violating processes \cite{dine,pomoral,
dvali,nelson2,nelson3,nelson,riotto,dimopoulos}. In 
this scenario, some of the first two generation 
scalars have masses $M_S \gg m_Z$. 
To introduce some notation,
$n_5$ $(n_{10})$ will denote the number of ${\bf \bar{5}}$ $({\bf 10})$
scalars of the MSSM particle content
that are very heavy. \footnote{It is assumed that the heavy scalars 
form complete $SU(5)$ multiplets to avoid a large Fayet-Illiopoulus $D-$
term at the one loop level \cite{nelson2,dimopoulos}.}
We will refer to these scalars as ``heavy'' scalars and the
other scalars as ``light''scalars.
Thus at energy scales $E \ll M_S$ the particle
content is that of the MSSM, minus the $n_5$ ${\bf \bar{5}}$ and 
$n_{10}$ ${\bf 10}$
scalars.
In the literature this is often
referred to as `The More Minimal Supersymmetric
Standard Model' \cite{nelson2}.  

There are, however, other possible and {\it equally valid} 
sources of fine tunings. The measured 
value of the $Z$ mass is such an example \cite{barbieri1}. 
The minimum of the renormalized
Higgs potential determines the value of the $Z$ mass which is already known 
from experiment. The vev of the Higgs field is, in turn, a function of 
the bare parameters of the theory. The relation used here, valid at the 
tree level, is 
\begin{equation}
\frac{1}{2}m^2_Z = - \mu^2
+ \frac{m^2_{H_d}(\mu_G)-m^2_{H_u}(\mu_G) \tan^2 \beta}
{\tan^2 \beta-1}.
\label{muEW}
\end{equation} 
It is clear from this equation that requiring 
correct electroweak symmetry breaking 
relates the value of the soft Higgs masses at the weak scale, 
$m^2_{H_d}(\mu_G)$ and $m^2_{H_u}(\mu_G)$,  
to the supersymmetric Higgs mass $\mu$.
A numerical computation determines the dependence of
$m^2_{H_u}(\mu_G)$ and $m^2_{H_d}(\mu_G)$ 
on the bare parameters 
$M_{A,0}$, $m^2_{\tilde{t},0}$ and $M_S$. 
In the MSSM, the cancellation required 
between
the bare parameters of the theory so that it is not 
excluded by the $Z$ mass 
increases as the scale of supersymmetry 
breaking is increased. The bare mass of the gluino and stops, 
and the first two generation
squarks must typically 
be less than a few TeV and ten TeV, respectively, 
so that successful electroweak symmetry breaking is not 
fine tuned at more than the one per cent level \cite{barbieri1,anderson,
dimopoulos}.

These two potential 
fine tuning problems - the supersymmetric flavor problem and that 
of electroweak symmetry breaking - are not completely independent, for they 
both relate to the size of supersymmetry breaking \cite{dimopoulos,nima}. 
Thus the consistency of  
any theoretical framework that attempts to resolve one fine tuning 
issue can be tested by requiring that it not
reintroduce any comparable 
fine tunings in other sectors of the theory. 
This is the situation for the case under consideration 
here. Raising the masses of the first two generation scalars 
can resolve the supersymmetric flavor problem. 
As discussed in \cite{dimopoulos}, this results in
a fine tuning of $m_Z$ through the two loop dependence of 
$m^2_{H_u}(\mu_G)$ on $M_S$.
There is, however, another source of fine tuning of
$m_Z$ due to the heavy scalars:
these large masses require that the bare masses 
of the light scalars, in particular the stop, be typically larger than 
a few TeV to keep the soft 
(mass)$^2$ positive at the weak scale \cite{nima}. This large value for 
the bare stop mass 
prefers a large value for vev of the Higgs field, thus 
introducing a fine tuning 
in the electroweak sector. Further, this fine tuning is typically 
not less than the 
original fine tuning in the flavor sector.
This is the central issue of this chapter.

\subsection{$\Delta m_K$ {\it Constraints}}
\label{mk}
At the one loop level the 
exchange of gluinos and squarks generates a 
$\Delta S = 2$ operator (see Fig.\ref{kk}). In the limit
$M_3 << M_S$ (where $M_3$ is the gluino mass)
that we are interested in, the
$\Delta S = 2$ effective Lagrangian at the scale
 $M_S$ obtained by 
integrating out
the squarks
is
\begin{equation}
{\cal L}_{eff} = \frac{ \alpha _S ^2 (M_S) }{216 M_S^2} 
\left(C_1 {\cal O} _1 +
\tilde{C}_1 \tilde{ {\cal O} }_1 +
C_4 {\cal O} _4 + C_5 {\cal O} _5+\hbox{h.c.}\right).
\end{equation}
Terms that are $O(M^2_3 /M^2_S)$ are subdominant and 
are neglected. We  
expand the exact result in powers of $\delta_{LL,RR}=s_{L,R} c_{L,R} \eta_{L,R} (\tilde{m}^2_1-
\tilde{m}^2_2)_{L,R}/
\tilde{m}^2_{AV,L,R}$, where $\tilde{m}^2_{AV}$ is the average mass 
of the scalars, and where $\eta_{L,R}$ is the phase 
and $s_{L,R}$ is the $1-2$ 
element of the $W_{L,R}$ matrix that appears at the 
gluino-squark-quark vertex.\footnote{In this chapter 
only 1-2 generation mixing is considered. Direct $L-R$ mass mixing 
is also neglected.} This 
approximation underestimates the magnitude of the 
exact result, so our analysis is conservative \cite{nima}. 
The coefficients $C_i$ to 
leading order in $\delta_{LL}$, $\delta_{RR}$, are
\begin{eqnarray}
C_1 & = & -22 \delta^d _{LL}, \nonumber \\
C_4 & = & 24 \delta^d _{LL} \delta^d _{RR}, 
\nonumber \\
C_5 & = & -40 \delta^d _{LL} \delta^d _{RR}.
\end{eqnarray}
The coefficient $\tilde{C}_1$ is obtained from $C_1$ with 
the replacement $\delta^d_{LL} \rightarrow \delta^d_{RR}$.
The operators ${\cal O}_i$ are
\begin{eqnarray}
{\cal O}_1 & = & \bar{d} ^a _L \gamma _{\mu}s_{L,a} 
\bar{d} ^b _L \gamma ^{\mu}
s_{L,b},
\nonumber \\
{\cal O}_4 & = & \bar{d} ^a _R s_{L,a} \bar{d} ^b _L s_{R,b}, 
\nonumber \\
{\cal O}_5 & = & \bar{d} ^a _R s_{L,b} \bar{d} ^b _L s_{R,a}
\end{eqnarray}
and $\tilde{ {\cal O} }_1$ is obtained from ${\cal O}_1$
with the replacement $L \rightarrow R$.
The Wilson coefficients, $C_1 - C_5$, are RG scaled 
from the scale of the
squarks, $M_S$, to 900 MeV
using the anomalous dimensions of the operators, ${\cal O}_1 
- {\cal O}_5$.
The anomalous dimension of ${\cal O}_1$ is well known
\cite{gaillard} and is $\mu d C_1 /d\mu =\alpha_s C_1/  \pi $.
We have
computed the other anomalous dimensions
and our result agrees with that of \cite{bagger} (see this
reference for a more general analysis
of QCD corrections to the SUSY contributions to $K -\bar{K}$
mixing). 
These authors , however, choose to RG scale to $\mu_{had}$, 
defined by $\alpha_s(\mu_{had})$=1. The validity 
of the pertubation expansion  
is questionable at this scale; we choose instead 
to RG scale to 900 MeV, where
$\alpha_s(\hbox{900 MeV}) \sim 0.4$. 
The result is
\begin{eqnarray}
C_1 (\mu _{had}) & = & \kappa _1 C_1 (M_S), 
\nonumber \\
\tilde{C}_1 (\mu _{had}) & = & \kappa _1 \tilde{C}_1 (M_S), 
\nonumber \\
C_4 (\mu _{had}) & = & \kappa _4 C_4 (M_S) + \frac{1}{3}
(\kappa _4 - \kappa _5) C_5 (M_S), \nonumber \\
C_5 (\mu _{had}) & = & \kappa _5 C_5 (M_S),
\end{eqnarray}
where
\begin{eqnarray}
\kappa _1 & = & \left( \frac{\alpha _s (m_c)}
{\alpha _s (\hbox{900 MeV})} \right)^{6/27}
\left( \frac{\alpha _s (m_b)
}{\alpha _s (m_c)} \right)^{6/25} \left( \frac{\alpha _s (m_t)}
{\alpha _s (m_b)} \right)^{6/23}
 \left( \frac{\alpha _s
(\mu _G)}{\alpha _s (m_t)} \right)^{6/21} \nonumber \\ 
 & & \times \left( \frac{\alpha _s (M_S)}
{\alpha _s
(\mu _G)} \right)^{ 6/( 9 + (n_5 + 3 n_{10})/2 ) }, \nonumber \\
\kappa _4 & = & \kappa _1 ^{-4}, \nonumber \\
\kappa _5 & = & \kappa _1 ^{1/2}.
\end{eqnarray}
The effective Lagrangian at the hadronic 
scale is then
\begin{eqnarray}
{\cal L} _{eff} & = &\frac{ \alpha _s ^2 (M_S) }{216 M_S^2}
\left( -22 (\delta^d_{LL})^2 \kappa _1 {\cal O}_1 - 22 
(\delta^d_{RR})^2 \kappa _1
\tilde{{\cal O}}_1
\right. \nonumber \\
 & & \left. + \delta^d _{LL} \delta^d _{RR} \left( \frac{8}{3} 
(4 \kappa _4 + 5 \kappa _5)
 {\cal O}_4 
- 40 \kappa _5 {\cal O}_5 \right)+\hbox{h.c.}\right) . 
\end{eqnarray}
The SUSY contribution to the $K - \bar{K}$ mass difference is
\begin{equation}
\Delta m_{K,\hbox{SUSY}} = 2 \hbox{Re} < K | {\cal L}_{eff} | \bar{K} >.
\end{equation}
The relevant matrix elements (with bag factors set to 1) are
\begin{eqnarray}
< K | {\cal O}_1 | \bar{K} > & = & \frac{1}{3} m_K f^2_K, 
\nonumber \\
< K | {\cal O}_4| \bar{K} > & = & \left ( \frac{1}{24} 
+ \frac{1}{4} \left(
\frac{m_K}{m_s + m_d} \right)^2 \right ) m_K f^2_K, \nonumber \\
< K | {\cal O}_5 | \bar{K} > & = & \left ( \frac{1}{8} + 
\frac{1}{12} \left(
\frac{m_K}{m_s + m_d} \right)^2 \right ) m_K f^2_K
\end{eqnarray}
in the vacuum insertion approximation. We use \cite{pdg} 
$m_K =497$ MeV, $f_K =160$ MeV, $m_s =150$ MeV
, $(\Delta m_K)_{exp}=3.5 \times 10^{-12}$ MeV, and 
$\alpha _s (M_Z) = 0.118$. This gives $\alpha_s(m_b)=0.21$, 
$\alpha_s(m_c)=0.29$ and $\alpha_s(\hbox{900 MeV})=0.38$ using 
the one loop RG evolution. 
A minimum value for $M_S$ is gotten, 
once values for $(n_5,n_{10},\delta^d_{LL},\delta^d_{RR})$ are
specified, by requiring $\Delta m_{K,SUSY} \ltap (\Delta m_K)_{exp}$. 
In the case that both $\delta_{RR}\neq0$ and $\delta_{LL} \neq 0$, 
we assume that both the left-handed and right-handed squarks are
heavy, so that $(n_5,n_{10})=(2,2)$. 
In this case we require that only the
dominant contribution to $\Delta m_K $, which
is $\sim \delta^d_{LL} \delta^d_{RR}$,
equals the measured value of $\Delta m_K$. 
If $\delta_{RR} \neq0$ and $\delta_{LL}=0$, we assume that 
only the right-handed squarks are heavy, and thus $(n_5,n_{10})=(2,0)$.
Similarly, if $\delta_{LL} \neq0$ and $\delta_{RR}=0$ then 
$(n_5,n_{10})=(0,2)$.
Limits are given in Tables \ref{mktable0} and \ref{mktable}
for some
choices of these parameters. These results agree with reference 
\cite{bagger} for the same choice of input parameters. 
For comparison, the limits gotten by 
neglecting the QCD corrections are also 
presented in Tables \ref{mktable0} and \ref{mktable}.
We consider $\delta^d_{LL}$ $(\delta^d_{RR})=
(i)$ $1$, $(ii)$ $0.22$, $(iii)$ $0.1$, and $(iv)$ $0.04$.
These correspond to: $(i)$ no mixing and no degeneracy;
$(ii)$ mixing of the order of the Cabibbo angle 
($\lambda$), {\it i.e.,} $\sim 0.22$; $(iii)\; O(\lambda)$ mixing
and
$\sim 0.5$ degeneracy; and $(iv)\;O(\lambda)$ mixing and 
$O(\lambda)$ degeneracy.
We expect only cases $(i)$, $(ii)$ and $(iii)$ to be  
relevant if the 
supersymmetric flavor problem is resolved by 
decoupling the first two generation
scalars.  
From Table \ref{mktable} we note that for $(n_5,n_{10})=
(2,0)$, $M_S$ must 
be larger than $\sim$ 30 TeV if it is assumed there is 
no small mixing or degeneracy  
$(\delta^d_{RR}=1)$ between the first two generation 
scalars. 

\renewcommand{\arraystretch}{0.9}
\begin{table}
\begin{center}
\vspace{0.2in}
\begin{tabular}{||l|l|l||} \hline
$\sqrt{\hbox{Re}(\delta^d_{LL}\delta^d_{RR})}$ & 
$(n_5,n_{10})=(2,2)$ & $(n_5,n_{10})=(2,2)$\\ \hline
 & QCD incl. & no QCD \\ \hline
$1$ & $\hbox{182 TeV}$ & $\hbox{66 TeV}$ \\ \hline
$0.22$ & $\hbox{40 TeV}$ & $\hbox{15 TeV}$ \\ \hline
$0.1$ & $\hbox{18 TeV}$ & $\hbox{7.3 TeV}$ \\ \hline
$0.04$ & $\hbox{7.3 TeV}$ & $\hbox{3.1 TeV}$  \\ \hline
\end{tabular}
\end{center}
\vspace{-0.2in}
\caption{Minimum values for heavy scalar masses 
$M_S$ obtained from
the measured value of $\Delta m_K$ assuming 
$M^2_3/M^2_S\ll 1$. The limits labeled `QCD incl.' include 
QCD corrections as discussed in the text. Those labeled as `no QCD' 
do not.}
\protect\label{mktable0}
\vspace{0.3in}
\end{table}

\renewcommand{\arraystretch}{0.7}
\begin{table}
\begin{center}
\vspace{0.2in}
\begin{tabular}{||l|l|l||}\hline
$\hbox{Re}(\delta^d_{RR}) \hbox{ }(\delta^d_{LL}=0)$
& $(n_5,n_{10})=(2,0)$ & $(n_5,n_{10})=(2,0)$\\
 \hline
& QCD incl. & no QCD \\ \hline
$1$ & $\hbox{30 TeV}$ & $\hbox{38 TeV}$ \\ \hline
$0.22$ & $\hbox{7.2 TeV}$ & $\hbox{8.9 TeV}$ \\ \hline
$0.1$ & $\hbox{3.4 TeV}$ & $\hbox{4.1 TeV}$\\ \hline
$0.04$ & $\hbox{1.4 TeV}$ & $\hbox{1.7 TeV}$ \\ \hline
\end{tabular}
\end{center}
\vspace{-0.25in}
\caption{Minimum values for heavy scalar masses
$M_S$ obtained from
the measured value of $\Delta m_K$ assuming
$M^2_3/M^2_S \ll 1$. The limits labeled as `QCD incl.'
include QCD
corrections as discussed in the text. Those labeled as `no QCD'
do not.
The limits for
$(n_5,n_{10})=(0,2)$ obtained by $\delta_{LL}^d
\leftrightarrow \delta_{RR}^d$ are similar
and not shown.}
\label{mktable}
\vspace{0.25in}
\end{table}

The limits gotten from the measured rate of $CP$ violation are now briefly
discussed.
Recall that the $CP$ violating parameter $\epsilon$ is approximately
\begin{equation}
|\epsilon| \sim \frac{|\hbox{Im}<K|{\cal{L}}_{eff}|\bar{K}>|}
{\sqrt{2} \Delta m_K}
\label{cp}
\end{equation}   
and its measured value is 
$|\epsilon| \sim |\eta_{00}|
=$2.3$\times$10$^{-3}$ \cite{pdg}. In this case, the 
small value of $\epsilon$ implies either that the 
phases appearing in the soft scalar mass matrix 
are extremely tiny, or that the masses of the heavy scalars are 
larger than the limits given in Tables \ref{mktable0} 
and \ref{mktable}. In the case where the phases are $O(1)$,\\ 
$\hbox{Im}<K|{\cal{L}}_{eff}|\bar{K}> \sim 
\hbox{Re}<K|{\cal{L}}_{eff}|\bar{K}>
$ and thus the stronger constraint on $M_S$ is 
obtained from $\epsilon$ and not $\Delta m_K$, 
for the same choice of input parameters. 
In particular, the constraint from $CP$ violation 
increases the minimum allowed value of $M_S$ by a 
factor of $1/ \sqrt{2 \sqrt{2} \epsilon} \sim$12.5. 
This significantly increases the minimum value of the 
initial light scalar masses that is allowed by the positivity requirement. 

\subsection{{\it RGE analysis}}
\label{rgeanal}
The values of the soft masses at the weak scale are determined by
the RG evolution.
In the $\overline{DR}'$ scheme \cite{dred,epscalar,drbarp}, the RG
equations for the light scalar masses, including the
gaugino, the trilinear term - $\lambda_t A_t H_u \tilde{q}_3 \tilde{u}^c_3$ 
and $\lambda_t$ contributions at
the one loop level and the heavy scalar
contribution at the two loop level \cite{twolooprge}, are

\begin{eqnarray}
\frac{d}{dt}m_{i}^2(t=\ln \mu)&=& -\frac{2}{\pi}\sum_{A} \alpha_A(t)
C_A^i M^2_A(t)
+\frac{4}{16 {\pi}^2} \sum_{A} C_A^i \alpha^2_A(t)
(n_5 m^2_5+3 n_{10} m^2_{10}) \nonumber \\
& &+\frac{8}{16 {\pi}^2}\frac{3}{5} Y_{i} \alpha_1(t)
\left(\frac{4}{3}
\alpha_3(t)-\frac{3}{4}\alpha_2(t)-\frac{1}{12} \alpha_1(t)\right) \nonumber \\ 
 & & \times (n_5 m^2_5-
n_{10} m^2_{10}) \nonumber \\
& & +\frac{\eta_i \lambda_t^2(t)}{8 {\pi} ^2}
\left(m^2_{H_u} (t)+m^2_{\tilde{u}^c_3} (t) +
m^2_{\tilde{q}_3}(t)+A_t(t)^2\right),
\label{rg1}
\end{eqnarray}
with $\eta=(3,2,1)$ for $\tilde{f}_i =H_u$, $\tilde{t}^c$,
$\tilde{t}$, respectively, and zero otherwise.
For simplicity it is assumed that
$M_{A,0}/\alpha_{A,0}$ are all equal at $M_{SUSY}$.
The initial value of the gluino mass, $M_{3,0}$, is then chosen to be
the independent parameter.
To avoid a large Fayet-Illiopoulus $D$-term at
the one loop level, we assume that the heavy scalars form complete
$SU(5)$ representations \cite{nelson2,dimopoulos}.
There is still the contribution, in the above RGE, of 
the Fayet-Illiopoulus $D$-term
due to the light scalars
$\sim \alpha _1 / (4 \pi) \; Y_i\;\sum_j Y_j \tilde{m}^2_j$. We do not include
it for two reasons. The first is that this contribution
depends on the soft masses of {\it all} the light scalars
which is clearly very model dependent. Also, we have checked
that, if all light scalar masses at the boundary
are roughly the same, this contribution
changes the constraints 
on the initial scalar masses by at most only a few percent,
for example, it changes the coefficient
of $m^2_{\tilde{f},0}$ in the numerical solutions, Eqns.(\ref{m2}),
(\ref{stopsol}),
(\ref{highscale}) and (\ref{highstop}),
by a few percent.
We use $SU(5)$ normalisation for the $U(1)_Y$ coupling constant and
$Q=T_3+Y$. Finally,
$C_A^i$ is the quadratic Caismir for the gauge group
$G_A$ that is
$4/3$ and $3/4$ for the fundamental representations
of $SU(3)$ and $SU(2)$, and $3/5\;Y^2_i$
for the $U(1)_Y$ group.
The cases
$(n_5,n_{10})$= (I) $(2,2)$, (II) $(2,0)$, (III) $(0,2)$
are considered.
The results for the case $(3,0)$ is obtained, to a good approximation,
 from Case (II) by a
simple scaling, and it is not discussed any further.

Inspection of Eqn.(\ref{rg1})
reveals that in RG scaling from a high scale to a smaller scale
the two loop gauge contribution of the heavy scalars
to the soft (mass)$^2$ is negative, and
that of the gauginos is positive.
The presence of the
large $\lambda_t$ Yukawa coupling in the RGE drives the value
of the stop soft (mass)$^2$ even more
negative. This effect increases the
bound on the initial value for the stop
soft masses and is included in our analysis.

In the MSSM there is an extra parameter, $\tan\beta$, which is the
ratio of the vacuum expecations values of the Higgs fields that
couple to the up-type and down-type quarks respectively.
Electroweak symmetry breaking then
determines the top quark mass to be
$m_t=\lambda_t/\sqrt{2} v\sin \beta$ with $v\sim$ 247 GeV.
In our analysis we consider the regime of small to moderate
$\tan\beta$, so that all Yukawa couplings other than $\lambda_t$
are neglected in the RG evolution.
In this approximation
the numerical results for $\tilde{f}_i\neq \tilde{t}$ or
$\tilde{t}^c$ are independent of $\tan\beta$.
In our numerical analysis we considered $\tan\beta$=2.2.
We have also checked the results for $\tan\beta$=10, and have
found that they differ
by less than $10\%$ percent.

In the case of low energy supersymmetry breaking,
the scale $M_{SUSY}$
is not much larger than
the mass scale of the heavy scalars. Then the
logarithm $\sim$$\ln(M_{SUSY}/M_S)$ that
appears in the
solution to the previous RG equations is only $O(1)$.
In this case the finite parts of the
two loop diagrams may not be negligible and should be included in
our analysis. We use these finite parts to {\em estimate} the size
of the two loop heavy scalar contribution in an actual model.

The full two loop expression for the soft scalar mass at a
renormalization scale $\mu_R$ is
$m^2_{full}(\mu_R)=m^2_{\overline{DR}'}(\mu_R)+m^2_{finite}(\mu_R)$, where
$m^2_{\overline{DR}'}(\mu_R)$ is the solution to the RG equation in
$\overline{DR}'$ scheme, and $m^2_{finite}(\mu_R)$ are the finite parts
of the one and 
two loop diagrams, also computed in $\overline{DR}'$ scheme.
The finite parts of the two loop diagrams
are computed in appendix \ref{appB} and the details are given therein.
The answer is (assuming all heavy scalars are degenerate with common
mass $M_S$)
\begin{eqnarray}
m^2_{i,finite}(\mu_R)&=&-\frac{1}{8}\left(\ln (4 \pi)-\gamma
+ \frac{\pi^2}{3}-2-
\ln\left(\frac{M^2_S}{\mu_R^2}\right)\right)
\times \sum_{A}
{\left(\frac{\alpha_A(\mu_R)}{\pi}\right)}^2 \nonumber \\
 & & \times (n_5+3 n_{10}) \; C_A^i
 M^2_S 
-\frac{3}{5}\frac{1}{16 {\pi}^2}
 \alpha_1(\mu_R)(n_5-n_{10}) Y_i \nonumber \\
 & & \times \left(6-\frac{2}{3} {\pi}^2+2 (\ln (4 \pi)-\gamma)
-4 \ln\left(\frac{M_S^2}{\mu_R^2}\right)\right)
\nonumber \\
& & \times
\left(\frac{4}{3} \alpha_3(\mu_R)-\frac{3}{4}
\alpha_2(\mu_R)-\frac{1}{12}\alpha_1(\mu_R)\right) M_S^2,
\label{finite}
\end{eqnarray}
where the gaugino and fermion masses are neglected.
Since we use the $\overline{DR}^{\prime}$ scheme to compute 
the finite parts of the soft scalar masses, the limits we obtain
on the initial masses 
are only valid, strictly speaking, in this scheme. This is 
especially relevant for the case of low scale SUSY breaking. 
So 
while these
finite parts should be viewed as semi-quantitative, they should
suffice for a discussion of the fine tuning that results from the 
limit on the bare stop mass.
For the case of high scale SUSY breaking, the RG logarithm
is large and so the finite parts are not that important.

Our numerical analysis for either low energy or
high energy supersymmetry breaking is described as follows.

The RG equations are evolved from the scale $M_{SUSY}$
to the scale at which the heavy scalars are decoupled. 
This
scale is denoted by $\mu_S$ and should be $O(M_S)$. 
The RG scaling  
of the heavy scalars is neglected.
At this scale the finite parts of the two loop diagrams
are added to $m^2_{\tilde{f}_i}(\mu_S)$. We note that since the two loop
information included in our RG analysis is the leading $O(M^2_S)$ effect,
it is sufficient to only use tree level matching at the scale
$\mu_S$. Since the heavy scalars are not
included in the effective theory
below $M_S$ and do not contribute to the gauge coupling beta
functions, the numerical results
contain an
implicit dependence
on the number of heavy scalars. This results in a value for
$\alpha_3(\mu_S)$ that is smaller
than the case in which all sparticles are at 
$\sim 1$ TeV. This tends to weaken the constraint, and so it
is included in our analysis.\footnote{This is the origin of a small 
numerical discrepancy of $\sim 10\%$ between our results and the 
analysis of \cite{nima} in the approximation $\lambda_t=0$.}
The soft masses are then evolved using the one loop
RGE to the mass scale at which the
gluinos are decoupled. This scale is fixed to be $\mu_G$=1 TeV. 

A constraint on the initial value of the soft
masses is obtained by requiring 
that at the weak scale the physical
scalar (mass)$^2$ are positive.
The experimental limit is $\sim$ 70 GeV for charged or 
colored scalars \cite{LEP}. The physical mass of a scalar is equal 
to the sum of  
the $\overline{DR} ^{\prime}$
soft scalar mass, the electroweak $D-$term, the supersymmetric 
contribution, and 
the finite one loop and two loop
contributions. The finite one loop contributions
are proportional to the gaugino and other light scalar masses, and are smaller
than the corresponding 
RG logarithm that is summed in $m^2_{\overline{DR} ^{\prime}}
(\mu_R)$. So
we neglect these finite one loop parts. Further, the electroweak $D-$terms 
are less than 70 GeV. For the scalars other than the stops, the 
supersymmetric contribution is negligible. 
In what follows then, we will require that 
$m^2_{\tilde{f}_i}(\mu_G)>0$
(including the finite two loop parts)\footnote{
As mentioned earlier, in the case of high scale SUSY breaking,
the finite two loop parts are also smaller than the RG logarithm
and thus are not so important.}
for scalars other than the 
stops.
The discussion with the stops is complicated by both the large
supersymmetric contribution, $m^2_t$, to the physical mass and
by the $L-R$ mixing between the gauge eigenstates. This mixing
results
in a state with (mass)$^2$ less than
$\hbox{min}(m^2_{\tilde{t}}+m^2_t,
m^2_{\tilde{t}^c}+m^2_t)$, so it is a conservative
assumption to
require that for both gauge eigenstates the value of
$m^2_{\tilde{t}_{i}}+m^2_t$ is larger than the
experimental limit. 
This implies that 
$m^2_{\tilde{t}_{i}}\gtap ($70 GeV$)^2-$$($175 GeV$)^2
=-($160 GeV$)^2$.
In what follows we require instead that
$m^2_{\tilde{t}_{i}}\geq 0$.
This results in an error that is
$ (160 $GeV$)^2/2m_{\tilde{t}_{i,0}}
\approx 26 $ GeV if
the constraint obtained by neglecting $m_t$ is $\sim$ 1 TeV.
For the parameter range of interest it will be shown that the
limit on the initial squark masses
is $\sim$ 1 TeV, so this approximation is consistent.

We then combine the above two analyses as follows.
The $\Delta m_K$ constraints of section \ref{mk}
determine a minimum value for $M_S$
once some theoretical preference for
the $\delta$'s is given.
Either a natural
value for the $\delta$'s is predicted by some model, 
or the $\delta$'s are
arbitrary and chosen solely by naturalness considerations.
Namely, in the latter case the fine tuning to suppress 
$\Delta m_K$ is roughly $2/\delta$. 
Further, a model may also predict the ratio $M_3 /M_S$. 
Otherwise, Eqns.(\ref{ft}) and (\ref{muEW}) may be used as a rough
guide to determine an upper value for $M_3$, based upon naturalness
considerations of the $Z$ mass.
Without such a limitation, the
positivity requirements are completely irrelevant if the 
bare gluino mass is suffuciently large; 
but then the $Z$ mass is fine tuned.
Using these values of $M_3$ and $M_S$, 
the RGE analysis gives a minimum value for the initial
stop masses which is consistent with $\Delta m_K$ and 
positivity of the soft (mass)$^2$.
This translates into some fine tuning of the $Z$ mass,
which is then roughly quantified by
Eqns.(\ref{ft}) and (\ref{muEW}).

\section{Low Energy Supersymmetry Breaking}
\label{lowsusy}
 In this section we investigate the 
positivity requirement within a 
framework that satisifes both of the 
following: (i)
supersymmetry breaking is communicated to the visible sector 
at low energies; and (ii) multi-TeV scale soft masses, $M_S$, 
are generated for some of the first two generation scalars.
This differs from the usual low energy supersymmetry breaking 
scenario in that we assume $M^2_S \gg m^2_{\tilde{t}_i,0}$. 
In the absence of a specific model, however, 
it is difficult to obtain from the positivity 
criterion 
robust constraints on the 
scalar spectra for the following reasons. 
At the scale $M_{SUSY}$ it is expected that, 
in addition to the heavy scalars of the MSSM, there are 
particles
that may have SM quantum numbers and supersymmetry 
breaking mass parameters. All these extra states contribute 
to the soft scalar masses of the light particles. 
The sign of this
contribution depends on, among other things,
whether the soft (mass)$^2$ for these
additional particles is positive or negative - clearly very 
model dependent. The total two loop 
contribution to the light scalar 
masses is thus a sum of a model dependent
part and a model independent part.
By considering only the model independent 
contribution we have only isolated one particular contribution
to the total value of the soft scalar masses near the supersymmetry
breaking scale. We will, however, use these results to 
{\em estimate}  
the typical size of the finite parts in an actual model. 
That is, if in an actual model 
the sign of the finite parts is negative and 
its size
is of the same magnitude as in
Eqn.(\ref{finite}), the constraint in that model 
is identical
to the constraint that we obtain. The constraint for other values 
for the finite parts is then obtained from our results by a simple scaling.

Before discussing the numerical results, the 
size of the finite contributions are estimated
in order to
illustrate the problem. Substituting $M_S\sim$ 25
TeV, $\alpha_3($25 TeV$)\sim$ 0.07 
and $\alpha_1($25 TeV$)\sim$ 0.018 into Eqn.(\ref{finite})
gives
\begin{equation}
\delta m^2_{\tilde{q}}\approx-(\hbox{410 GeV})^2(n_5+3 n_{10})
\left(\frac{M_S}{\hbox{25 TeV}}\right)^2
\end{equation}
for squarks, and
\begin{equation}
\delta m^2_{\tilde{e}^c}\approx
-\left((n_5+3 n_{10})
(\hbox{70 GeV})^2
+(n_5-n_{10})(\hbox{100 GeV})^2\right)
\left(\frac{M_S}{\hbox{25 TeV}}\right)^2
\end{equation}
for the right-handed selectron. The negative contribution is
large if $M_S\sim$ 25 TeV. For example,
if $n_5=n_{10}=2$ then $\delta m^2_{\tilde{e}^c}\approx
-($200 GeV$)^2$ and $\delta m^2_{\tilde{q}}\approx-($1.2 TeV$)^2$.
If $n_5=2$, $n_{10}=0$, then $\delta m^2_{\tilde{e}^c}\approx
-($170 GeV$)^2$ and $\delta m^2_{\tilde{q}}\approx-($580 GeV$)^2$.

In this low energy supersymmetry breaking scenario, it is 
expected that $M_{SUSY}\sim M_S$. In our numerical 
analysis we will set $M_{SUSY}=\mu_S$ since the actual
messenger scale is not known. The scale $\mu_S$ is chosen 
to be 50 TeV. At the scale $\mu_S=$50 TeV the 
$\mu _S$-independent parts
of Eqn.$($\ref{finite}$)$ are added to the initial 
value of the soft scalar masses.  
The soft masses are then 
evolved using the RG equations (not including the two loop
contribution) to the scale $\mu_G$= 1TeV.

\begin{figure}
\vspace{-1.0in}
\centerline{\epsfxsize=1.2\textwidth \epsfbox{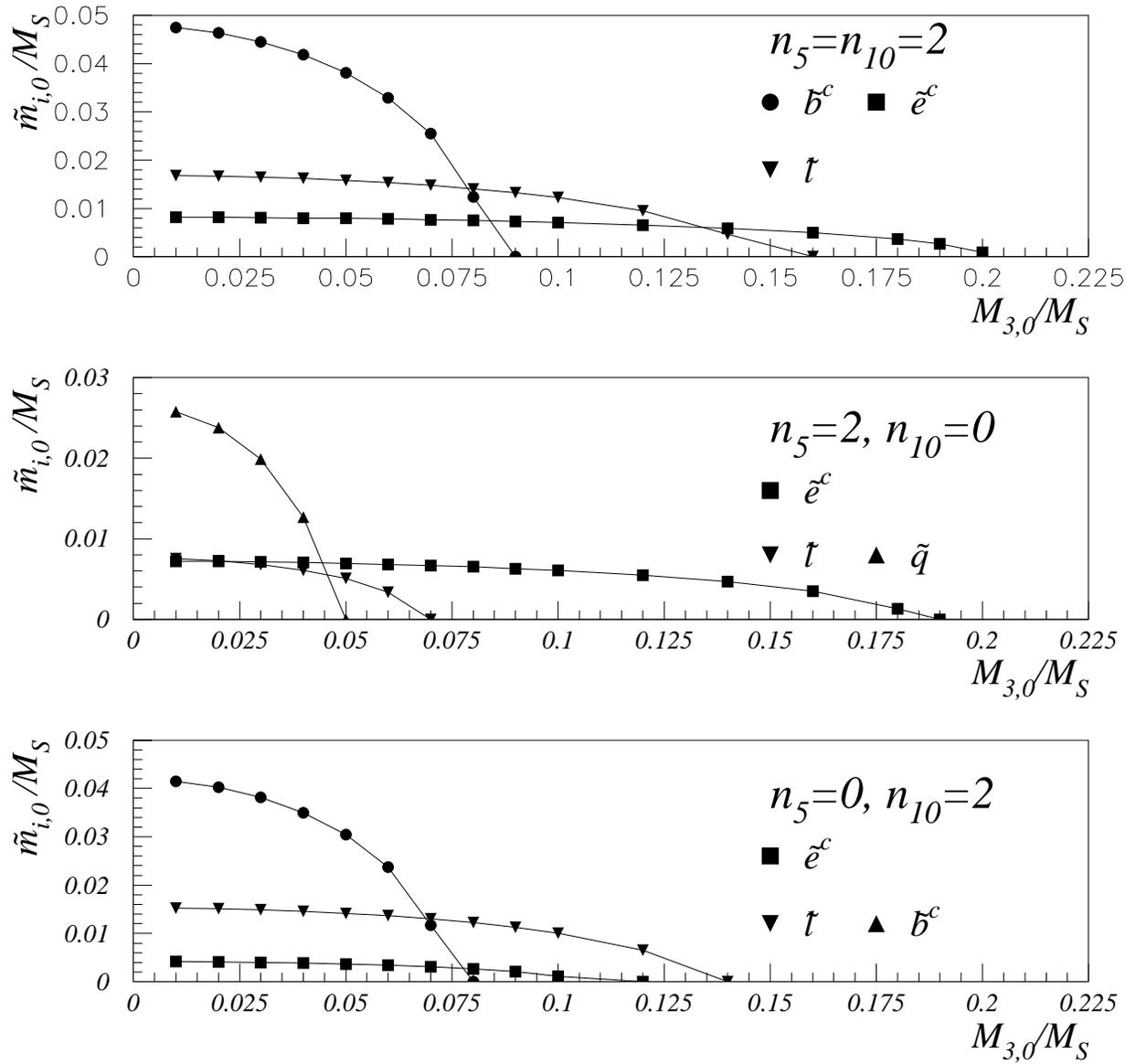}} 
\vspace{-0.2in}
\caption{Limits for $m_{\tilde{f}_i,0}/M_S$ 
from the requirement that the (mass)$^2$ are
positive at the weak scale, for low energy 
supersymmetry breaking. The regions below the curves
are excluded. For the case (2,0),
the limits for the other squarks are very similar to
that for $\tilde{q}$ and are therefore not shown.}
\protect\label{m2l}
\end{figure}

First we discuss the constraints the positivity requirement 
implies for $\tilde{f}_i\neq \tilde{t}_L$ or $\tilde{t}_R$.
In this case $m^2_{\tilde{f}_i}$ is
renormalised by
$M^2_{3,0}$, $M^2_S$ and $m^2_{\tilde{f}_i,0}$. We find
\begin{eqnarray}
m^2_{\tilde{f}_i}(\mu_G) & = & m^2_{\tilde{f}_i,0}
+(0.243 C_3^i+0.0168 C_2^i+0.00156 Y^2_i) M^2_{3,0}
\nonumber \\
& & -(0.468 C_3^i+0.095 C_2^i+0.0173 Y^2_i)
\frac{1}{2}(n_5+3 n_{10})\times 10^{-3} M^2_S \nonumber \\
& & -0.0174 (n_5-n_{10}) Y_i \times 10^{-3} M^2_S,
\label{m2}
\end{eqnarray}
where the strongest dependence on $(n_5,n_{10})$ has been 
isolated. The numerical coefficients in 
Eqn.(\ref{m2}) also depend on $(n_5,n_{10})$ and the numbers presented
in Eqn.(\ref{m2}) are for $(n_5,n_{10})=($2,0$)$. This 
sensitivity is, however, only a few percent between the 
three cases under consideration here.\footnote{This dependence 
is included in Fig.\ref{m2l}.} 
Requiring
positivity of the soft scalar (mass)$^2$ directly constrains
$m^2_{\tilde{f}_i,0}/M^2_S$ and $M^2_{3,0}/M^2_S$.

The positivity requirement $m^2_{\tilde{f}_i}(\mu_G) >0$ for
$\tilde{f}_i\neq \tilde{t}$ or $\tilde{t}^c$ is given
in Fig.\ref{m2l} for different values of $n_5$ and $n_{10}$.
That is, in Fig.\ref{m2l}  
the minimum value of $m_{\tilde{f}_i,0}/M_S$ required to 
keep the soft (mass)$^2$ positive at the scale $\mu_G$ is plotted 
versus $M_{3,0}/M_S$.
We conclude from these figures that the positivity criterion is
weakest for $n_5$=2 and $n_{10}$=0. This is expected since in
this case the heavy particle content is the smallest. We note 
that even
in this `most minimal' scenario the negative contributions to the (mass)$^2$
are rather large. In particular, we infer from Fig.\ref{m2l} 
that for $n_5=2,n_{10}=0$ and
$M_S\sim$ 25 TeV, $\delta m^2_{\tilde{e}^c} \approx-($190 GeV$)^2$
for $M_{3,0}$ as large as 1 TeV.
In this case it is the two loop contribution from the
hypercharge $D$-term
that is responsible for the large negative (mass)$^2$.
In the case $(n_5,n_{10})$=$(2,2)$, 
we obtain from Fig.\ref{m2l} that
for $M_S\sim$ 25 TeV, 
$\delta m^2_{\tilde{e}^c} \approx-($200 GeV$)^2$
and $\delta m^2_{\tilde{b}^c} \approx-($1.2 TeV$)^2$ for 
$M_{3,0}$ as large as 1 TeV.

We now apply the positivity requirement to the stop sector. 
In this case it is not possible to directly constrain the
boundary values of the stops for the following simple reason.
There are only two positivity constraints, whereas the
values of $m^2_{\tilde{t}}(\mu_G)$
and $m^2_{\tilde{t}^c}(\mu_G)$ are functions of the three
soft scalar masses $m^2_{\tilde{t},0}$,
$m^2_{\tilde{t}^c,0}$ and $m^2_{H_u,0}$. To obtain a
limit some theoretical assumptions must be made to relate
the three initial soft scalar masses.

The numerical solutions to the RG equations 
for $\tan\beta$=2.2 and $(n_5,n_{10})=(2,0)$ are:
\begin{eqnarray}
m^2_{\tilde{t}}(\mu_G) & = & -0.0303 A_t^2+ 0.00997 A_t M_{3,0}
+ 0.322 M^2_{3,0}  \nonumber \\
& & - 0.0399 (m^2_{H_u,0}+m^2_{\tilde{t}^c,0})
+ 0.960 m^2_{\tilde{t},0} - 0.00064 c_{L} M^2_S, \nonumber  \\
m^2_{\tilde{t}^c}(\mu_G) & = & -0.0606 At^2  + 0.0199 A_t M_{3,0} 
+ 0.296 M^2_{3,0}  \nonumber  \\
& & 0.920 m^2_{\tilde{t}^c,0}- 0.0797 (m^2_{H_u,0}+m^2_{\tilde{t},0})
- 0.000495 c_{R} M^2_S, \nonumber \\
m^2_{H_u}(\mu_G) & = & -0.0909 A_t^2  + 0.0299 A_t M_{3,0} 
- 0.0298 M^2_{3,0}  \nonumber \\
& & + 0.880 m^2_{H_u,0} -0.119 (m^2_{\tilde{t},0}
+m^2_{\tilde{t}^c,0})+0.0000748 c_{H} M^2_S.
\label{stopsol}
\end{eqnarray}
The numerical coefficients other than that of $M_S$ do not 
vary more than a few percent between the 
different values for $(n_5,n_{10})$. 
For $M_S$, we find that  $(c_L,c_R,c_H)$ is 
$(1,1,1)$, $(3.6,3.8,4.5)$, 
$(2.8,3,3.65)$, for $(n_5,n_{10})=($2$,$0$)$,$(2,2)$
and $(0,2)$, respectively. 
We find from Eqns.(\ref{ft}) and (\ref{muEW}) 
that to keep $m^2_Z$ fine tuned
at less than $1\%$ ($c \leq 100$) in each of the bare parameters,
we must have:
$\mu \ltap$ 500 GeV; $M_{3,0} \ltap 3.7
$ TeV; $m_{\tilde{t},0} \ltap 1.8$ TeV; and $M_S \ltap 
47$ {TeV for $(n_5,n_{10})=(2,2)$. 
Finally, for other values of these parameters
the fine tuning increases as
$c= 100 \times \tilde{m}^2/ \tilde{m}^2_0$, where $\tilde{m}_0$
is the value
of $\tilde{m}$ that gives $c=100$.

To constrain the initial values of the stop masses
we will only consider gauge mediated supersymmetry breaking (GMSB) 
mass relations between
the stop and Higgs boundary
masses. From Eqn.(\ref{stopsol}) we see that to naturally 
break electroweak symmetry a small hierarchy 
$m^2_{\tilde{t}_i,0} > m^2_{H_u,0}$ is required. 
This is naturally 
provided by gauge mediated boundary conditions.\footnote{In fact, low energy
gauge mediated supersymmetry breaking provides ``too
much'' electroweak symmetry breaking \cite{gmfinetune}.}  
The relations between the soft scalar masses when
supersymmetry breaking is communicated to the visible 
sector by gauge messengers
are \cite{gm} 
\begin{equation}
m^2_{i,0}
=\frac{3}{4}\sum_{A}C_A^i
\frac{\alpha^2_A(M_{SUSY})}
{\alpha^2_3(M_{SUSY})+\alpha^2_1(M_{SUSY})/5}
 m^2_{\tilde{t}^c,0}.
\label{GMBC}
\end{equation}
Substituting these relations into Eqn.(\ref{stopsol})
and assuming $A_{t,0}=$0 
determines $m^2_{\tilde{t}}(\mu_G)$ and 
$m^2_{\tilde{t}^c}(\mu_G)$ as a function of $M_{3,0}$, $M^2_S$ 
and $m^2_{\tilde{t}^c,0}$. In Fig.\ref{m2lgm} we 
have plotted the minimum value of $m_{\tilde{t}^c,0}/M_{3,0}$ 
required to maintain both $m^2_{\tilde{t}}(\mu_G)\geq0$ and
$m^2_{\tilde{t}^c}(\mu_G)\geq0$. 

Another interesting constraint on these class
of models is found if it is assumed that the 
initial masses of {\it all} the 
light scalars are related at the supersymmetry breaking scale 
by some gauge mediated supersymmetry breaking relations, as
in Eqn.(\ref{GMBC}). This ensures
the degeneracy, as required by the flavor changing constraints, of any 
light scalars of the first two generations. This is required if, 
for example, one of 
$n_5$ or $n_{10}$ are zero.  
Then in our previous limits on 
$m_{\tilde{f}_i,0}$ for $\tilde{f}_i\neq\tilde{t}$ or 
$\tilde{t}^c$, constraints on the initial 
value of $m_{\tilde{t}^c}$ are obtained 
by relating $m_{\tilde{f}_i,0}$ 
to  $m_{\tilde{t}^c,0}$ using Eqn.(\ref{GMBC}). 
In this case the slepton masses provide the 
strongest constraint and they are also shown in 
Fig.\ref{m2lgm}.
This result may be understood from the 
following considerations.
The two loop hypercharge $D$-term 
contribution to the soft mass 
is  $\sim Y_i (n_5-n_{10}) \alpha_1 \alpha_3 M^2_S$
and this has two interesting consequences. The first is that
for $n_5 \neq n_{10}$, 
the resulting $\delta \tilde{m}^2$ is 
always negative for one of $\tilde{e}^c$ or $\tilde{l}$. 
Thus in this case there 
is always a constraint on $m^2_{\tilde{t}^c}$ 
once gauge mediated boundary conditions are assumed.
That this negative contribution is large is 
seen as follows. The combined tree level mass and two loop 
contribution to the selectron mass 
is approximately $m^2_{\tilde{e}^c,0}-k \alpha_1 
\alpha_3 M^2_S$ where $k$ is a numerical factor. 
Substituting the gauge mediated relation 
$m^2_{\tilde{e}^c,0}\sim \alpha^2_1/
\alpha^2_3 m^2_{\tilde{t}^c,0}$, the combined selectron mass is 
$\alpha^2_1/\alpha^2_3 (m^2_{\tilde{t}^c,0}-k (\alpha_3/ \alpha_1) 
\alpha^2_3 M^2_S)$. Since the combined mass of the stop is
$\sim m^2_{\tilde{t}^c,0}-k^{\prime} \alpha^2_3 M^2_S$, 
the limit for $m^2_{\tilde{t}^c,0}$ 
obtained from the positivity requirement for
$m^2_{\tilde{e}^c}$  
is comparable to or larger than the constraint 
obtained from requiring that $m^2_{\tilde{t}^c}$ remains 
positive.
For example, with $n_5=2$, $n_{10}=0$ and $M_S \sim 25$ TeV,
the right-handed slepton constraint 
requires that $m_{\tilde{t}^c,0}\sim$ 1.1 TeV. For $n_{10}$=2, 
$n_5$=0 and $M_S \sim 25$ TeV, 
$\tilde{l}$ is driven negative and 
implies that $m_{\tilde{t}^c,0}\sim$ 1 TeV. From 
Fig.\ref{m2lgm} we find that these results 
are comparable
to the direct constraint on $m_{\tilde{t}^c,0}$ obtained by
requiring that color is not broken.

\begin{figure}
\vspace{-0.7in}
\centerline{\epsfxsize=1.2\textwidth \epsfbox{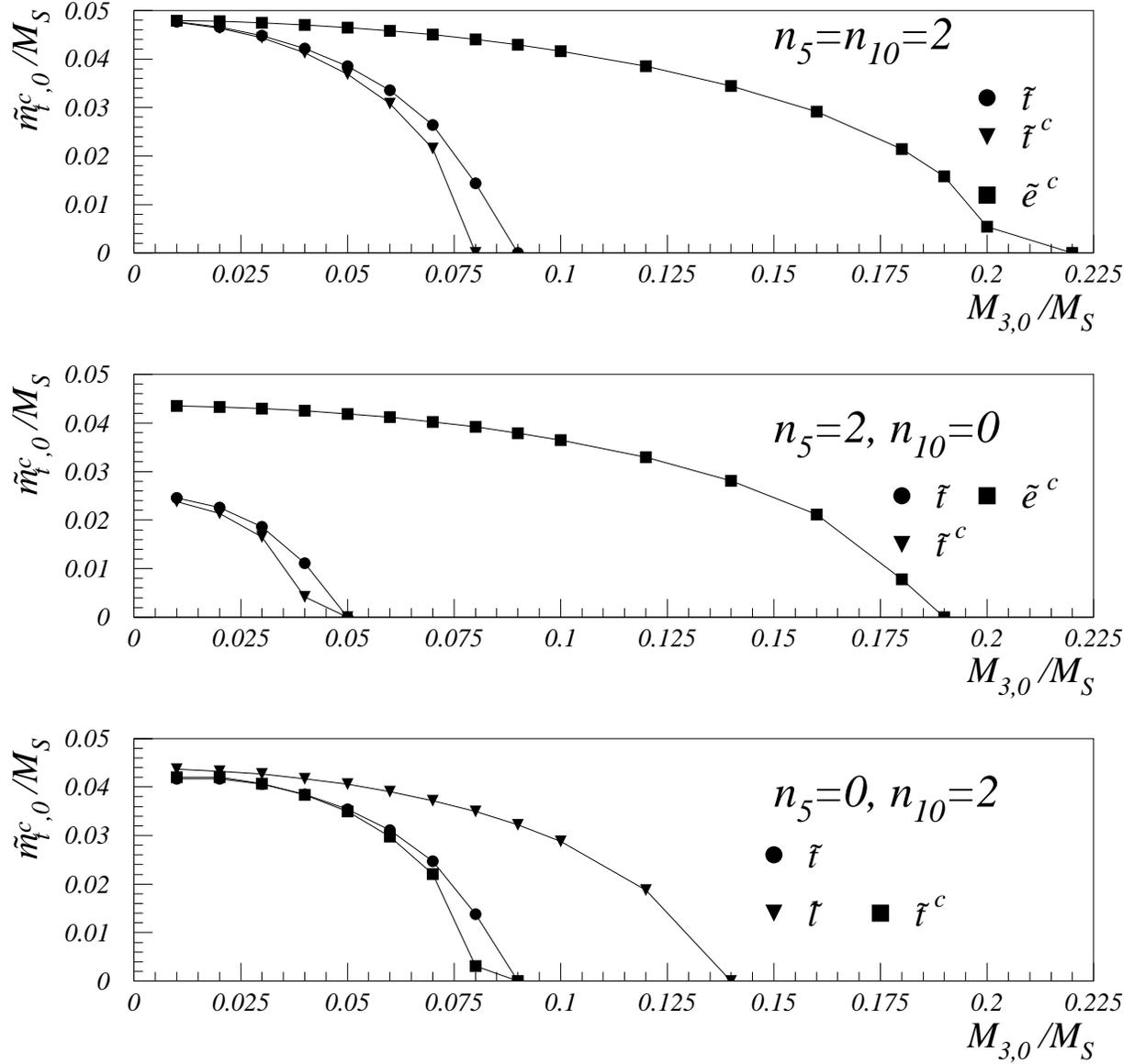}}
\caption{Limits for $m_{\tilde{t}^c,0}/M_S$
from the requirement that the stop and slepton (mass)$^2$ are
positive at the weak scale.
The regions below the curves
are excluded.
Low energy 
gauge mediated
supersymmetry breaking and $\tan \beta=$2.2 
are assumed.}
\protect\label{m2lgm}
\end{figure}


The positivity analysis only constrains 
$m_{\tilde{t}_i,0}/M_S$ for a fixed value 
of $M_{3,0}/M_S$.
To directly limit the initial scalar masses some 
additional information is 
needed. 
This is provided by the measured value of $\Delta m_K$. 
If some mixing 
and degeneracy between the first two generation scalars
is assumed, parameterized by 
$(\delta_{LL},\delta_{RR})$, 
a  
minimum value for $M_S$ is 
obtained by requiring that the supersymmetric contribution 
to $\Delta m_K$ does not exceed the measured value. 
We use the results
given in section \ref{setup} to calculate this minimum 
value. 
This result together with 
the positivity analysis then 
determines a minimum value for $m_{\tilde{t}^c,0}$ for a 
given initial gluino mass $M_{3,0}$. 
The RG analysis 
is repeated with $\mu_S=M_S$, rather than $\mu_S$=50 TeV. 
We only present the 
results found by assuming GMSB mass relations between the scalars. 
These results are 
shown in Fig.\ref{m2lgmmk}. 
The mass limits for other $\tilde{f}_i$ are easily obtained from the
information provided in Fig.\ref{m2l} 
and Table \ref{mktable} and are not shown.
From Fig.\ref{m2lgmmk} we find that for $(n_5,n_{10})$
$=(2,2)$ and a large range of $M_{3,0}$, $m_{\tilde{t}^c,0}$ 
must be larger than $7$ TeV for 
$\sqrt{\delta_{LL} \delta_{RR}}=1$, and larger than $2$ TeV for 
$\sqrt{\delta_{LL} \delta_{RR}}=0.22$. 
This results in $c(m_Z^2,m_{\tilde{t} ,0}^2)$ of
1500 and 100, respectively.
In this case both the squark and selectron limits for $m_{\tilde{t}^c,0}$ 
are comparable.
The limits for other choices for $\sqrt{\delta_{LL} \delta_{RR}}$ are 
obtained from Fig.\ref{m2lgmmk} by a simple scaling, since to a good
approximation $\Delta m_K \sim \delta_{LL} \delta_{RR}/M^2_S$. 
For the cases $(n_5,n_{10})=(2,0)$ and $(0,2)$, the corresponding limits 
are much weaker. In the case $(n_5,n_{10})=(2,0)$, 
for example, only for $\delta_{RR}\sim 1$ does the 
constraint that the selectron (mass)$^2>0$
require that $m_{\tilde{t}^c,0} \sim $1 TeV. The limits for a smaller 
value of $\delta$ are not shown.

We conclude with some comments about how these results change if $CP$ 
violation is present in these theories with $O(1)$ phases.
Recall from section \ref{setup} that for the same choice of 
input parameters, the limit 
on $M_S$ and hence, if the gluino mass is small, the limit on 
the initial stop mass increases by about a 
factor of 12. This may be interpreted in one of two ways. 
Firstly, this constrains those models that were relatively 
unconstrained by the $\Delta m_K$ limit. We concentrate 
on the models with $n_5=2$ and $n_{10}=0$, since this case 
is the most weakly constrained by the combined $\Delta m_K$ and 
positivity analysis. The conclusions for other models will be 
qualitatively the same.  We find from Fig.\ref{m2lgmmk} the 
limit $m_{\tilde{t}^c,0}>$1 TeV \footnote{For GMSB relations only. 
The direct constraint on the stop masses is slightly weaker.} 
is true only if $\delta_{RR} \sim O(1)$. Smaller values 
of $\delta_{RR}$ do not require large initial stop masses. 
From the $CP$ violation constraint, however, smaller values 
for $\delta_{RR}$ are now constrained. For example,  
if $\delta_{RR}\sim$ 0.1 and $O(1)$ phases are present, 
then $m_{\tilde{t}^c,0}>$1 TeV is 
required. Secondly, the strong constraint from $\epsilon$ 
could partially or completely compensate a weakened constraint 
from the positivity analysis. This could occur, for example, 
if in an actual model the negative two loop contribution 
to the stop (mass)$^2$ for the same initial input parameters 
is smaller than the estimate used here. For example, if the 
estimate of the two loop contribution in an actual model 
decreases by a factor       
of $\sim (12.5)^2$ and $O(1)$ phases are present, the 
limit in this case from $\epsilon$ for the same 
$\delta$ is identical to the values presented in Fig.\ref{m2lgmmk}.

\begin{figure}
\vspace{-0.5in}
\centerline{\epsfxsize=1.1\textwidth \epsfbox{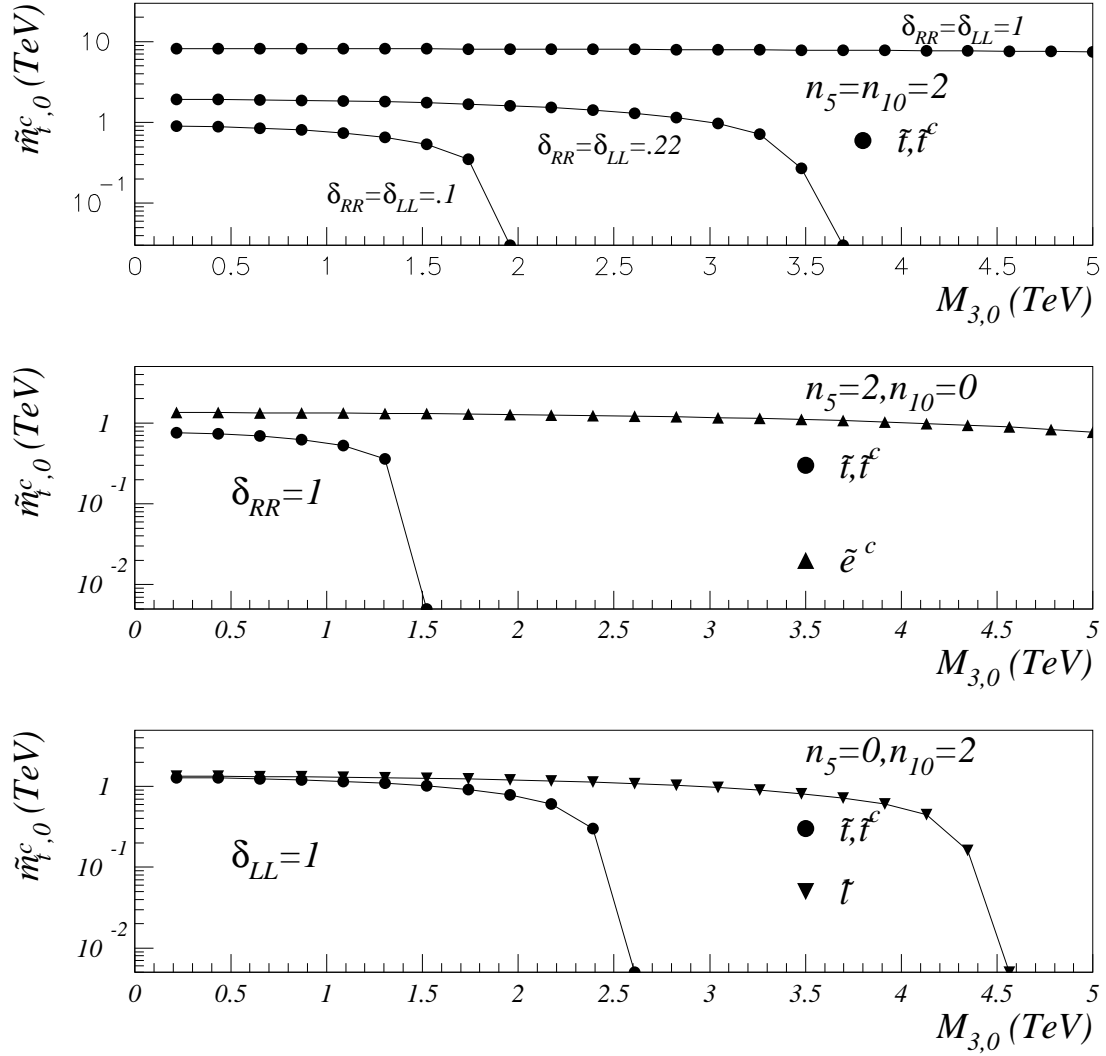}}
\caption{Limits for $m_{\tilde{t}^c,0}$ 
from the requirement that the stop and slepton (mass)$^2$ are
positive at the weak scale while suppressing
$\Delta m_K$, for different
values of $(n_5,n_{10})$, and $(\delta_{LL},\delta_{RR})$. 
The regions below the curves
are excluded.
Low energy
gauge mediated
supersymmetry breaking mass relations and $\tan \beta=$2.2 
are assumed.}
\protect\label{m2lgmmk}
\end{figure}  


\section{High Scale 
Supersymmetry Breaking}
\label{highsusy}
In this section, we consider the case in which SUSY 
breaking is communicated to the
MSSM fields at a high energy scale, that is taken to 
be $M_{GUT} = 2 \times
10^{16}$ GeV. 
In this case, the negative contribution 
of the heavy scalar
soft masses to the soft (mass)$^2$ of the light 
scalars is enhanced by $\sim \ln (M_{GUT}/\hbox{50 TeV})$,
since the heavy scalar soft masses contribute to the 
RGE from $M_{GUT}$ to
mass of the heavy scalars. It is clear that as the scale of SUSY 
breaking is lowered the
negative contribution of the heavy scalar
soft masses reduces.

This scenario was investigated in reference \cite{nima}, and we  
briefly discuss the difference between that analysis and the results 
presented here.
In the analysis of reference \cite{nima}, the authors made the 
conservative choice of neglecting $\lambda_t$ in the 
RG evolution.  The large 
value of $\lambda_t$ can change the analysis, and 
it is included here. 
We find  
that for some pattern of 
initial stop and up-type Higgs scalar masses, for example, universal scalar
masses, this effect increases the constraint on the stop 
masses by almost 
a factor of two.  
This results in an increase of a factor of $\sim 3-4$ in the amount of 
fine tuning required to obtain the correct $Z$ mass. 
Further, in combining the positivity 
analysis with the constraints from the 
$\Delta m_K$ analysis, the QCD corrections to the Flavour Changing
Neutral Current (FCNC) operators have been 
included, as discussed in section \ref{setup}. In the case $(n_5,n_{10}) 
=(2,2)$, this effect alone 
increases the limit on $M_S$ and hence the limit on the stop mass
by a factor of $\sim 2-3$. 
The combination of these two elements implies that the positivity constraints 
can be quite severe. 

We proceed as follows. First, we solve the RGE's from 
$M_{GUT}$ to $\mu _S$ where
the heavy scalars are decoupled. At this scale, we add 
the finite parts of
the two loop diagrams. Next, we RG scale  
(without
the heavy scalar terms in the RGE's) from
$\mu _S$ to $\mu _G$ using these new boundary conditions.
Except where stated otherwise, 
the scales $\mu_S$ and $\mu_G$ are fixed to be 50 TeV and 1 TeV, 
respectively.
  
For $\tilde{f} _i \neq \tilde{t}$, $\tilde{t}^c$ we find,
\begin{eqnarray}
m^2_{\tilde{f} _i}(\mu _G) &=&  m^2_{\tilde{f} _{i,0}} +
( 2.84 C^i_3 + 0.639 C_2^i + 0.159 Y^2_i )
M^2_{3,0} \nonumber \\
 & & -( 4.38 C^i_3 + 1.92 C^i_2 + 0.622 Y^2_i )
\frac{1}{2} (n_5 +3n_{10}) \times 10^{-3} M^2_S
 \nonumber \\
 & & -  0.829 (n_5 - n_{10}) Y_i \times 10^{-3} M^2_S.
\label{highscale}
\end{eqnarray}
These results agree with reference \cite{nima} for the same 
choice of input parameters.
As in the previous section, the numerical coefficients in 
Eqn.(\ref{highscale}) depend on
$(n_5,n_{10})$ through the gauge coupling evolution, and the 
numbers in
Eqn.(\ref{highscale}) are for $(n_5,n_{10}) = (2,0)$.
\footnote{The numerical results presented in Fig.\ref{hm2} 
include this dependence.}
In Fig.\ref{hm2} we plot the values of $m_{\tilde{f} _{i,0}}/M_S$ 
that determine 
$m^2_{\tilde{f} _i}(\mu _G) = 0$ as a function of $M_3/M_S$,
for $\tilde{f}_i = \tilde{l}_i$, $\tilde{q}_i$, 
$\tilde{u}^c_i$, $\tilde{d}^c_i$ 
and $\tilde{e}^c_i$. We emphasize that the results presented in 
Fig.\ref{hm2} are independent
of any further limits that FCNC or fine tuning considerations may 
imply, and are thus  
useful constraints on any model building attempts.

\begin{figure}
\vspace{-0.7in}
\centerline{\epsfxsize=1.2\textwidth \epsfbox{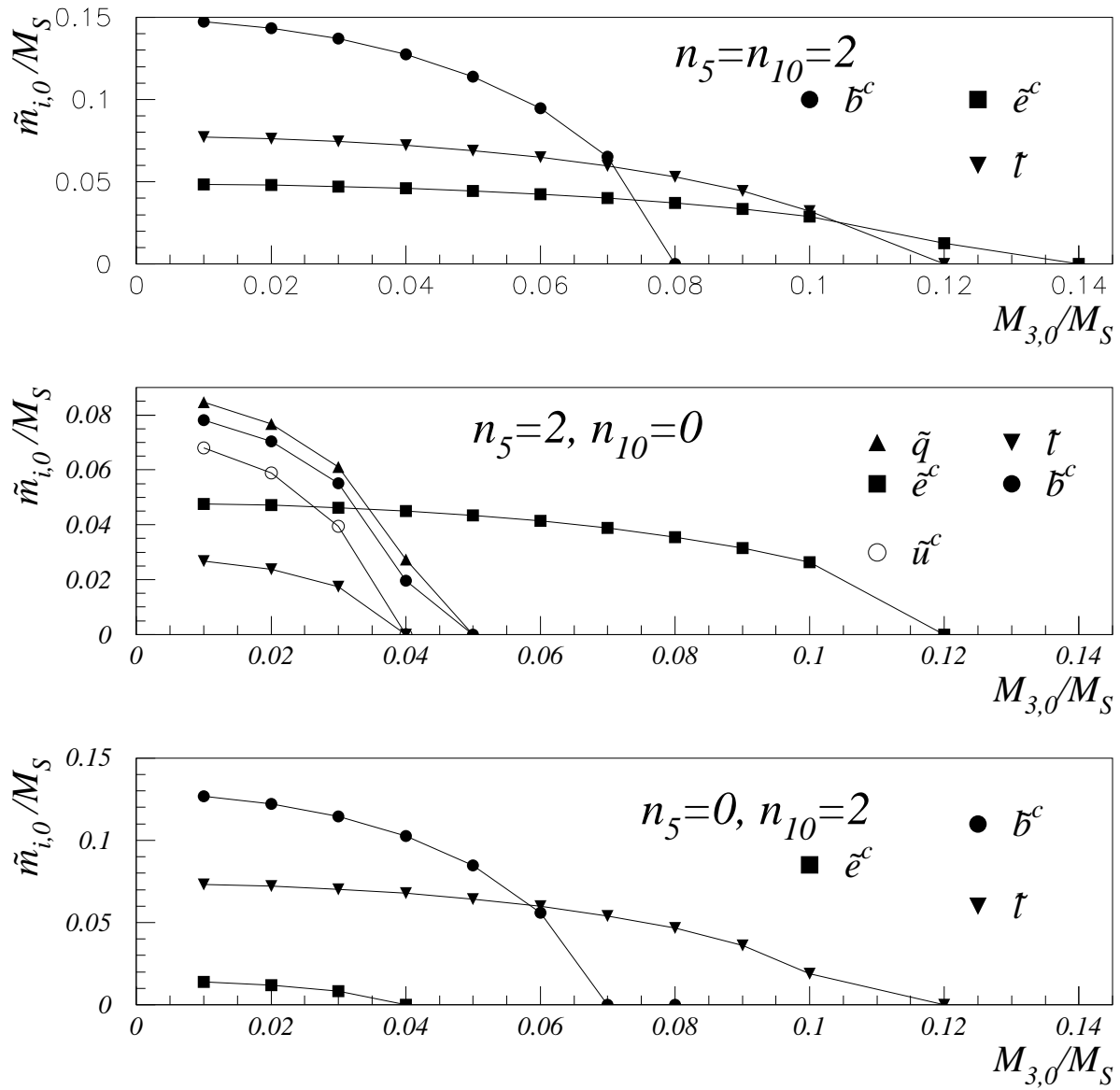}}
\vspace{-0.2in}
\caption{Limits for $m_{\tilde{f}_i,0}$ for different values of
$(n_5,n_{10})$ from the requirement
that the (mass)$^2$ are positive at the weak scale,
assuming 
a supersymmetry breaking scale of $M_{GUT}$. The allowed region lies above 
{\em all} the lines.}
\protect\label{hm2}
\end{figure}


For the stops, the numerical solutions to the RGE's for $\tan \beta=2.2$ are
\begin{eqnarray}
m^2_{\tilde{t}}(\mu_G) & = & - 0.021 A_t^2 +0.068 A_t M_{3,0} +
 3.52 M^2_{3,0} \nonumber \\
 & & - 0.142 (m^2_{H_u,0}+ m^2_{\tilde{t}^c,0}) +
0.858 m^2_{\tilde{t},0} -
 c_L 0.00567 M^2_S, \nonumber \\
m^2_{\tilde{t}^c}(\mu_G) & = & - 0.042 A_t^2 + 0.137 A_t M_{3,0} 
+ 2.35 M^2_{3,0} 
 \nonumber \\
 & & - 0.283 (m^2_{H_u,0}+ m^2_{\tilde{t},0}) +
0.716 m^2_{\tilde{t}^c,0} -
 c_R 0.00259M^2_S, \nonumber \\
m^2_{H_u}(\mu_G) & = & - 0.063 A_t^2 + 0.206 A_t M_{3,0} - 
1.73  M^2_{3,0}
\nonumber \\
 & & - 0.425 (m^2_{\tilde{t},0}+ 
m^2_{\tilde{t}^c,0})+
0.574 m^2_{H_u,0} +
 c_H 0.00218 M^2_S,
\label{highstop}
\end{eqnarray}
where $(c_L,c_R,c_H) = (1,1,1)$, $(3.9,4.7,4.5)$,
$(3,4,3.6)$ for $(n_5,n_{10}) = 
(2,0)$, $(2,2)$ and $(0,2)$, respectively.
The  
mixed two loop contribution to the RG evolution is $\propto (n_5-n_{10})$ 
and is not negligible. 
Thus there
is no simple relation between the $c$'s for different values of 
$n_5$ and $n_{10}$. 
From Eqns.(\ref{muEW}) and (\ref{ft}) we find that to keep $m^2_Z$ fine tuned 
at less than $1\%$ ($c \leq 100$) in each of the bare
 parameters, 
we must have: 
$\mu \ltap$ 500 GeV; $M_{3,0} \ltap 500
$ GeV; $m_{\tilde{t}_i,0} \ltap 1$ TeV; and $M_S \ltap 
7$ TeV for $(n_5,n_{10})=(2,2)$. The fine tuning 
of the $Z$ mass with respect to the heavy scalars is discussed in 
\cite{dimopoulos}. Finally, for other values of these parameters 
the fine tuning increases as 
$c= 100 \times \tilde{m}^2/ \tilde{m}^2_0$, where $\tilde{m}_0$ 
is the value 
of $\tilde{m}$ that gives $c=100$.

As in section \ref{lowsusy}, some relations 
between $m^2_{\tilde{t},0}$,
$m^2_{\tilde{t}^c,0}$ and $m^2_{H_u,0}$ are needed to obtain 
a constraint from
Eqn.(\ref{highstop}), using 
$m^2_{\tilde{t}}(\mu _G) >0$ and
$m^2_{\tilde{t}^c}(\mu _G) >0$. 
We discuss both model dependent and model independent constraints on 
the initial values of the stop masses. 
The outline of the rest of this section 
is as follows. First, we assume universal boundary conditions. These 
results are presented in Fig.\ref{m0}. 
Model independent constraints are obtained
by the following.
We assume that 
$m^2_{H_u,0}=0$
and choose $A_{t,0}$ to maximize the value of the stop masses at the weak 
scale. These results are presented in Fig.\ref{mhu0}. We further 
argue that these constraints represent minimum constraints as long 
as $m^2_{H_u,0} \geq 0$. 
To obtain another set of model independent constraints, we use  
the electroweak symmetry breaking
relation to eliminate $m^2_{H_u,0}$ in favor of $\mu$.
Then we present the positivity limits
for different values of $\tilde{\mu}/M_S$, where $\tilde{\mu}^2=\mu^2+
\frac{1}{2}m^2_Z$, and 
assume that \\
$m^2_{H_d,0}=0$ to 
minimize the value of $\mu$.\footnote{Strictly speaking, this last 
assumption is unnecessary. Only the combination 
$\tilde{\mu}^2_H \equiv 
\tilde{\mu}^2-m^2_{H_d,0}/ \tan ^2 \beta$ appears in our analysis. 
Thus for $m^2_{H_d,0} \neq 0$ our results are unchanged if the 
replacement $\tilde{\mu} \rightarrow \tilde{\mu}_H$ is made.}
These limits are model independent and are presented in 
Fig.\ref{mu},
for the case $n_5 = n_{10} =2$. 
We then combine these analyses with the limits on $M_S$ obtained 
from $\Delta m_K$. We conclude with some discussion about the anomalous 
$D-$term solutions to the flavor problem.

 We first consider universal boundary
conditions for the stop and Higgs masses. 
That is, we assume that $m^2_{\tilde{t},0} 
= m^2_{\tilde{t}^c,0}=
m^2_{H_u,0}= \tilde{m}^2_0$. In Fig.\ref{m0} we plot for 
$\tan \beta=2.2$ the 
minimum value of $\tilde{m}_0/M_S$ required to
maintain $m^2_{\tilde{t}}(\mu _G) >0$ and
$m^2_{\tilde{t}^c}(\mu _G) >0$. This value of $\tan \beta$ 
corresponds to $\lambda_t(M_{GUT})=0.88$, in 
the case that $(n_5,n_{10})=(2,0)$. 
For comparison, the results 
gotten assuming $\lambda_t=0$ may be found in reference \cite{nima}. 
For 
$n_5=n_{10}=2$ we note from Fig.\ref{m0} that if $M_S=$ 20 TeV 
and the gaugino masses are small, the limit on the stop mass is 
$m_{\tilde{t}^c,0} \geq$ 6 TeV. 
This limit is weakened to 5.6 TeV if $M_{3,0}\ltap$ 300 GeV 
is allowed. 
Even in this case, 
this large initial stop mass requires a fine tuning  
that is $c\sim (\hbox{5.6 TeV})^2/m^2_Z \sim$ 3700, {\it i.e.}, 
a fine tuning of $\sim 10^{-3}$ is needed to obtain the correct $Z$ mass.     
 
We now assume $m^2_{H_u,0}=0$ and choose the initial 
value of $A_{t,0}$ to {\em maximize} the value of 
$m^2_{\tilde{t}_i}(\mu_G)$.
The values  
of $m^2_{\tilde{t},0}$ and $m^2_{\tilde{t}^c,0}$ are 
chosen such that  
$m^2_{\tilde{t}}(\mu _G) >0$ and
$m^2_{\tilde{t}^c}(\mu _G)>0$. We note that in this case 
the constraint is 
weaker because the $\lambda_t$ contribution to the RG 
evolution of the stop (mass)$^2$ is less negative. 
These results are plotted in Fig.\ref{mhu0}. 

We discuss this case in some more detail and argue that 
the minimum value of $m_{\tilde{t}_i,0}$ obtained in this way will 
be valid for all $m^2_{H_u} \geq0$ and all
$A_{t,0}$. 
Eliminate the $A_{t,0}$ term by choosing $A_{t,0}=
k M_{3,0}$ such that the $A_t$ contributions to $m^2_{\tilde{t}_i}(\mu_G)$ 
is maximized. Other choices for $A_{t,0}$ require larger values for 
$m^2_{\tilde{t}_i,0}$ to maintain $m^2_{\tilde{t}_i}(\mu_G)=0$. 
The value of $k$ is determined by the following. A general 
expression for the value of the soft masses of the stops at the weak scale 
is 
\begin{equation}
 m^2_{\tilde{t}}(\mu_G)=-a A^2_{t,0}+b A_{t,0} M_{3,0}+c M^2_{3,0}+\cdots,
\end{equation}
\begin{equation}
 m^2_{\tilde{t}^c}(\mu_G)=-2a A^2_{t,0}+2b A_{t,0} M_{3,0}+d M^2_{3,0}+\cdots,
\end{equation}       
with $a$, $c$ and $d$ positive. The maximum value of 
$m^2_{\tilde{t}_i}(\mu_G)$ 
is obtained by choosing $A_{t,0}=bM_{3,0}/2a$.  
The value of the stops masses at this 
choice of $A_{t,0}$ are  
\begin{equation}
 m^2_{\tilde{t}}(\mu_G)=(c+\frac{b^2}{4 a}) M^2_{3,0}+\cdots,
\end{equation}
\begin{equation}                                                           
m^2_{\tilde{t}^c}(\mu_G)=(d+ 2 \frac{b^2}{4 a}) M^2_{3,0}+\cdots.
\end{equation}
An inspection of Eqn.(\ref{highstop}) gives 
$b=0.068$ and $a=0.021$ for 
$\tan \beta =2.2$. In this case the `best' value for $A_{t,0}$ is 
$A^B_{t,0}\sim 1.6 M_{3,0}$. It then follows 
that the quantity $b^2/4a=0.055$ is a small correction 
to the coefficient of the gaugino contribution in Eqn.(\ref{highstop}).
Thus the difference between the minimum initial stop masses 
for $A_{t,0}=0$ and $A_{t,0}$= 
$A^B_{t,0}$ is small.
Next assume that $m^2_{H_u,0}=0$.
Requiring  both 
$m^2_{\tilde{t}}(\mu _G)=0$ and $m^2_{\tilde{t}^c}(\mu _G)=0$ 
determines a minimum value for $m^2_{\tilde{t},0}$ and 
$m^2_{\tilde{t}^c,0}$.   
Now since the $m^2_{H_u,0}$ contribution to both the
the stop soft (mass)$^2$ is negative (see Eqn.(\ref{highstop})), the
minimum values for $m^2_{\tilde{t}_i,0}$ 
found by the preceeding procedure are also minimum values
if we now allow any $m^2_{H_u,0} >0$. 

We conclude that for all $A_{t,0}$ and all $m^2_{H_u,0} \geq0$, 
the limits presented in Fig.\ref{mhu0} represent 
lower limits on the initial stop masses if we require that the soft
(mass)$^2$ remain positive at the weak scale. Further, the limits in this 
case are quite 
strong. For example, from Fig.\ref{mhu0} we find that if 
$M_S \sim$ 20 TeV and $M_{3,0} \sim $ 200 GeV 
(so that $M_{3,0}/M_S \sim$10$^{-2})$,
then the initial stop masses must be greater than 3.5 TeV
in the case that $(n_5,n_{10})=(2,2)$
The results are stronger in a more realistic 
scenario, {\it i.e.,} with
$m_{H_u,0}^2 > 0$.  If, for example, $m^2_{H_u,0} \sim 
m^2_{\tilde{t}^c,0}/9$ the constraints are larger by only a few percent. 
In the case that $m^2_{H_u,0}=m^2_{\tilde{t}^c,0}=m^2_{\tilde{t},0}$, 
presented in Fig.\ref{m0}, however, 
the constraint on the initial $\tilde{t}^c$ mass
increases by almost a factor of two. 

To obtain constraints on the initial stop masses we have thus 
far had to assume some relation between $m^2_{H_u,0}$ and 
$m^2_{\tilde{t}^c,0}$, for example, $m^2_{H_u,0}=0$ or 
$m^2_{H_u,0}=m^2_{\tilde{t}^c,0}$. Perhaps a better approach is 
to use the EWSB relation, Eqn.(\ref{muEW}), to
eliminate $m^2_{H_u,0}$ in favor of $\mu ^2$. This has the advantage of
being model independent.   
It is also a useful reorganization of independent parameters since 
the amount of fine tuning required to obtain the correct
$Z$ mass increases as $\mu$ is increased. 
To obtain some limits we choose $m_{H_d ,0}^2 = 0$ 
\footnote{This assumption is unnecessary. 
See the previous footnote.} to minimize the value of $\mu^2$, 
and require that $m^2_{H_u,0}$ is positive. The minimum value of 
$m_{\tilde{t}^c,0}/M_S$ and $m_{\tilde{t},0}/M_S$ 
for different choices of $\tilde{\mu}/M_S$ are gotten by 
solving $m^2_{\tilde{t}^c}(\mu _G) =0$ and 
$m^2_{\tilde{t}}(\mu _G) =0$. These results are presented in 
Fig.\ref{mu}. 
In this Figure the positivity constraints terminate at that
value of $M_{3,0}$ which gives
$m^2_{H_u,0}=0$. 

As discussed above, reducing the value of $m^2_{H_u,0}$ 
decreases the positivity limit on $m_{\tilde{t}_i,0}$. Consequently  
the fine tuning of $m_Z$ with respect to $m_{\tilde{t}_i,0}$ is also 
reduced. But using Eqns.(\ref{highstop}) and (\ref{muEW}),
it can be seen that
decreasing $m_{H_u ,0}^2$ while
keeping $m_{\tilde{t}^c}^2(\mu _G) =0$ and
$m_{\tilde{t}}^2(\mu _G) =0$ 
results in a larger $\mu$, thus increasing
the fine tuning
with respect to
$\mu$. 
This can also be seen from Fig.\ref{mu}. We find, for example, that if  
$M_{3,0}/M_S \sim 0.01$, 
the small value $\tilde{\mu}/M_S=0.01$ requires 
$m_{\tilde{t}_i,0}/M_S \sim 0.25$. For $M_S=10$ TeV, this corresponds to
$\mu \sim$ 100 GeV and $m_{\tilde{t}_i,0}\geq$ 2.5 TeV. 
A further
inspection of Fig.\ref{mu} shows that for the same 
value of $M_{3,0}/M_S$, 
a value of 
$m_{\tilde{t},0}/M_S =0.17$
is allowed (by reducing
$m^2_{H_u,0}$) only if $\tilde{\mu}/M_S$ is increased to 0.14.  
This corresponds to $\mu=1.4$ TeV for
$M_S=10$ TeV; this implies that $c(m^2_Z; \mu) \sim 930$. 
We find that the limit on the initial
stop masses can only be decreased at the expense of increasing $\mu$.

Finally, the limits become weaker if
$m^2_{H_u,0}<0$. This possibility is theoretically unattractive on two
accounts.
Firstly, a nice feature of supersymmetric extensions to the SM
is that the dynamics of the model, through the
presence of the large top quark Yukawa coupling, naturally leads
to
the breaking of the electroweak symmetry \cite{ross}. This is lost if
electroweak symmetry breaking is already present at the
tree level.
Secondly, the fine tuning required to obtain the correct $Z$ mass is increased.
From Fig.\ref{mu} we infer that while reducing $m^2_{H_u,0}$ below zero
 does 
reduce the limit on the initial stop masses, the value of $\mu$ increases 
beyond the values quoted in the previous paragraph, thus 
futher increasing the 
fine tuning of the $Z$ mass. 
This scenario is not discussed any further.

\begin{figure}
\vspace{-0.7in}
\centerline{\epsfxsize=1.2\textwidth \epsfbox{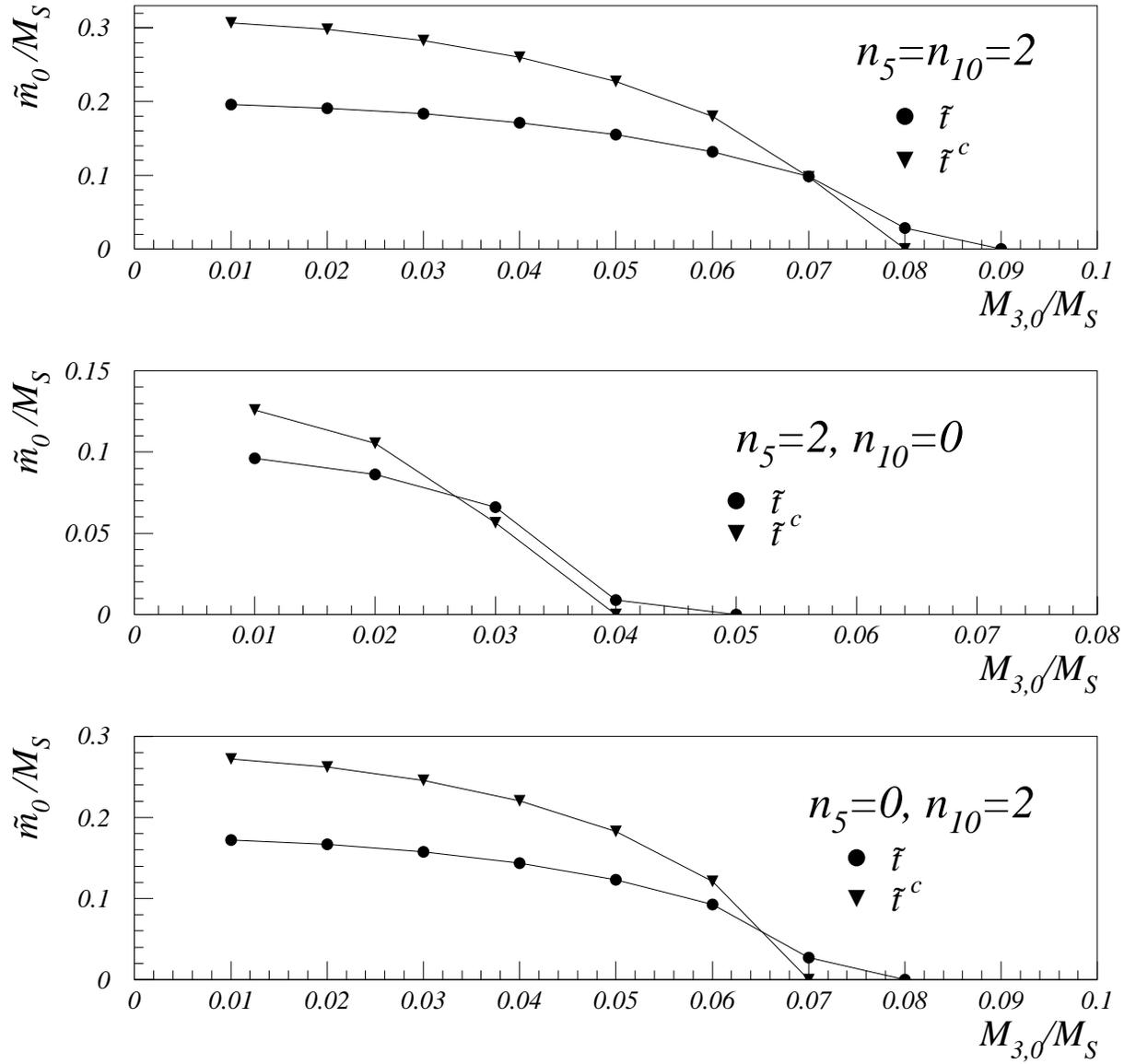}}
\vspace{-0.2in}
\caption{Limits for $\tilde{m}_0/M_S$ 
from the requirement
that the stop (mass)$^2$ are positive at the weak scale,
for 
$\tan \beta=2.2$, $A_{t,0}=0$ and assuming universal scalar masses 
at $M_{GUT}$ for the stop and Higgs scalars. 
The region below the curves are
excluded.}
\protect\label{m0}
\end{figure}        

\begin{figure}
\vspace{-0.7in}
\centerline{\epsfxsize=1.2\textwidth \epsfbox{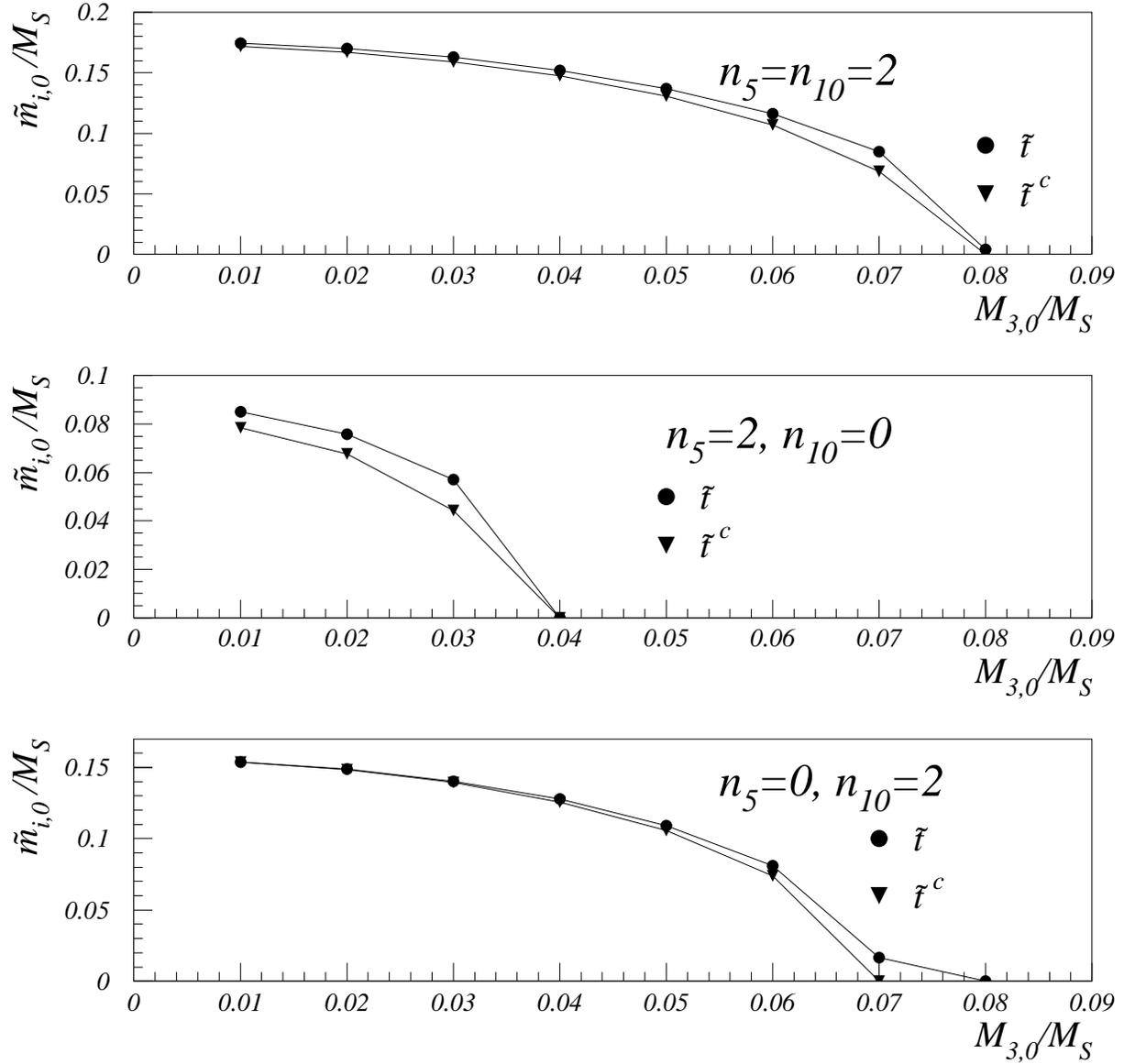}}
\vspace{-0.2in}
\caption{Limits for $m_{\tilde{t},0}/M_S$, $m_{\tilde{t}^c,0}/M_S$,
from the requirement
that the stop (mass)$^2$ are positive at the weak scale, 
for $M_{SUSY}=M_{GUT}$, $\tan \beta=2.2$ and assuming that 
$m^2_{H_u,0}=0.$ The value of $A_{t,0}$ is chosen to maximize the 
value of the stop soft masses at the weak scale. 
The region below the curves are
excluded.}  
\protect\label{mhu0}
\end{figure}

\begin{figure}
\centerline{\epsfxsize=1.0\textwidth \epsfbox{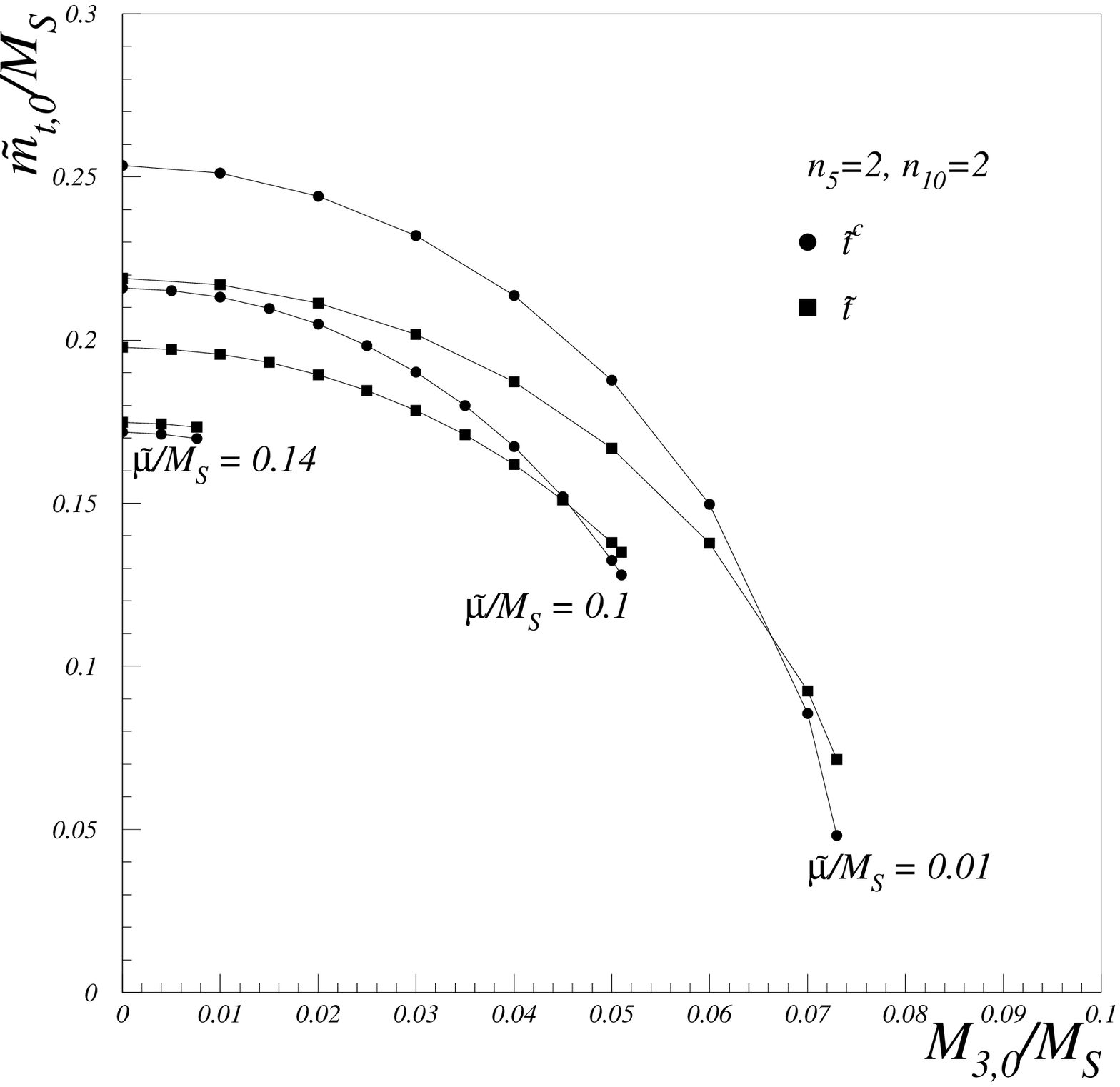}}
\caption{Limits for $m_{\tilde{t},0}/M_S$, $m_{\tilde{t}^c,0}/M_S$,
from the requirement
that the stop (mass)$^2$ are positive at the weak scale,
for $(n_5,n_{10})=(2,2)$, 
$M_{SUSY}=M_{GUT}$, $\tan \beta=2.2$, and different 
values of $\tilde{\mu}/M_S$. The contours end at that value of 
$M_{3,0}/M_S$ that gives $m_{H_u,0}/M_S=0$.
The value of $A_{t,0}$ is chosen to maximize the
value of the stop soft masses at the weak scale.}
\protect\label{mu}
\end{figure}  

We now combine the positivity analysis of this section with the results 
of section \ref{setup} to place lower limits on the soft 
scalar masses.
For given values of $\delta_{LL}, \delta_{RR}$, 
a minimum value of $M_S$, $M_{S,min}$, is 
found using the results of section \ref{setup}. 
This is combined with the 
positivity analysis in Fig.\ref{mhu0}, to produce 
the results shown in Fig.\ref{mhu0mk}. We also show other 
limits gotten by assuming $m^2_{H_u,0}=m^2_{\tilde{t}^c,0}$. 
These results are presented in Fig.\ref{mhu3mk}.
In Fig.\ref{mu2} we also present the stop mass limits for 
different values of $\mu$, and restrict to 
$m^2_{H_u,0} \geq0$ and $\sqrt{\delta_{LL} \delta_{RR}}=0.04$. 
In all cases the heavy scalars 
were decoupled at $M_{S,min}$, rather than 
50 TeV, and so the positivity analysis was 
repeated. The value of $A_{t,0}$ was chosen 
to maximize the value of the stop masses at the weak scale. 
For completeness, the results for the 
cases $(n_5,n_{10})=(2,0)$ and 
$(0,2)$ and $m^2_{H_u,0}=0$ are presented in Fig.\ref{mhu0mk2}.  
We repeat that the minimum allowable values for the 
stop masses consistent with 
$m^2_{H_u,0}>0$, gotten by setting $m^2_{H_u,0}=0$, are given in 
Figs.\ref{mhu0mk} and \ref{mhu0mk2}.

We next briefly discuss some consequences of this numerical analysis. 
We concentrate on the case $n_5=n_{10}=2$, since this is the relevant
case to consider if the supersymmetric flavor
problem is solved by decoupling
the heavy scalars.
Other choices for $n_5$ and $n_{10}$ require
additional physics to explain the required degeneracy or
alignment of any
light non-third generation scalars. 
From Figs.\ref{mhu0mk} and \ref{mhu3mk} we find that for 
$\sqrt{\delta_{LL} \delta_{RR}}=0.22$ and $M_{3,0} \leq$ 1 TeV, 
$m_{\tilde{t}_i,0} >$7 TeV is required.  
If instead we 
restrict both $c(m^2_Z;M^2_S)$ and $c(m^2_Z; M^2_{3,0})$ 
to be less than 100,  
then we must have
$M_S \ltap$ 7 TeV and $M_{3,0} \ltap$ 500 GeV. 
To not be excluded by 
$\Delta m_K$, we further require 
that $\sqrt{\delta_{LL} \delta_{RR}}\leq 0.04$ 
which leads to a fine tuning of one part in
$\sim 2/\delta$
, {\it i.e.,} $\sim 50$.
An inspection of Figs.\ref{mhu0mk} 
and \ref{mhu3mk} implies that for $\sqrt{\delta_{LL} \delta_{RR}}
\approx 0.04$,
$m_{\tilde{t},0}$ must be larger than 0.9$-$1.3 TeV, depending on the 
value of $m^2_{H_u,0}$. Alternatively, if we also 
restrict $\mu \leq$ 500 GeV, 
then from Fig.\ref{mu2} we find that $m_{\tilde{t}_i,0} \geq $ 800 
GeV. 
Thus $c(m^2_Z; m^2_{\tilde{t}_i,0})=64-170$. This fine tuning 
can be reduced only by increasing the $c(m^2_Z)$'s for the 
other parameters to more than 100 (or by increasing the fine tuning of
$\delta$ to more than one part in 50). 
We conclude that unless $\sqrt{\delta_{RR} \delta_{LL}}$ is 
naturally small, decoupling the 
heavy scalars does not provide a natural
solution to the flavor problem. 

\begin{figure}
\vspace{-1in}
\centerline{\epsfxsize=1.1\textwidth \epsfbox{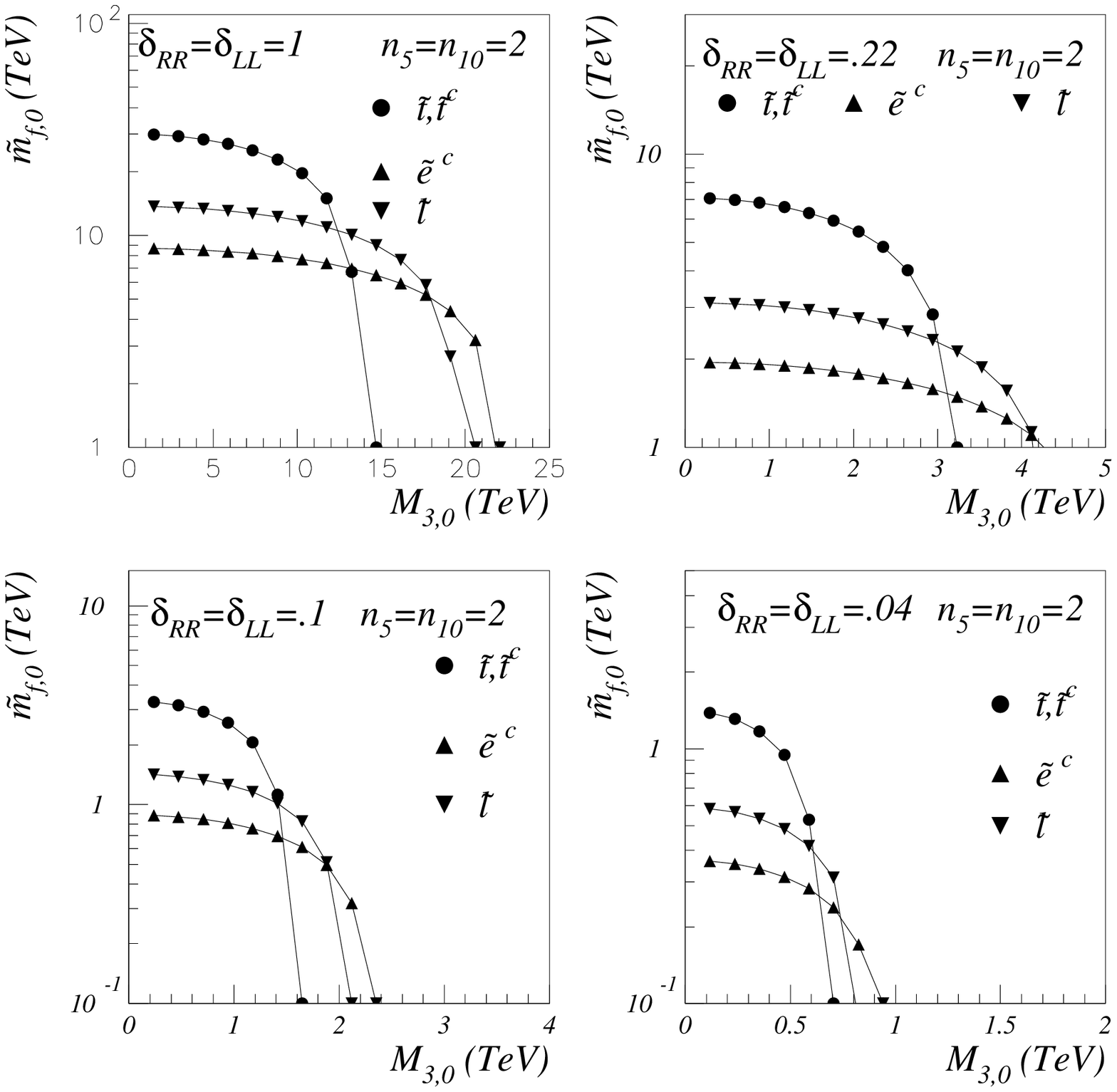}}
\vspace{-0.2in}
\caption{
Limits for $m_{\tilde{t},0}$ and $m_{\tilde{t}^c,0}$,
$m_{\tilde{e}^c}$, and $m_{\tilde{l}}$
from the requirement
that the (mass)$^2$ are positive at the weak scale
while suppressing $\Delta m_K$.
It was assumed that $M_{SUSY}=M_{GUT}$, $\tan \beta=2.2$ and that
$m^2_{H_u,0}=0.$ The value of $A_{t,0}$ was chosen to maximize the
value of the stop soft masses at the weak scale. The heavy scalars were 
decoupled at the minimum value allowed by $\Delta m_K$.}
\protect\label{mhu0mk}
\end{figure}

\begin{figure}
\vspace{-.8in} 
\centerline{\epsfxsize=1.2\textwidth \epsfbox{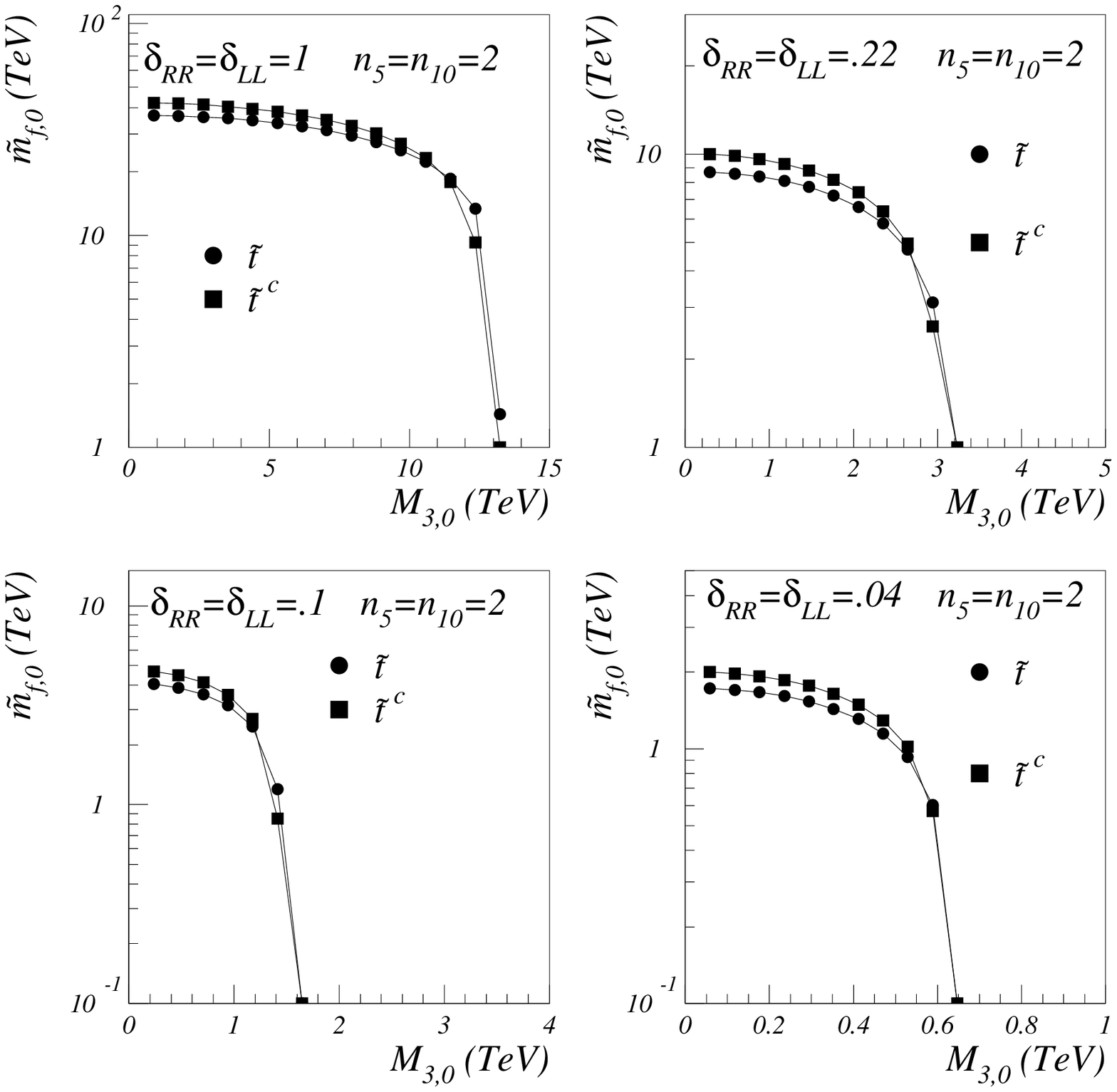}}
\vspace{-0.2in}
\caption{Limits for $m_{\tilde{t},0}$ and $m_{\tilde{t}^c,0}$
from the requirement
that the stop (mass)$^2$ are positive at the weak scale
while suppressing $\Delta m_K$.
It was assumed that $M_{SUSY}=M_{GUT}$, $\tan \beta=2.2$ and that
$m^2_{H_u,0}=m^2_{\tilde{t}^c,0}.$ 
The value of $A_{t,0}$ was chosen to maximize the
value of the stop soft masses at the weak scale. The heavy scalars were
decoupled at the minimum value allowed by $\Delta m_K$.}
\protect\label{mhu3mk}
\end{figure}

\begin{figure}
\centerline{\epsfxsize=1\textwidth \epsfbox{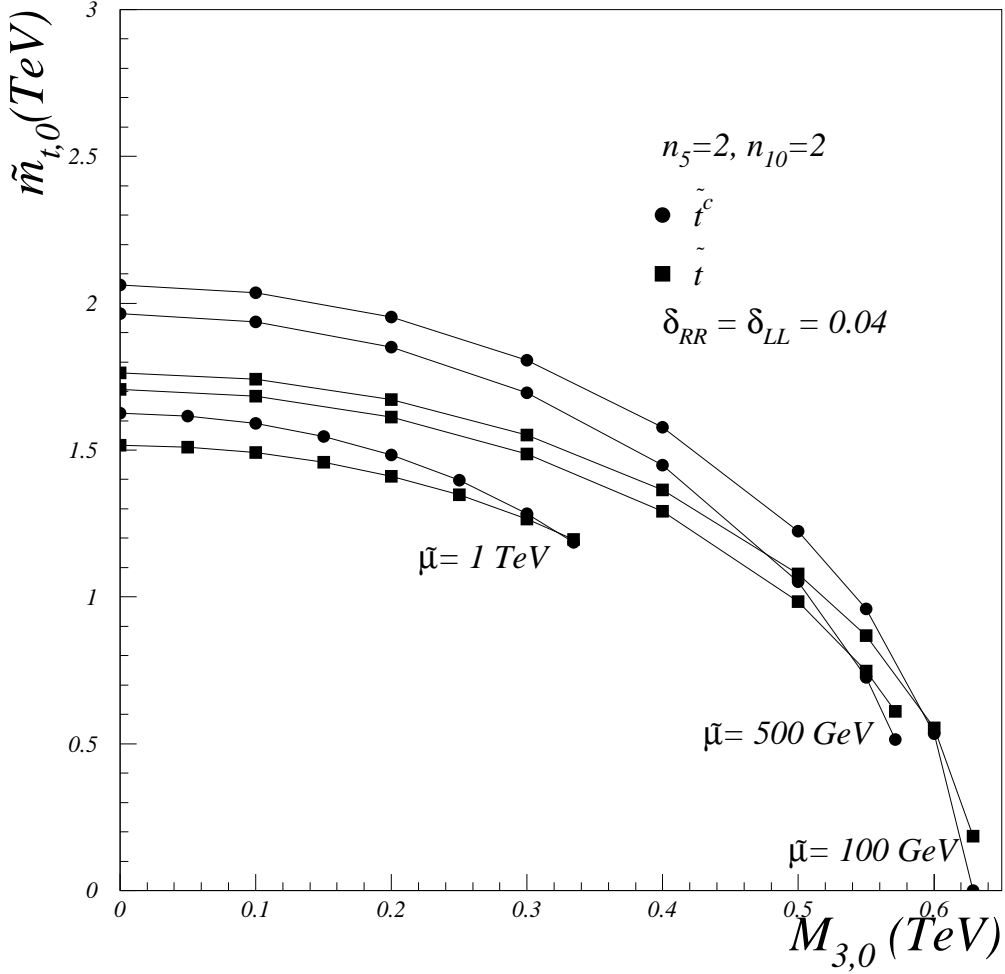}}
\caption{Limits for $m_{\tilde{t},0}$ and $m_{\tilde{t}^c,0}$
from the requirement
that the stop (mass)$^2$ are positive at the weak scale
while suppressing $\Delta m_K$,
for $(n_5,n_{10})=(2,2)$, $\protect\sqrt{\delta_{LL} \delta_{RR}}=0.04$,
and different values of $\mu$.
The contours terminate at $m^2_{H_u,0}=0$.
It was assumed that $M_{SUSY}=M_{GUT}$ and $\tan \beta=2.2$. 
The value of $A_{t,0}$ was chosen to maximize the
value of the stop soft masses at the weak scale. 
The heavy scalars were
decoupled at the minimum value allowed by $\Delta m_K$.
}
\protect\label{mu2}
\end{figure}    
 
\begin{figure}
\vspace{-0.7in}
\centerline{\epsfxsize=1.1\textwidth \epsfbox{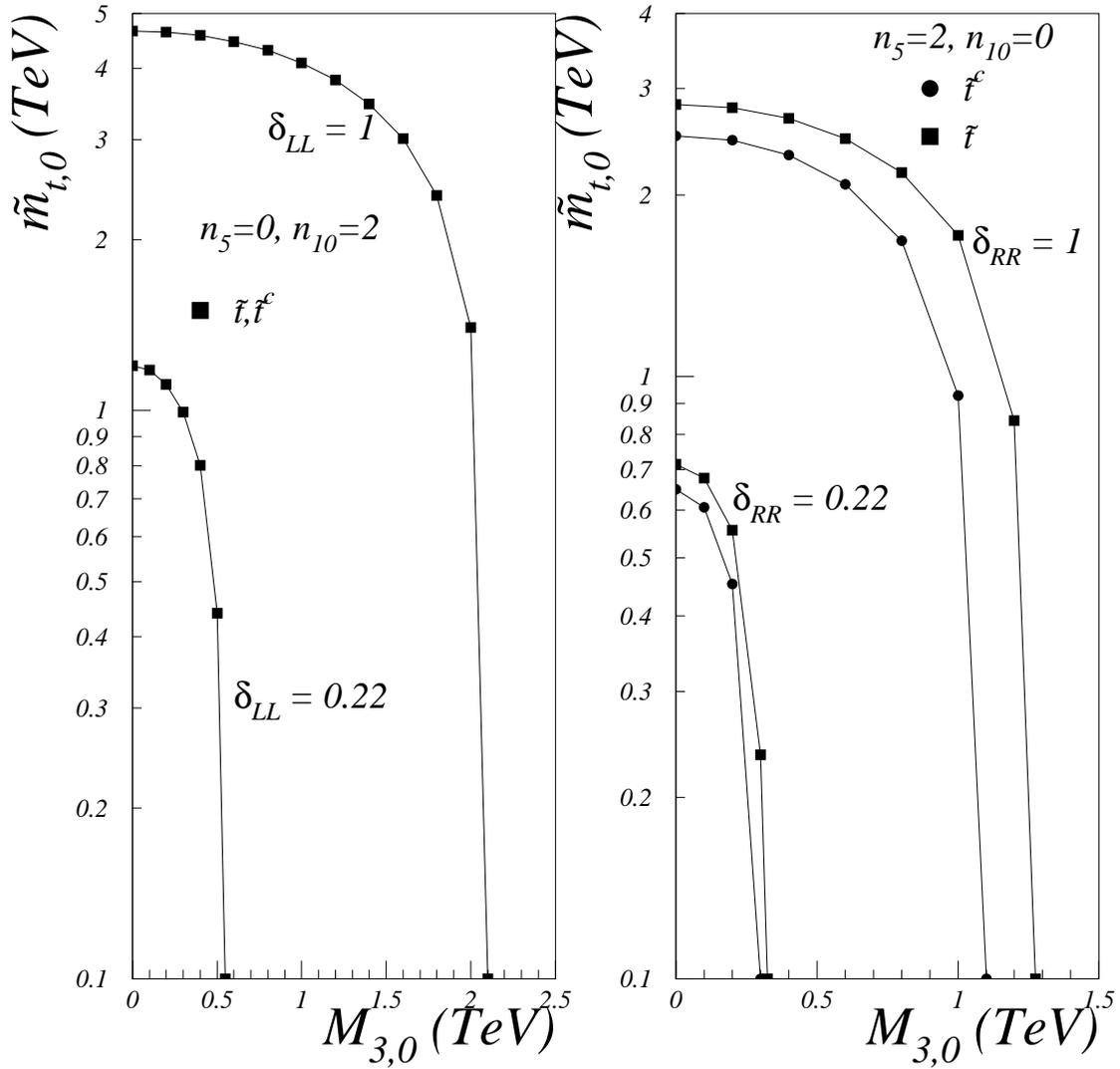}}
\vspace{-0.2in}
\caption{Limits for $m_{\tilde{t},0}$, $m_{\tilde{t}^c,0}$
from the requirement
that the stop (mass)$^2$ are positive at the weak scale
while suppressing $\Delta m_K$,
for the cases $(n_5,n_{10})=(2,0)$ and $(0,2)$.
It was assumed that $M_{SUSY}=M_{GUT}$, $\tan \beta=2.2$ and that
$m^2_{H_u,0}=0.$ The value of $A_{t,0}$ was chosen to maximize the
value of the stop soft masses at the weak scale. The heavy scalars were
decoupled at the minimum value allowed by $\Delta m_K$.}
\protect\label{mhu0mk2}
\end{figure}      

To conclude this section we 
discuss the constraint this analysis implies for those models 
which generate
a split mass spectrum between the third and the first two
generations
through the $D$-term contributions of an anomalous
$U(1)$
gauge symmetry \cite{dvali,nelson,riotto}.
These models
can ``explain'' the hierarchy of the
Yukawa couplings.
In the model of set D of \cite{nelson},
the two ${\bf \bar{5}}$'s of the first two generations are
at $7$ TeV and $6.1$ TeV and
the two ${\bf 10}$'s are at $6.1$ and $4.9$ TeV, respectively, so that
$\Delta m_K$ is suppressed. 
These values must be increased 
by a factor of 2.5 
to correct for the QCD enhancement of the SUSY 
contribution to $\Delta m_K$, as discussed in section
\ref{setup}.   
To obtain a conservative 
bound on the initial stop masses from the positivity requirement, 
we first assume that all the 
heavy scalars have a common mass $M_S=2.5 \times 5$TeV$=12.5$ TeV.
(It would have been 5 TeV without the QCD correction.)
Then assuming a weak scale value of the gluino mass which is less than 
$1.5$ TeV 
(so that $c(m^2_Z,M^2_{3,0})$ is less
than 100) and setting $m^2_{H_u,0}=0$ $(m^2_{\tilde{t}^c,0})$, we find from 
Fig.\ref{mhu0} $($\ref{m0}$)$ that $m_{\tilde{t} ,0}\geq 
2.0$ $(3.4)$ TeV is required.  
This leads to $c(m^2_Z; m^2_{\tilde{t} ,0}) \geq
400$ $ (1100)$.
To obtain a better bound, we repeat our 
analysis using $n_5 m^2_5+\hbox{3} n_{10} m^2_{10}=((\hbox{7 TeV})^2+ 
(\hbox{6.1 TeV})^2+\hbox{3}\times (\hbox{6.1 TeV})^2+
\hbox{3}\times (\hbox{4.9 TeV})^2)\times (\hbox{2.5})^2$. It is possible to 
do this since only this 
combination appears in the RG analysis for $(n_5,n_{10})=(2,2)$. 
We find (assuming
$m_{H_u ,0}^2 =0$ and the gluino mass at the weak scale
is less than $1.5$ TeV) that $m_{\tilde{t} ,0}
\stackrel{>} \sim 3$ TeV.
In the model of \cite{riotto},
$\delta_{RR} \approx \delta_{LL} \approx 0.01$. To obtain a limit on the 
initial stop masses, we use the bound obtained from either 
Figs.\ref{mhu0mk} or \ref{mhu3mk} for 
$\delta_{RR}=\delta_{LL} \approx 0.04$, 
and divide the limit by a factor of 4.
By inspecting these Figures we find that this model is only weakly 
constrained, even if $ m^2_{H_u,0} \sim m^2_{\tilde{t},0}$. We now 
discuss the limits in this model  
when $O(1)$ $CP$ violating phases are present. 
To obtain the minimum value of 
$M_S$ in this case, 
we should multiply the minimum value of $M_S$ obtained from 
the $\Delta m_K$ constraint 
for $\delta_{LL}=\delta_{RR}=0.04$ by 12.5/4; dividing 
by 4 gives the result for
$\delta_{LL}=\delta_{RR}=0.01$ and 
multiplying by 12.5 gives the constraint on $M_S$ from  
$\epsilon$. 
The result is $M_S\gtap$ 23 TeV. 
Next, we assume that $M_{3,0}$ is less than 600 GeV, so that the 
value of the gluino mass at the weak scale is less than $\sim$ 1.5 TeV. 
This gives $M_{3,0}/M_S \leq0.026$. Using these values of $M_{3,0}$ and 
$M_S$, an inspection of Figs.\ref{m0} and \ref{mhu0} implies that 
$m_{\tilde{t},0}$ must be larger 
than 3.9 TeV to 6.7 TeV, depending on the value 
of $m^2_{H_u,0}$. This gives 
$c(m^2_Z; m^2_{\tilde{t} ,0}) \geq 1600$.
In the model of \cite{dvali},
$M_{3,0}/M_S \approx 0.01$ and $m_{\tilde{f} ,0}/M_S
\approx 0.1$.
Inspecting Figs.\ref{m0} and \ref{mhu0} we find that these values are
excluded for $(n_5,n_{10})=(2,2)$ and $(0,2)$. The case $(2,0)$ is 
marginally allowed.
The model of \cite{dvali} with 
$(n_5,n_{10})=(2,2)$ and $\lambda_t=0$ was also excluded 
by the analysis of reference \cite{nima}.  

\section{Conclusions}
\label{end}
In this chapter we have studied whether the SUSY flavor problem
can be solved by
making the scalars of the first and second
generations heavy, with masses $M_S$
($\stackrel{>} \sim
$few TeV), without destabilising the weak scale. If the
scale, $M_{SUSY}$, at which SUSY breaking is mediated to the SM scalars
is close to the GUT scale, then the heavy scalars
drive the light scalar (in particular the stop) (mass)$^2$
negative through two loop RG evolution. In order
to keep the (mass)$^2$ at the weak scale positive,
the initial value
of the stop (and other light scalar) soft masses, $m_{\tilde{f}_i,0}$,
must typically be
$\stackrel{>} \sim 1$ TeV, leading
to fine tuning in EWSB. We included two new effects in this
analysis:
the effect of $\lambda _t$ in the RGE's which makes
the stop (mass)$^2$
at the weak scale more negative and hence makes
the constraint                           
on the initial value stronger, and the QCD corrections
to the
SUSY box diagrams which contribute to $K-\bar{K}$
mixing.

Some results of our analysis for $M_{SUSY} = M_{GUT}$
can be summarized as follows. We restrict the gluino
mass (at the weak
scale) to be less than about $1.5$ TeV, so that the fine
tuning of
$m^2_Z$ with respect to the bare gluino mass, $M_{3,0}$, is not worse
than $1\%$. This requires that $M_{3,0} \ltap 600$ GeV.
We also assume that $m_{H_u ,0}^2 = 0$ to 
maximize the value of the stop masses at the weak scale.
We find that 
if $\sqrt{\delta_{LL} \delta_{RR}}=0.22$ then $M_S \geq$ 40 TeV 
is required to be consistent with $\Delta m_K$.  With
these assumptions, 
this implies that 
for $M_{3,0}$ less than 1 TeV, 
$m_{\tilde{t}_i,0} >$ 6.5 TeV is needed 
to not break color and charge at the 
weak scale.
Even for $\sqrt{\delta _{LL} \delta _{RR}} = 0.04$,
we find that
we need $M_S \stackrel{>} \sim 7$ TeV.
This implies that $m_{\tilde{t},0} >1$ TeV is required 
if $M_{3,0}\leq $ 500 GeV. 
This results in
a fine tuning of $1\%$.        
For $\delta_{LL} = 1$ and $\delta _{RR} = 0$,
we find that $M_S \stackrel{>} \sim 30$ TeV and
$m_{\tilde{t},0} > 4$ TeV.
For $\delta_{LL} = 0.22$ and $\delta _{RR} = 0$,
we find that $M_S \stackrel{>} \sim 7$ TeV and
$m_{\tilde{t},0} > 1$ TeV
(this holds for an initial gluino mass less than about $300$ GeV).
For $\delta_{LL} = 0$ and $\delta _{RR} = 1$,
we find that $M_S \stackrel{>} \sim 30$ TeV and
$m_{\tilde{t}^c,0} > 2$ TeV.
The constraints are weaker for smaller values of $\delta$.
In a realistic model,
$m_{H_u ,0}^2$
might be comparable to $m_{\tilde{t},0}^2$ and the
constraints
on $m_{\tilde{t},0}$ in this case 
are stronger. This is also discussed.
We note that
independent of the constraint from
$K-\bar{K}$ mixing, our analysis can be used to check the
phenomenological viability of any model that has
heavy scalars. We also discuss the phenomenological viability of 
the anomalous $D-$term solution, and find it to be problematic.

We then considered the possibility that $M_{SUSY} = M_S$.
In this case, there is no RG log enhancement of the
negative contribution of the heavy scalar masses to
the light scalar (mass)$^2$.
For this case, we computed the   
finite parts of the two loop diagrams
and used these results as estimates of the two loop
contribution
of the heavy scalars to the light scalar soft (mass)$^2$.
We then combined these results with the constraints
from $K-\bar{K}$ mixing
to obtain lower limits on the boundary values of the stop mass.
As an example, we assumed gauge mediated SUSY breaking boundary conditions
for the light scalars. If $n_5 \neq n_{10}$ then one of the 
selectron masses, 
rather than the stop masses, provides the stronger constraint 
on $m_{\tilde{t}_i,0}$ once gauge mediated boundary 
conditions are used to relate $m_{\tilde{e}^c,0}$ and $m_{\tilde{l},0}$
to   
$m_{\tilde{t}_i,0}$. Some of our results
can be summarized as follows. We restrict the gluino
mass at the weak scale
to be less than about $3$ TeV, again
to avoid more than $2\%$ fine tuning of $m^2_Z$
with respect to the gluino mass. 
For $\sqrt{\delta_{LL} \delta_{RR}}=0.22$ we find that 
$m_{\tilde{t}_i,0} \geq$ 1 TeV is required. 
The fine tuning of $m^2_Z$ with respect to the stop
mass is $\sim 3\%$ in this case.
For the cases $\delta_{LL} = 0$ and $\delta _{RR} = 1$,
and $\delta_{LL} = 1$
and $\delta _{RR} = 0$ we find that $m_{\tilde{t},0}
\stackrel{>} \sim 1$ TeV.
As before, the constraints on $m_{\tilde{t},0}$ for smaller
values of
$\delta$ are weaker than $\sim 1$ TeV. Again, we emphasize 
that the constraints in an 
actual model of this low energy supersymmetry breaking scenario 
could be different, and our results should
be treated as estimates only. Finally, we also briefly discuss 
the  
$CP$ violating constraints from $\epsilon$, and find that 
all these limits increase
by a factor of $\sim 12$ if $O(1)$ phases are present.

\chapter{Summary}
In this thesis, we studied some fine tuning and 
naturalness issues in the supersymmetric Standard Model (SSM).
SUSY solves the gauge hierarchy problem of the SM,
if the superpartners of the SM particles are at the weak scale:
the Higgs (mass)$^2$ is 
stabilized at the scale of the SUSY breaking
masses of the superpartners and is negative
due to the large top quark Yukawa coupling.
Therefore, electroweak symmetry breaking occurs
naturally. 
However, we argued that
constraints from phenomenology (the ones we discuss all
come from FCNC's) require that, unless we add some global symmetries
to the SSM, there is some degree of fine tuning/unnaturalness
in some (other) sector of the SSM (in some cases, the fine tuning 
of the weak scale is reintroduced).

We showed that supersymmetric $R$-parity breaking ($\not \! \! R_p$)
interactions always result in Flavor Changing Neutral Current
(FCNC) processes.
Within a single coupling scheme, these processes can be avoided in
either the charge $+2/3$ or the charge $-1/3$ quark sector,
but not both.
These processes were used to place constraints on $\not \!\! R_p\,$
couplings.
The constraints on the first and the second generation couplings are better
than
those existing in the literature. Thus, we have to
either impose $R$-parity
or tolerate some unnaturalness in the form of small values of
the $R$-parity violating couplings.

Non-degenerate squarks and sleptons of especially
the first two generations lead to FCNC's; this is the
SUSY flavor problem.
If SUSY is mediated by gravity (supergravity theories),
then we have to 
either
fine tune the scalar masses 
to give the required degeneracies or 
introduce flavor symmetries or
quark-squark alignment.
Another way of communicating SUSY breaking to the sparticles
is by the SM gauge intercations.
In this case, scalars with the same gauge quantum numbers
are degenerate leading to very small
SUSY contributions to
FCNC's. 

However,
the models of low energy
gauge mediation predict a large hierarchy in the scalar mass spectrum
resulting in a large and negative value for the Higgs soft (mass)$^2$
at the weak scale. This means that
the $\mu$ term has to be fine tuned to give the correct $Z$ mass. We
found that if LEP2 does not discover
SUSY, then these models would lead to a $7 \%$ fine tuning.
We constructed a model with a non-minimal messenger sector (more
messenger $SU(2)_w$ doublets than $SU(3)_c$ triplets)
which reduced the fine tuning to $\sim 40 \%$.
Our model has some extra vector-like quarks (to maintain
gauge coupling unification) which get a mass at the weak scale
from a coupling to a singlet. We used the same singlet to generate
the $\mu$ and $B\mu$ Higgs masses by coupling it
to the Higgs doublets. This model requires
$\sim 25 \%$ fine tuning.
We showed that these models with the split $({\bf 5} + {\bf \bar{5}})$
messenger fields can be derived from a $SU(5) \times SU(5)$
GUT 
using a doublet-triplet splitting mechanism.

The SUSY flavor problem can also be solved by making the scalars
of the first two generations heavy (with mass $M_S
\stackrel{>} \sim
$few TeV). A priori, this does not result in any fine tuning
in EWSB since only the stop mass has to be smaller than
$\sim 1$ TeV to get the weak scale naturally.
However, the heavy scalars drive the light scalar (in particular
the stop) (mass)$^2$ 
negative through two loop
Renormalization Group Equations (RGE), if the scale at which
SUSY breaking is mediated to the sparticles ($M_{SUSY}$)
is high
(say the GUT scale). Thus, the boundary value of the 
stop mass has to be large to avoid negative stop
(mass)$^2$ at the weak scale, in turn, leading to fine tuning
in EWSB. Two new effects were included in our analysis:
the effect of the top quark Yukawa coupling in the RGE 
which makes the constraint on the stop mass stronger 
since it makes the stop (mass)$^2$ more negative,
and the QCD
corrections to the SUSY contributions to $K-\bar{K}$ mixing. 
Even with a 
degeneracy between the
squarks of the first two generations of the order
of the Cabibbo angle, {\it i.e.,} $\sim 0.22$, these squarks
must be heavier than $\sim 40$ TeV
to suppress $\Delta m_K$. This implies, in the
case of a high scale of supersymmetry breaking, that
the boundary value of the stop mass has to be
greater than
$\sim 7$ TeV to keep the stop (mass)$^2$
positive at the weak scale. 

We also studied the case where
$M_{SUSY}$ is of the order of the mass of the heavy scalars.
We computed the finite parts of the same two loop diagrams
and used these as estimates of the two loop contribution of the heavy scalar
masses to the stop (mass)$^2$.
It was found that for
mixing between the squarks of the order of the Cabibbo angle,
the stop mass
at the boundary needs to be larger than $\sim 2$ TeV
to avoid negative stop (mass)$^2$ at the weak scale.
Thus, for both cases, the large boundary value
of the stop masses ($\stackrel{>}{\sim}1$ TeV) reintroduces
fine tuning 
in electroweak symmetry breaking.


\appendix
\addcontentsline{toc}{chapter}{Appendices}

\chapter{Fine Tuning Functions}
\label{appA}  



In this appendix the Barbieri-Giudice fine tuning parameters for
both the MSSM and NMSSM in a gauge mediated 
SUSY breaking scenario are presented.

In an MSSM with gauge mediated SUSY breaking, the
fundamental parameters of the theory (in the 
visible sector) are: $\Lambda_{mess}$,
$\lambda_t$, 
$\mu$, and $\mu^2_3$.
Once electroweak symmetry breaking occurs, the 
extremization conditions determine both $m^2_Z$ and 
$\tan\beta$ as a function of these parameters. To 
measure the sensitivity of $m^2_Z$ to one of the
fundamental parameters $\lambda_i$, 
we compute the variation
in $m^2_Z$ induced by a small change in one of 
the $\lambda_i$. The 
quantity 
\begin{equation}
\frac{\delta m^2_Z}{m^2_Z}\equiv c(m^2_Z;\lambda_i)
\frac{\delta\lambda_i}{\lambda_i},
\end{equation}
where 
\begin{equation}
c(m^2_Z;\lambda_i)=\frac{\lambda_i}{m^2_Z}
\frac{\partial m^2_Z}{\partial\lambda_i},
\end{equation}
measures this sensitivity \cite{barbieri1}. 
In the case of gauge mediated 
SUSY breaking models, there are four functions 
$c(m^2_Z;\lambda_i)$ to be computed. 
They are:

\begin{equation}
c(m^2_Z;\mu^2)=\frac{2\mu^2}{m^2_Z}\left(1+
\frac{\tan^2\beta+1}{(\tan^2\beta-1)^2}\frac{
4\tan^2\beta(\tilde{\mu}^2_1-\tilde{\mu}^2_2)}
{(\tilde{\mu}^2_1-\tilde{\mu}^2_2)(\tan ^2 \beta + 1)
-m^2_Z(\tan^2\beta-1))
}\right),
\end{equation}

\begin{eqnarray}
c(m^2_Z;\mu^2_3)&=&4\tan^2\beta\frac{\tan^2\beta+1}
{(\tan^2\beta-1)^3}\frac{\tilde{\mu}^2_1-\tilde{\mu}^2_2}
{m^2_Z} \nonumber \\
 & \approx & \frac{4}{\tan ^2 \beta}
\frac{\tilde{\mu}^2_1-\tilde{\mu}^2_2}
{m^2_Z}, \; \; \hbox{for large} \; \tan \beta,
\end{eqnarray}

\begin{eqnarray}
c(m^2_Z;\lambda_t)&=& 2 \frac{\lambda _t ^2}{m_Z^2} 
\frac{\partial 
m_Z^2}{\partial m_{H_u}^2}\frac{\partial m_{H_u}^2}
{\partial \lambda _t^2}
\nonumber \\
 & = &\frac{4}{m^2_Z}
 \lambda^2_t \frac{\tan^2\beta}
{\tan^2\beta-1}\frac{\partial m^2_{H_u}}{\partial 
\lambda^2_t}\left(1+2\frac{\tilde{\mu}^2_1-
\tilde{\mu}^2_2}{\tilde{\mu}^2_1+\tilde{\mu}^2_2}
\frac{\tan^2\beta+1}{(\tan^2\beta-1)^2}\right) \nonumber \\
&\approx&\frac{4}{m^2_Z}\frac{\partial m^2_{H_u}}
{\partial \lambda^2_t}, \; \; \; \hbox{for large} \; 
\tan\beta.
\end{eqnarray}
This measures the sensitivity of $m^2_Z$ to the 
electroweak scale value of $\lambda_t$, 
$\lambda_t(m_{weak})$. The Yukawa coupling 
$\lambda_t(m_{weak})$ is not, however, a
 fundamental parameter of the theory. The 
fundamental parameter is the value of the
coupling at the cutoff $\Lambda^0=M_{GUT}$ 
or $M_{Pl}$ of
the theory.
We really 
should be computing the sensitivity of 
$m^2_Z$ to this value of $\lambda_t$. 
The measure of sensitivity is then correctly 
given by
\begin{eqnarray}
c(m^2_Z;\lambda_t(\Lambda^0))&=&
\frac{\lambda_t(\Lambda^0)}{\lambda_t(m_{weak})}
c(m^2_Z;\lambda_t(m_{weak}))
\frac{\partial \lambda_t(m_{weak})}{\partial 
\lambda_t(\Lambda^0)}. 
\end{eqnarray}
We remark that for the model discussed in the 
text with 
three $l+\bar{l}$ and one $q+\bar{q}$ messenger 
fields,
the numerical value of 
$(\lambda_t(\Lambda^0)/\lambda_t(m_{weak}))
\partial \lambda_t(m_{weak})/ 
\partial \lambda_t(\Lambda^0)$ is typically
$\sim 0.1$ because 
$\lambda_t(m_{weak})$ is attracted to its 
infra-red fixed point. This results in a smaller 
value for $c(m^2_Z;\lambda_t)$ than is obtained 
in the absence of these considerations.

With the assumption that $m_{H_u}^2$ and $m_{H_d}^2$ 
scale with 
$\Lambda _{mess} ^2$, we get 
\begin{eqnarray}
c(m_Z^2; \Lambda _{mess} ^2) & = & c(m_Z^2;m_{H_u}^2) 
+ c(m_Z^2;m_{H_d}^2) 
\nonumber \\
 & = & 1 + 2 \frac{\mu ^2}{m_Z^2} 
-\frac{\tan^2\beta+1}{(\tan^2\beta-1)^2} \times 
\nonumber \\
 & & \frac{
4\tan^2\beta (m^2_{H_u}+m^2_{H_d})
(\tilde{\mu}^2_1-\tilde{\mu}^2_2)/m^2_Z}
{(\tilde{\mu}^2_1-\tilde{\mu}^2_2)
(\tan^2\beta+1)-m^2_Z(\tan^2\beta-1)}.
\end{eqnarray}
The Barbieri-Giudice functions for $m_t$ are 
similarly computed. They are

\begin{equation}
c(m_t;\mu^2_3)=\frac{1}{2}c(m^2_Z;\mu^2_3)
+\frac{1}{1-\tan^2\beta},
\end{equation}
\begin{equation}
c(m_t;\mu^2)=\frac{1}{2}c(m^2_Z;\mu^2)
+2\frac{\mu^2}{\tilde{\mu}^2_1+\tilde{\mu}^2_2}
\frac{1}{\tan^2\beta-1},
\end{equation}
\begin{equation}
c(m_t;\lambda_t)=1+\frac{1}{2}c(m^2_Z;\lambda_t)
+\frac{\lambda_t}{\tan^2\beta-1}
\frac{1}{\tilde{\mu}^2_1+\tilde{\mu}^2_2}
\frac{\partial m^2_{H_u}}{\partial \lambda_t},
\end{equation}
\begin{equation}
c(m_t; \Lambda _{mess} ^2) = \frac{1}{2}c(m_Z^2; 
\Lambda _{mess} ^2) 
- \frac{( \tilde{\mu} _1^2 + \tilde{\mu} _2^2 - 
2 \mu ^2 )}
{(1-\tan ^2 \beta) ( \tilde{\mu} _2^2 +\tilde{\mu} _1^2)}.
\end{equation}

Since 
$m_Z$ and $m_t$ are measured, two
of the four fundamental parameters may be eliminated. 
This leaves two free parameters, which for conveinence 
are chosen to be $\Lambda_{mess}$ and $\tan \beta$.

In a NMSSM with gauge mediated SUSY breaking, the 
scalar 
potential for 
$N, \; H_u$ and $H_d$ at the weak scale is specified by the 
following six
parameters: $\lambda _i = m_N^2, m_{H_u}^2, m_{H_d}^2$, 
the $N H_u H_d$ coupling $\lambda _H$,
the scalar $N H_u H_d$ coupling $A_H$, 
and the $N^3$ coupling, $\lambda _N$.
In minimal gauge mediated SUSY breaking, the
trilinear soft SUSY breaking term $NH_uH_d$ is
zero at tree level and is generated at one loop by
wino and bino exchange. In this case,
 $A_H(\lambda_i)=\lambda_H \tilde{A}(\lambda_i)$.
Since the trilinear 
scalar term
$N^3$ is generated at two loops, it is small 
and is 
neglected.
The extremization conditions which determine 
$m^2_Z = g^2_Z v^2/4 \; (v= \sqrt{v_u^2 + v_d^2}), 
\tan \beta =
v_u/v_d$ and $v_N$ as a function of these 
parameters are given 
in section
\ref{NMSSM}. Eqn.(\ref{vn})  can be written, 
using $\mu = \lambda _H v_N/\sqrt{2}$ as
\begin{equation}
m_N^2 + 2 \frac{\lambda _N^2}{\lambda _H^2} \mu ^2 
- \lambda _H \lambda _N \frac{1}{2} 
v^2 \sin 2 \beta + \frac{1}{2} \lambda _H^2 v^2 - 
\frac{1}{4\mu}
A_H v^2 \lambda _H \sin 2 \beta = 0.
\label{NMSSM1A}
\end{equation}
Eqn.(\ref{NMSSM1}) is
\begin{equation}
\frac{1}{8} g_Z^2 v^2 + \mu ^2 - m_{H_u}^2 
\frac{\tan ^2 \beta}{ 1 
- \tan ^2 \beta} 
+ m_{H_d}^2 \frac{1}{1- \tan ^2 \beta} = 0.
\label{NMSSM2A}
\end{equation}
Substituting $v_N^2$ from Eqn.(\ref{vn}) in 
Eqn.(\ref{Bmu}) and 
then using this expression for $\mu _3^2$ in 
Eqn.(\ref{NMSSM2}) gives
\begin{equation}
(m_{H_u}^2 + m_{H_d}^2 + 2 \mu ^2) \sin 2 \beta 
+ \frac{\lambda _H}{\lambda _N} \left( m_N^2 
+ \frac{1}{2} \lambda _H^2 v^2\right) + 
A_H \left(-\frac{2 \mu}{\lambda _H}
- \frac{1}{4} \frac{v^2 \lambda _H ^2 
\sin 2 \beta}{\mu \lambda _N}\right) = 0.
\label{NMSSM3A}
\end{equation}
The quantity $c = (\lambda_i/m^2_Z)
(\partial m^2_Z / \partial\lambda_i) $
measures the sensitivity of $m_Z$ to these
parameters. This can be computed by differentiating 
Eqns.(\ref{NMSSM1A}),
(\ref{NMSSM2A}) and (\ref{NMSSM3A})
with respect to these parameters to obtain, after 
some algebra, 
the following 
set of linear 
equations:
\begin{equation}
(A+A_{A_H}) X^i = B^i+B^i_{A_H},
\end{equation}
where
\begin{eqnarray}
A & = & \left( \begin{array} {ccc}
\frac{1}{2} & 1 & \frac{\mu^2_1 - \mu^2_2}{v^2} 
\frac{2 \tan \beta}
{(1 - \tan ^2 \beta )^2} \\
\frac{\lambda _H^3(\lambda _H -\lambda _N \sin2\beta)}
{g_Z^2 \lambda _N^2} 
& 1 & -\frac{1}{2} \frac{\lambda _H^3}{\lambda _N} 
\frac{1 -\tan ^2 \beta}{(1 +\tan ^2 \beta )^2} \\
\frac{v^2}{g_Z^2 (\mu^2_1+\mu^2_2)}
\frac{\lambda _H^3}{\lambda _N} & 
\frac{\sin2\beta v^2}{\mu^2_1+\mu^2_2} & 
\frac{1 -\tan ^2 \beta}{(1 +\tan ^2 \beta )^2}
\end{array} \right) ,\\
A_{A_H}&=&\frac{A_H}{\mu} \times \\
 & & \left(\begin{array}{ccc}
0&0&0\\
-\frac{\lambda^3_H\sin2\beta}{2 g^2_Z \lambda^2_N} &
\frac{\lambda^3_H\sin2\beta}{16 \lambda^2_N}\frac{v^2}{\mu^2}
& \frac{\tan^2\beta-1}{(1+\tan^2\beta)^2}
\frac{\lambda^3_H}{4 \lambda^2_N} \\
-\frac{\lambda^2_H}{2 g^2_Z\lambda_N}
\frac{v^2\sin2\beta}{\mu^2_1+\mu^2_2}
&\frac{v^2}{\mu^2_1+\mu^2_2}(
\frac{\lambda^2_H\sin2\beta}{16 \lambda_N}\frac{v^2}{\mu^2}
-\frac{1}{2\lambda_H})
&\frac{\tan^2\beta-1}{(1+\tan^2\beta)^2}
\frac{\lambda^2_H}{4\lambda_N}
\frac{v^2}{\mu^2_1+\mu^2_2} 
\end{array} \right) ,\nonumber \\
X^{\lambda_H,\lambda_N} & = & \left( \begin{array} {c}
\frac{1}{v^2} \frac{\partial m^2_Z}{\partial \lambda_i} \\
\frac{1}{v^2} \frac{\partial \mu^2}{\partial \lambda_i} \\
\frac{\partial \tan \beta}{\partial \lambda_i} \\
\end{array} \right) ,\\
X^{m^2_i} & = & \left( \begin{array} {c}
\frac{\partial m^2_Z}{\partial m^2_i} \\
\frac{\partial \mu^2}{\partial m^2_i} \\
v^2\frac{\partial \tan \beta}{\partial m^2_i} \\
\end{array} \right) , (i=u,d,N),
\end{eqnarray}
with $\lambda_i=m^2_N,m^2_{H_u},m^2_{H_d},\lambda_H,
\lambda_N$,
and
\begin{eqnarray}
B^{m^2_N}+B^{m^2_N}_{A_H} & = & \left( \begin{array} {c}
0 \\
- \frac{1}{2} \frac{\lambda _H^2}{\lambda _N^2}\\
-\frac{\lambda _H}{\lambda _N} \frac{v^2}{2(\mu^2_1+
\mu^2_2)}\\
\end{array} \right) ,\\
B^{m^2_{H_u}}+B^{m^2_{H_u}}_{A_H} & = & \left( \begin{array} {c}
\frac{\tan ^2 \beta}{1-\tan^2\beta}\\
0 \\
-v^2 \frac{\sin 2 \beta}{2(\mu^2_1+\mu^2_2)}\\
\end{array} \right) ,\\
B^{m^2_{H_d}}+B^{m^2_{H_d}}_{A_H}  & = & \left( \begin{array} {c}
\frac{1}{\tan^2\beta-1} \\
0 \\
-v^2\frac{\sin 2 \beta}{2(\mu^2_1+\mu^2_2)}\\
\end{array} \right) ,\\
B^{\lambda_H} & = & \left( \begin{array} {c}
0\\
-\frac{\lambda _H^3}{\lambda _N^2}
+\frac{3}{4} \frac{\lambda _H^2\sin 2 \beta}{\lambda _N}  
-\frac{\lambda _H}{\lambda _N^2} \frac{m_N^2}{v^2}\\
-\frac{1}{(\mu^2_1+\mu^2_2)} \left( \frac{1}{2}
\frac{m_N^2}{\lambda _N} + \frac{3}{4} v^2 \frac{
\lambda _H^2}{\lambda _N} \right) \\
\end{array} \right) ,\\
B^{\lambda_H}_{A_H}& = & \frac{A_H}{\mu}
\left( \begin{array} {c}
0\\
\frac{\lambda^2_H\sin2\beta}{2 \lambda^2_N} \\
\frac{3}{8}\frac{\lambda_H}{\lambda_N}
\frac{v^2\sin2\beta}{\mu^2_1+\mu^2_2} \\
\end{array} \right) ,\\
B^{\lambda_N} & = & \left( \begin{array} {c}
0\\
-\frac{1}{4} \frac{\lambda _H^3 \sin 2 \beta}{\lambda _N^2}  
+ \frac{\lambda _H^2}{\lambda _N^3} \frac{m_N^2}{v^2}
+\frac{1}{2} \frac{\lambda _H^4}{\lambda _N^3} \\
\frac{1}{2(\mu^2_1+\mu^2_2)} 
\frac{\lambda _H}{\lambda ^2_N} (m_N^2 + \frac{1}{2} v^2 
\lambda _H^2) 
\end{array} \right),\\
B^{\lambda_N}_{A_H}& = & 
\frac{A_H}{\mu}\left( \begin{array} {c}
0\\
-\frac{\lambda^3_H\sin2\beta}{4 \lambda^3_N} \\
-\frac{\lambda^2_H}{8 \lambda^2_N}\frac{v^2\sin2\beta}
{\mu^2_1+\mu^2_2} \\
\end{array} \right). 
\end{eqnarray}
In deriving
these equations
$A_H(\lambda_i)=\lambda_H \tilde{A}(\lambda_i)$ was assumed 
and $\partial \tilde{A} /\partial \lambda_H$ was neglected.
Inverting these set of equations gives the $c$ functions.
We note
that these expressions for 
the various $c$ functions are valid for
any NMSSM in which the $N^3$ scalar term 
is negligible and the $N H_u H_d$ scalar
term is proportional to $\lambda _H$. 
In general, these 6
parameters might, in turn, depend on some fundamental 
parameters,
$\tilde{\lambda} _i$.
Then, the sensitivity to these fundamental parameters is:
\begin{eqnarray}
\tilde{c_i} & \equiv& \frac{\tilde{\lambda}_i}{m^2_Z}
\frac{\partial m^2_Z}{\partial \tilde{\lambda}_i} 
\nonumber \\
 & = & \frac{\tilde{\lambda}_i}{m^2_Z} \sum_{j} 
\frac{\partial \lambda 
_j}{\partial \tilde{\lambda}_i}
\frac{\partial m^2_Z}{\partial \lambda _j} \nonumber \\
& =& \sum_{j}\frac{\tilde{\lambda}_i}{\lambda_j}
c(m^2_Z;\lambda_j)\frac{\partial \lambda_j}
{\partial \tilde{\lambda}_i} .
\end{eqnarray}
For example, in the NMSSM of section \ref{NMSSM}, 
 the fundamental 
parameters are $\Lambda _{mess}, \lambda _H, \lambda _N, 
\lambda _t$
and $ \lambda _q $ ($A_H$ is a function of $
\lambda _H$ and $\Lambda_{mess}$). 
Fixing $m_Z$ and $m_t$ leaves 3 free parameters, 
which we choose to be
$\Lambda _{mess}, \lambda _H$ and $\tan \beta$.
As explained in that section, the effect of 
$\lambda _H$ in the 
RG scaling of $m_{H_u}^2$ and $m_{H_d}^2$ 
was neglected, whereas the sensitivity of 
$m_N^2$ to 
$\lambda _H$ could be non-negligible.
Thus, we have       
\begin{equation}
\tilde{c}(m_Z^2;\lambda _H) = c(m_Z^2;\lambda _H) + 
c(m_Z^2;m_N^2)
 \frac{\lambda _H}{m_N^2} \frac{\partial m_N^2}
{\partial \lambda _H}.
\end{equation}
We find, in our model, that 
$c(m_Z^2;m_N^2)$ is smaller than $c(m_Z^2;\lambda _H)$ 
by a factor
 of $\sim 2$. Also, using approximate analytic and 
also numerical
 solutions to the RG equation
 for $m_N^2$, we find that $(\lambda _H/m_N^2)
(\partial m_N^2/
\partial \lambda _H)$ is $\ltap \; 0.1$.
Consequently, in the analysis of section \ref{NMSSM}
the additional contribution to $\tilde{c}
(m^2_Z;\lambda_H)$
due to the dependence of $m_N^2$ on $\lambda _H$ was 
neglected. A similar conclusion is true for $\lambda _N$.
Also,
\begin{equation}
\tilde{c}(m_Z^2;\lambda _q) = c(m_Z^2;m_N^2) 
\frac{\lambda _q}{m_N^2} \frac{\partial m_N^2}
{\partial \lambda _q}.
\end{equation}
We find that $(\lambda _q/m_N^2) 
(\partial m_N^2/\partial \lambda _q)$ is $\approx 1$ 
so that $\tilde{c}(m_Z^2;\lambda _q)$ is smaller than
$\tilde{c}(m_Z^2;\lambda _H)$ by a factor of 2.

\chapter{Two Loop Calculation}
\label{appB}



In this appendix we discuss the two loop contribution of the 
heavy scalar soft masses to the
light scalar soft masses. These contributions can be divided into 
two classes. In the first class, a vev for the hypercharge $D$-term 
is generated at two loops. The Feynman diagrams for these contributions 
are given in Figure \ref{mixedtwoloopdiag} and 
are clearly $\sim \alpha_1 \alpha_i$.
These diagrams are computed in a later portion of this appendix. 
In the other class, the two loop diagrams are $\sim \alpha^2_i$. 
These have been
computed by Poppitz and Trivedi \cite{poppitz}. So, we will 
not give details of this
computation which can be found in their paper. However, 
our result for the
finite parts of these diagrams differs slightly from theirs 
and we discuss the
reason for the discrepancy. When one regulates the theory 
using dimensional
reduction \cite{dred,epscalar} (compactifying to $D < 4$ dimensions), the 
vector field decomposes
into a $D$-dimensional vector and $4-D$ scalars, 
called $\epsilon$-scalars, in
the adjoint representation of the gauge group. 
Thus the number of Bose and Fermi degrees of freedom 
in the vector multiplet remain equal.
The $\epsilon$-scalars receive,
at one loop, a divergent contribution to their mass, 
proportional to the
supertrace of the mass matrix of the matter fields. 
Neglecting the fermion masses,
this contribution is
\begin{equation}
\delta m^2_{\epsilon} = - \frac{\alpha}{4 \pi}
\left(\frac{2}{\epsilon}+\ln 4\pi-\gamma\right) (n_5 + 3 n_{10}) M_S^2.
\end{equation}
In our notation $D=4-\epsilon$.
Poppitz and Trivedi choose
the counterterm to cancel this divergence in the 
$MS$ scheme, {\it i.e.}, the
counterterm consists only of the divergent part, 
proportional to $1/\epsilon$.
When this counterterm is inserted in a one loop 
$\epsilon$-scalar graph with SM
fields (scalars) as the external lines, one  
obtains a divergent contribution to the
SM scalar soft masses (the $1/\epsilon$ of the counterterm 
is cancelled after
summing over the $\epsilon$ adjoint scalars running 
in the loop). Poppitz and
Trivedi use a cut-off, $\Lambda _{UV}$, to regulate this 
graph, giving a
contribution from this graph that is:
\begin{equation}
m^2_i = - \sum_{A} (n_5 + 3 n_{10})
C^i_A \frac{1}{16} \left(\frac{\alpha _A}{\pi}\right)^2
M_S^2 \hbox{ln} \Lambda ^2 _{UV},
\end{equation}
with no finite part. We, on the other hand, 
choose the $\epsilon$-scalar mass
counterterm in the $\overline{MS}$ scheme, {\it i.e.}, 
proportional to $2/\epsilon -
\gamma + \hbox{ln}4 \pi$ (where $\gamma \approx 0.58$ 
is the Euler constant) and
use dimensional reduction to regulate the graph with the 
insertion of the
counterterm. This gives a contribution
\begin{eqnarray}
m^2_i & = & - \sum_{A} (n_5 + 3 n_{10})
C^i_A \frac{1}{16} \left(\frac{\alpha _A}{\pi}\right)^2
M_S^2 \left(\frac{2}{\epsilon} -
\gamma + \hbox{ln}4 \pi\right)^2 \epsilon \nonumber \\
 & = & - \sum _{A} (n_5 + 3 n_{10})
C^i_A \frac{1}{8} \left(\frac{\alpha _A}{\pi}\right)^2
M_S^2 \left(2/\epsilon -
2 \gamma + 2 \;\hbox{ln}4 \pi\right).
\end{eqnarray}
In the first line the 
first factor of $(2/\epsilon -
\gamma + \hbox{ln}4 \pi)$ is from the counter-term 
insertion, the second factor 
is the result of the loop integral, and 
the over-all factor of $\epsilon$ counts the number of 
$\epsilon$-scalars running in the loop. 
In the $\overline{MS}$ scheme, {\it i.e.}, after
subtracting $2/\epsilon -
\gamma + \hbox{ln}4 \pi$, we are left with a finite 
part\footnote{The same finite part is obtained in the $MS$ 
scheme, regulated with ${DR}^{\prime}$ .} proportional to $-
\gamma + \hbox{ln}4 \pi$. The remaining diagrams together 
give a finite result
and we agree with Poppitz and
Trivedi on this computation. Our result for the finite part
of the two loop diagrams (neglecting the fermion masses) is
\begin{eqnarray}
m^2_{i,finite} (\mu) & = & -\frac{1}{8} 
\left( \hbox{ln} (4 \pi) - \gamma +
\frac{\pi^2}{3} - 2 - \hbox{ln} \left(\frac{M^2_S}{\mu ^2}\right) \right) 
\nonumber \\
 & & \times \sum_{A} \left( \frac{\alpha _A(\mu)}
{\pi} \right) ^2 (n_5 +
 3 n_{10}) C^i_A M^2_S,
\end{eqnarray}
whereas the Poppitz-Trivedi result does not have 
the $\hbox{ln} (4 \pi) -
\gamma$ in the above result. This result was 
used in Eqn.(\ref{finite}). 
The computation of the 
two loop hypercharge
$D$-term, which gives contribution to the soft scalar 
(mass)$^2$ proportional
to $\alpha _1 \alpha _s$ and $\alpha _1 \alpha _2$ 
({\it i.e.}, the ``mixed''
two loop contributon) is discussed below in detail.



{\it Two-loop hypercharge D-term}

\begin{figure}
\vspace{-0.6in}
\centerline{\epsfxsize=0.6\textwidth \epsfbox{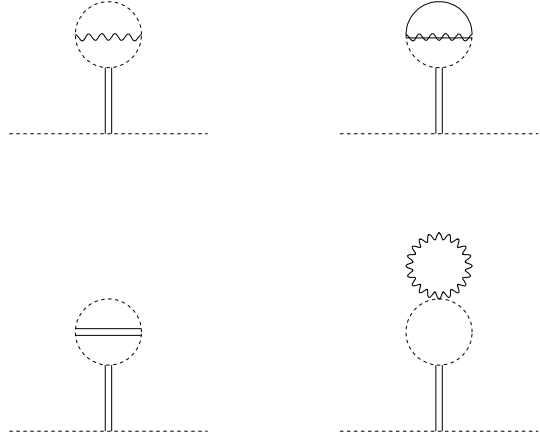}}
\vspace{-0.7in}
\caption{Mixed two loop corrections to the scalar mass. Wavy lines, 
wavy lines with a straight line
through them, solid lines, and dashed lines denote gauge boson, 
gaugino, fermion and scalar propagators, respectively. 
The double line denotes the
hypercharge $D$-term propagator.}
\protect\label{mixedtwoloopdiag}
\vspace{.25in}
\end{figure}

We compute the two loop diagrams of
Figure \ref{mixedtwoloopdiag} in the Feynman gauge and set all 
fermion and gaugino masses to zero. It is 
convienent to calculate in this gauge 
because both the
scalar self-energy and the $D_Y$-term
vertex corrections are finite at one loop 
and thus require no 
counter-terms. We have also computed the
two loop diagrams in the Landau gauge and have found that 
the result agrees with the calculation in the Feynman gauge. 
The calculation in the Landau gauge 
requires counter-terms 
and is more involved, and hence
the discussion is not included. 
Finally, 
in the calculation a global $SU(5)$ symmetry is 
assumed so that a hypercharge $D$-term is not 
generated at one loop \cite{nelson2,dimopoulos}. 

The sum of the four Feynman diagrams in Figure \ref{mixedtwoloopdiag} 
is given 
in the Feynman gauge by
\begin{equation}
-i \tilde{\Pi}_{D,f}=i \frac{3}{5}
g^2_1 Y_f \sum_i Y_i\sum_{A} 
g_A^2 C_A^i\left(4 I_1(m^2_i)-4I_2(m^2_i)+I_3(m^2_i)\right),
\end{equation}
where the sum is over the gauge and flavour states of the 
particles in the loops. If the particles in the loop 
form complete ${\bf \bar{5}}$ and ${\bf 10}$ representations with 
a common mass $M_S$, the 
sum simplifies to
\begin{eqnarray}
-i \tilde{\Pi}_{D,f} & = &i \frac{3}{5} \; 16 \pi^2
\alpha_1 Y_f (n_5-n_{10})\left(\frac{4}{3} \alpha_3-\frac{3}{4} \alpha_2
-\frac{1}{12} \alpha_1\right) \nonumber \\
 & & \times \left(4 I_1(M^2_S)-4I_2(M^2_S)+I_3(M^2_S)\right).
\label{appeqn}
\end{eqnarray}    

The functions $I_1$, $I_2$ and $I_3$
are
\begin{eqnarray}
I_1(m^2) & = & \int \frac{d^D p}{(2\pi)^D} 
\int \frac{d^D k}{(2\pi)^D}\frac{1}{(p^2-m^2)^2}
\frac{(2p-k)^2}{k^2}\frac{1}{(p-k)^2-m^2}, \\
I_2(m^2) & = & \int \frac{d^D p}{(2\pi)^D}
\int \frac{d^D k}{(2\pi)^D}\frac{1}{(p^2-m^2)^2}
\frac{k^2-k\cdot p}{k^2}\frac{1}{(p-k)^2}, \\
I_3(m^2) & = & \int \frac{d^D k}{(2\pi)^D}
\frac{1}{(k^2-m^2)^2}
\int \frac{d^D q}{(2\pi)^D}\frac{1}{q^2-m^2}.
\end{eqnarray}

We now compute these functions.

{\it \underline{Evaluating $I_1$}}

After a Feynman parameterization and performing a
change of variables, $I_1=J_1+J_2$, where
\begin{equation}
J_1(m^2)=\Gamma(3)\int_{0}^{1} dx (1-x) \int \frac{d^D p}{(2\pi)^D}
\int \frac{d^D k}{(2\pi)^D} 4 \frac{p^2}{k^2}
\frac{1}{(p^2-(m^2-x(1-x)k^2))^3}
\end{equation}
and 
\begin{equation}
J_2(m^2)=\Gamma(3)\int_{0}^{1} dx (1-x) (2x-1)^2
 \int \frac{d^D p}{(2\pi)^D}
\int \frac{d^D k}{(2\pi)^D} 
\frac{1}{(p^2-(m^2-x(1-x)k^2))^3}.
\end{equation}


After some algebra we find that 
\begin{equation}
J_1(m^2)=\frac{\Gamma(3-D)}{(4\pi)^D}(m^2)^{D-3}
\frac{2 D}{D/2-1} B(2-D/2,3-D/2),
\end{equation}
%
%
\begin{eqnarray}
J_2(m^2)
& = & \frac{\Gamma(3-D)}{(4\pi)^D}(m^2)^{D-3}
\times
\left(4 B(3-D/2,2-D/2)-4 B(2-D/2,2-D/2) \right.\nonumber \\ 
 & & +\left.B(1-D/2,2-D/2)\right), 
\end{eqnarray}
where $B(p,q)=\Gamma[p] \Gamma[q]/ \Gamma[p+q]$ is the usual 
Beta function.

Combining these two results gives
\begin{equation}
I_1(m^2)
=\frac{\Gamma(3-D)}{(4\pi)^D} (m^2)^{D-3}
\frac{1-D}{D-2}B(3-D/2,2-D/2).
\end{equation}

{\it \underline{Evaluating $I_2$}}

\begin{eqnarray*}
I_2(m^2)&=&\int \frac{d^D p}{(2\pi)^D}
\int \frac{d^D k}{(2\pi)^D}\frac{1}{(p^2-m^2)^2}
\frac{k^2-k\cdot p}{k^2}\frac{1}{(p-k)^2} \\
& = & \frac{1}{(4 \pi)^D} 
\Gamma(3-D) (m^2)^{D-3} B(D/2,1-D/2).
\end{eqnarray*}

{\it \underline{Evaluating $I_3$}}

\begin{eqnarray*}
I_3(m^2) & = & \int \frac{d^D k}{(2\pi)^D} 
\frac{1}{(k^2-m^2)^2}
\int \frac{d^D q}{(2\pi)^D}\frac{1}{q^2-m^2} 
 \\
& = & \left(\frac{i}{(4\pi)^{D/2}} \Gamma(2-D/2) 
(m^2)^{D/2-2}\right)
\left(\frac{i}{(4\pi)^{D/2}}\frac{\Gamma(2-D/2)}
{D/2-1} (m^2)^{D/2-1}\right) \\
& = & -\frac{1}{(4\pi)^D}(\Gamma(2-D/2))^2 
\frac{1}{D/2-1}(m^2)^{D-3}.
\end{eqnarray*}

We may now combine $I_1$, $I_2$ and $I_3$ to obtain
\begin{eqnarray*}
T(m^2) & \equiv &
4I_1(m^2)-4I_2(m^2)+I_3(m^2) \\
& = & \frac{(m^2)^{D-3}}{(4 \pi)^D} \times \left(
4\left(\frac{1-D}{D-2}B(3-D/2,2-D/2)-B(D/2,1-D/2)\right) \right.\nonumber \\ 
 & & \left.\times \Gamma(3-D) 
 -\frac{1}{D/2-1} \Gamma(2-D/2)^2 \right).
\end{eqnarray*}
Writing $D=4-\epsilon$ and expanding in $\epsilon$ gives 
\begin{equation}
T(m^2)=\frac{1}{(16 \pi^2)^2}\left(\frac{4}{\epsilon}
+\left(6-\frac{2}{3} \pi^2+4(\ln (4 \pi)-\gamma)-4 \ln m^2\right) m^2
+O(\epsilon)\right).
\end{equation}
In the $\overline{MS}$ scheme the combination 
$2 \left(2/\epsilon+\ln(4 \pi)-\gamma\right)$ is subtracted 
out. The finite piece that remains is 
\begin{equation}
\frac{1}{(16 \pi^2)^2}
\left(6-\frac{2}{3} \pi^2+2(\ln (4 \pi)-\gamma)-4 \ln m^2\right) m^2.
\end{equation}
Thus in the $\overline{MS}$ scheme Eqn.(\ref{appeqn}) is 
\begin{eqnarray}
-i\tilde{\Pi}_{D,f} & = & i \frac{3}{5}\frac{1}{(16 \pi^2)}
 \alpha_1 Y_f  
\left(\frac{4}{3} \alpha_3-\frac{3}{4} \alpha_2
-\frac{1}{12} \alpha_1\right) (n_5 - n_{10})\nonumber \\
 & & \times \left(6-\frac{2}{3} \pi^2
+2(\ln (4 \pi)-\gamma)-4 \ln M_S^2\right) M_S^2,
\end{eqnarray}
which was used in Eqn.(\ref{finite}).


\begin{thebibliography}{99}

\bibitem{smref}S. L. Glashow, {\it Nucl. Phys.} {\bf 22} (1961) 579; 
S. Weinberg, {\it Phys. Rev. Lett.} {\bf 19} (1967) 1264; A. Salam in
{\it Elementary Particle Theory}, ed. N. Svartholm 
(Almqvist and Wiksell, Stockholm, 1968) p. 367.
\bibitem{GIM}S. L. Glashow, J. Iliopoulos, L. Maiani, {\it Phys. Rev.}
{\bf D2} (1970) 1285. 
\bibitem{QCD}D. Gross, F. Wilczek, {\it Phys. Rev. Lett.} {\bf 30}
(1973) 1343; H. D. Politzer, {\it Phys. Rev. Lett.} {\bf 30} (1973) 1346.
\bibitem{pdg}{\it Review of Particle Physics},
{\it Phys. Rev.} {\bf D54} (1996) 1.
\bibitem{kobayashi}
N. Cabibbo, {\it Phys. Rev. Lett.} {\bf 10} (1963) 531;
M. Kobayashi,
T. Maskawa, {\it Prog. Theor. Phys.} {\bf 49} (1973) 652.  
\bibitem{splitting}H. Georgi, S.L. Glashow,
{\it Phys. Rev. Lett.}
{\bf 32} (1974) 438.  
\bibitem{quinn}H. Georgi, H. Quinn, S. Weinberg, {\it Phys. Rev. Lett.}
 {\bf 33} 
(1974) 451.
\bibitem{langacker}See, for example, J. Ellis,
S. Kelley, 
D. V. Nanopoulos, {\it Nucl. Phys.} {\bf B373} (1992) 55;
P. Langacker, N. Polonsky,
{\it Phys. Rev.} {\bf D47} (1993) 4028. 
\bibitem{susskind}See, for example, L. Susskind,
{\it Phys. Rev.} {\bf
D20} (1979) 2619. 

\bibitem{nilles}For reviews of supersymmetry and 
supersymmetry phenomenology, see: \\
P. Fayet, S. Ferrara, {\it Phys. Rep.} {\bf 5} (1977) 249;
H.P. Nilles, {\it Phys. Rep.} {\bf 110} (1984) 1; 
M.F. Sohnius, {\it Phys. Rep.} {\bf 128} (1985) 2; 
I. Hinchliffe, {\it Ann. Rev. Nucl. Part. Sci.} {\bf 36} (1986) 505.
\bibitem{nonrenorm}  B. Zumino,
{\it Nucl. Phys.} {\bf B89} (1975)
535; P. West,
{\it Nucl. Phys.} {\bf B106} (1976) 219;
M. Grisaru, W. Siegel, M. Ro\v{c}ek,
{\it Nucl. Phys.} {\bf B159} (1979)
429.
\bibitem{ross} L. Ibanez, G. G. Ross, {\it Phys. Lett.} {\bf B110}
(1982) 215; K. Inoue, A. Kakuto, H. Komatsu, S. Takeshita,
{\it Prog. Theor. Phys. } {\bf 68} (1982) 927.
\bibitem{wise}L. Alvarez-Gaum\'e,
M. Claudson, M. Wise, {\it Nucl. Phys. }{\bf B207} (1982) 96.  
\bibitem{barbieri1}R. Barbieri, G. Giudice,
{\it Nucl. Phys.}
{\bf B306} (1988) 63.
\bibitem{anderson}G. Anderson, D. Casta$\tilde{n}$o,
{\it Phys. Lett.}
 {\bf B347} (1995) 300.

\bibitem{georgi}S. Dimopoulos, H. Georgi, {\it
Nucl. Phys.} {\bf B193} (1981) 150.
\bibitem{affleck}I. Affleck, M. Dine, N. Seiberg, 
{\it Nucl. Phys.} {\bf B256} (1985) 557.
\bibitem{lykken}L. Hall, J. Lykken, S. Weinberg,
{\it Phys.
 Rev.} {\bf D27} (1983) 2359.
\bibitem{gm} M. Dine, W. Fischler, M. Srednicki,
{\it Nucl. Phys. } {\bf B189} (1981) 575; C. Nappi, B. Ovrut,
{\it Phys. Lett.} {\bf B113} (1982) 175; M. Dine, W. Fischler,
{\it Nucl. Phys.} {\bf B204} (1982) 346; \cite{wise}.

\bibitem{suzuki}L. J. Hall, 
M. Suzuki, {\it Nucl. Phys.} {\bf B231} (1984) 419.
\bibitem{hall}S. Dimopoulos, 
L. J. Hall, {\it Phys. Lett.} {\bf B207} (1987) 210.
\bibitem{barger}V. Barger, G.F. Giudice, T. Han, {\it Phys. Rev.}
{\bf D40} (1989) 2987.
\bibitem{godbole}R. Godbole, P. Roy, X. Tata,
{\it Nucl. Phys.} {\bf B401} (1993) 67.
\bibitem{dawson}S. Dawson, {\it Nucl. Phys.} {\bf B261} (1985) 297.
\bibitem{choudhury} G. Bhattacharyya,
D. Choudhury, {\it Mod. Phys. Lett.} {\bf A10} (1995) 1699.
\bibitem{ellis} G. Bhattacharyya, J. Ellis,
K. Sridhar, {\it Mod. Phys. Lett.} {\bf A10} (1995) 1583.
\bibitem{mohapatra} R. Mohapatra,
{\it Phys. Rev.} {\bf D34} (1986) 3457.

\bibitem{gabbiani}F. Gabbiani, A. Masiero,
{\it Nucl. Phys.} {\bf B322}
(1989) 235; J. S. Hagelin, S. Kelley, T. Tanaka,
{\it Nucl. Phys.}
{\bf B415} (1994) 293; F. Gabbiani, E. Gabrielli,
A. Masiero,
L. Silvestrini, {\it Nucl. Phys.} {\bf B477}
(1996) 321. 
\bibitem{flavor} M. Dine, R. Leigh, A. Kagan, {\it Phys. Rev.}
{\bf D48} (1993) 4269; P. Pouliot, N. Seiberg, {\it Phys. Lett.}
{\bf B318} (1993) 169; D. B. Kaplan, M. Schmaltz, {\it Phys. Rev.}
{\bf D49} (1994) 3741; 
L. J. Hall, H. Murayama, {\it Phys. Rev. Lett.} {\bf 75}
(1995) 3985; N. Arkani-Hamed, H.-C. Cheng, L. J. Hall, {\it Phys. Rev.
}{\bf D54} (1996) 2242; R. Barbieri, L. J. Hall, {\it Nuovo Cim.}
{\bf 110A} (1997) 1; R. Barbieri, L. J. Hall, S. Raby, A. Romanino,
{\it Nucl. Phys.} {\bf B493} (1997) 3.
\bibitem{seiberg} Y. Nir, N. Seiberg, {\it Phys. Lett.}
{\bf B309} (1993) 337.
\bibitem{nirshirman}M. Dine, A. Nelson, Y. Nir,
Y. Shirman, {\it Phys. Rev.} {\bf D53} (1996) 2658.
\bibitem{arkani}N. Arkani-Hamed,
C. D. Carone, L. J. Hall, H. Murayama,
{\it Phys. Rev.} {\bf D54} (1996) 7032.
\bibitem{strumia}P. Ciafaloni, A. Strumia,
{\it Nucl. Phys.} {\bf B494} (1997) 41;
G. Bhattacharyya,
A. Romanino, {\it Phys. Rev.} {\bf D55} (1997) 7015.
\bibitem{alex}A. de Gouv\^ea, A. Friedland, H. Murayama, 
{\it Phys. Rev.} {\bf D57} (1998) 5676.
\bibitem{dine} M. Dine, A. Kagan, S. Samuel, {\it Phys. Lett.}
{\bf B243} (1990) 250.
\bibitem{pomoral} A. Pomoral, D. Tommansini, {\it Nucl. Phys.} 
{\bf B466} (1996) 3.
\bibitem{dvali}
G. Dvali, A. Pomarol, {\it Phys. Rev. Lett.} {\bf 77} 
(1996) 3728; G. Dvali, A. Pomarol,  hep-ph/9708364.
\bibitem{nelson2} A.G. Cohen, D.B. Kaplan,
A. Nelson, {\it Phys. Lett.} {\bf B388} (1996) 588.
\bibitem{nelson3}
S. Ambrosanio, A. Nelson, {\it Phys. Lett.} {\bf B411} (1997) 283.
\bibitem{nelson} A. Nelson, D. Wright, {\it Phys. Rev.} {\bf D56}
(1997) 1598.
\bibitem{riotto}R. N. Mohapatra, A. Riotto, {\it Phys. Rev.}
{\bf D55} (1997) 4262.
\bibitem{dimopoulos} S. Dimopoulos, G.F. Giudice,
{\it Phys. Lett.} {\bf B357} (1995) 573.
\bibitem{nima} N. Arkani-Hamed, H. Murayama,
{\it Phys. Rev.} {\bf D56} (1997) 6733.
\bibitem{bagger} J. Bagger, K. T. Matchev, R. Zhang,
{\it Phys. Lett.} {\bf B412} (1997) 77.      

\bibitem{barbieri} R. Barbieri, A. Masiero, {\it Nucl. Phys.} 
{\bf B267} (1986) 679.
\bibitem{leurer}
M. Leurer, {\it Phys. Rev. Lett.} {\bf 71} (1993) 1324.
\bibitem{lee}M. K. Gaillard, B. W. Lee, {\it Phys. Rev.} 
{\bf D10} (1974) 897.
\bibitem{lattice}J. Shigemitsu in {\it Proceedings of the
XXVII International Conference on High Energy
Physics}, Glasgow, Scotland, U.K., July 1994, eds.
P. J. Bussey and I. G. Knowles (Institute of Physics Publishing,
Bristol and Philadelphia, 1995).
\bibitem{bernard} C. Bernard,
{\it Nucl. Phys. (Proc. Suppl.)} {\bf B34} (1994) 47;
S. Sharpe, Lectures given at the Theoretical Advanced
Study Institute in Particle
Physics (TASI 94), Boulder, Colorado, U.S.A., 29 May - 24 Jun 1994.      
\bibitem{pdg1}{\it Review of Particle Properties},
{\it Phys. Rev.} {\bf D50} (1994) 1177.         
\bibitem{argus}
C. Albajar {\it et al.}, {\it Phys. Lett.} {\bf B186} (1987) 247; 
H. Albrecht {\it et al.}, {\it Phys. Lett.} {\bf B197} (1987) 452.
\bibitem{inami}T. Inami, C. S. Lim, {\it Prog. Theor. Phys.} 
{\bf 65} (1981) 292.
\bibitem{martinelli}G. Martinelli in {\it Proceedings of the
6th Rencontres De Bois}, Bois, France, 20 - 25 Jun 1994.     
\bibitem{atiya}M. S. Atiya {\it et al.}, 
{\it Phys. Rev. Lett.} {\bf 70} (1993) 2521; 
(erratum) M. S. Atiya 
{\it et al.}, {\it Phys. Rev. Lett.} {\bf 71} (1993) 305.
\bibitem{grossman} 
Y. Grossman, Z. Ligeti, E. Nardi, {\it Nucl.Phys.} {\bf B465} (1996) 369.

\bibitem{fayet} P. Fayet, {\it Nucl. Phys.} {\bf B90}
(1975) 104. 
\bibitem{barbieri2}R. Barbieri, G. Dvali, A. Strumia,
{\it Phys. Lett.}
 {\bf B333} (1994) 79.
\bibitem{dns}M. Dine, A. Nelson, Y. Shirman,
{\it Phys. Rev.} {\bf D51} (1995) 1362.
\bibitem{randall}I. Dasgupta, B. A. Dobrescu, 
L. Randall,
{\it Nucl. Phys.} {\bf B483} (1997) 95. 
\bibitem{unpublished}S. Dimopoulos, G. Giudice, 
A. Pomarol, {\it Phys. Lett.} {\bf B389} (1996) 37;
S. Martin,
{\it Phys. Rev.} {\bf D55} (1997) 3177.
\bibitem{thomas}See for example,
S. Dimopoulos, S. Thomas, J. Wells, {\it Nucl. Phys.} {\bf B488} 
(1997) 39 
and references
therein.
\bibitem{aleph}Talk presented by Glen Cowan (ALEPH 
collaboration) at
the special CERN particle physics seminar on physics 
results from the LEP
run at 172 GeV, 25 February, 1997.
\bibitem{cerngroup2}G. F. Giudice, M. L. Mangano, 
G. Ridolfi,
R. Ruckel ({\it convenors}), {\it Searches for 
New Physics},
hep-ph/9602207.
\bibitem{barr}S. M. Barr, {\it Phys. Rev.} {\bf D55} (1997) 6775.
\bibitem{wilczek}F. Wilczek, {\it Phys. Rev. Lett.}
{\bf 40} (1978) 279.
\bibitem{cerngroup}M. Carena, P. W. Zerwas 
({\it convenors}),
{\it Higgs Physics}, CERN Yellow Report CERN 96-01, 
hep-ph/9602250.
\bibitem{delphi}W. Adam {\it et al}, DELPHI collaboration,
CERN-PPE/96-119.
\bibitem{srednicki} H. P. Nilles, M. Srednicki, D. Wyler,
{\it Phys. Lett.} {\bf B124} (1983) 337; A. B. Lahanas,
{\it Phys. Lett.} {\bf B124} (1983) 341.

\bibitem{gaillard} M. K. Gaillard, B. W. Lee,
{\it Phys. Rev. Lett.} {\bf 33} (1974) 108.  
\bibitem{dred} W. Siegel, {\it Phys. Lett.} {\bf B84} (1979) 193.
\bibitem{epscalar} D. M. Capper, D. R. T. Jones,
P. van Nieuwenhuizen, {\it Nucl. Phys.} {\bf B167} (1980)
479.
\bibitem{drbarp} I. Jack, D. R. T. Jones, S. Martin,
M. Vaughn, Y. Yamada,  {\it Phys. Rev.} {\bf D50} (1994) 5481.
\bibitem{twolooprge} S. Martin, M. Vaughn,
{\it Phys. Rev.} {\bf D50} (1994) 2282. 
\bibitem{LEP} The Aleph Collaboration, pre-prints 856, EPS 619, 622,
submitted to the 1997 EPS-HEP Jerusalem Conference, 19-26 Aug.
\bibitem{gmfinetune} 
\cite{strumia}; 
chapter 3.
\bibitem{poppitz} E. Poppitz, S. Trivedi, {\it Phys. Lett.}
{\bf B401} (1997) 38.
\end{thebibliography}
\end{document}